\documentclass[fleqn,usenatbib]{mnras}
\usepackage{lmodern}

\usepackage{newtxtext,newtxmath}

\usepackage[T1]{fontenc}

\DeclareRobustCommand{\VAN}[3]{#2}
\let\VANthebibliography\thebibliography
\def\thebibliography{\DeclareRobustCommand{\VAN}[3]{##3}\VANthebibliography}

\usepackage{graphicx}    
\usepackage{amsmath}    
\usepackage{xcolor}    

\title[Learning the Coma Cluster]{Learning the Universe: Constrained simulations of the Coma galaxy cluster --  I. Radial X-ray and Compton-$y$ signatures}%
\author[U. P. Steinwandel et al.]{%
Ulrich P. Steinwandel,$^{1}$\thanks{E-mail: uli@mpa-garching.mpg.de}
Stuart McAlpine,$^{2}$
Richard Stiskalek,$^{3}$
R\"udiger Pakmor,$^{1}$
Volker Springel,$^{1}$
\newauthor%
Eugene Churazov,$^{1}$
Ildar Khabibullin,$^{4,1}$
Jens Jasche,$^{2}$
Guilhem Lavaux,$^{5}$
and Greg L. Bryan$^{6}$
\vspace*{0.1cm}\\%
$^{1}$Max Planck Institut f\"ur Astrophysik, Karl-Schwarzschildstr.~1, D-85748 Garching, Germany\\%
$^{2}$The Oskar Klein Centre, Department of Physics, Stockholm University, Albanova University Center, 106 91 Stockholm, Sweden\\%
$^{3}$Department of Physics, University of Oxford, Denys Wilkinson Building, Keble Road, Oxford, OX1 3RH, UK\\%
$^{4}$Rudolf Peierls Centre for Theoretical Physics, Department of Physics, University of Oxford, Clarendon Laboratory, Parks Rd, Oxford, OX1 3PU, United Kingdom\\%
$^{5}$CNRS \& Sorbonne Université, Institut d’Astrophysique de Paris (IAP), UMR 7095, 98 bis bd Arago, F-75014 Paris, France\\%
$^{6}$Department of Astronomy, Columbia University, 538 West 120 Street, New York, NY 10027, USA
}

\date{Accepted XXX. Received YYY; in original form ZZZ}

\pubyear{\the\year{}}

\begin{document}
\label{firstpage}
\pagerange{\pageref{firstpage}--\pageref{lastpage}}
\maketitle

\begin{abstract}
We present a suite of 50 high-fidelity simulations of Coma cluster analogues constructed from \texttt{BORG/Manticore} constrained initial conditions and evolved with the IllustrisTNG galaxy formation model. Regions predicted to form massive clusters comparable to Coma in mass and environment are selected and followed through cosmic time, producing realistic galaxy populations and intracluster medium properties. The ensemble captures both cosmic variance and uncertainties in the local initial conditions, providing a statistically robust framework for interpreting Coma in a cosmological context.
We focus on direct comparisons with observed thermodynamical profiles of the intracluster medium. Specifically, we extract X-ray surface brightness profiles from the simulated clusters and confront them with measurements from eROSITA, as well as compute the thermal Sunyaev--Zel'dovich effect via integrated Compton-$y$ profiles for comparison with Planck satellite data. The simulations reproduce the broad shape and normalisation of both observables, while also highlighting the range of scatter expected from environmental and assembly history differences. This enables us to assess how feedback processes, merger activity, and large-scale environment shape observable cluster properties.
Our results demonstrate that combining constrained cosmological initial conditions with state-of-the-art galaxy formation physics provides an effective strategy for generating targeted, observation-driven analogues of specific clusters. The resulting dataset offers a valuable resource for testing models of intracluster medium physics, calibrating scaling relations, and interpreting upcoming joint X-ray and Sunyaev--Zel'dovich observations of nearby massive clusters.
\end{abstract}




\section{Introduction}

Numerical simulations have become an essential tool in our understanding of cosmic structure formation and galaxy evolution. Over the past two decades, the field has progressed from pioneering dark matter–only simulations to fully hydrodynamical galaxy formation models that reproduce a wide range of observed galaxy and cluster properties. Early large-volume dark matter–only efforts such as the Millennium Simulation introduced by \citet{Springel2005} demonstrated the power of cosmological $N$-body techniques to trace the growth of structure in a $\Lambda$CDM universe. These simulations provided the first statistically significant ensembles of dark matter haloes, enabling studies of halo mass functions, clustering, and merger histories.
The Aquarius Project \citep{Springel2008} extended this work through high-resolution zoom-in simulations of individual dark matter haloes, enabling detailed investigations of substructure, tidal stripping, and the fine-grained assembly history of Milky Way–like haloes. Subsequent refinements, such as Millennium-II \citep{BoylanKolchin2009}, improved mass resolution over smaller volumes, allowing for more accurate modeling of small-scale structure and serving as benchmarks for hierarchical galaxy formation studies. Collectively, these early simulations established the computational and methodological foundations for the next generation of galaxy formation modeling.
The next major leap came with the incorporation of baryonic physics into cosmological simulations. First-generation hydrodynamical galaxy formation models such as OWLS and C-OWLS \citep[e.g.,][]{Schaye2010,LeBrun2014}, EAGLE \citep{Schaye2015}, Magneticum \citep{Hirschmann2014,Dolag2025}, and Illustris \citep{Vogelsberger2014,Vogelsberger2015,Genel2014,Sijacki2015} introduced sophisticated treatments of star formation, feedback, metal enrichment, and AGN activity. These simulations were able to self-consistently evolve gas, stars, and black holes within the cosmological context, providing predictions for galaxy stellar masses, star formation rates, morphologies, and the chemical composition of the interstellar medium (ISM) and circumgalactic medium (CGM).

By calibrating key subgrid parameters against observational datasets, these models achieved agreement with important galaxy scaling relations such as the stellar mass–halo mass relation, the galaxy size–mass relation, and the mass-metallicity relation, establishing a benchmark for subsequent studies.
The IllustrisTNG suite \citep[TNG;][]{Springel2018, Weinberger2017, Nelson2018, Pillepich2018, Nelson2019, Pillepich2019} refined these models with improved AGN feedback, magnetohydrodynamics, and numerical methods, addressing several of the limitations identified in Illustris. The TNG model introduced physically motivated galactic winds, refined black hole growth prescriptions, and updated chemical enrichment schemes, leading to more realistic galaxy morphologies, baryon fractions, and cluster-scale gas properties.

More recent efforts have expanded this approach to larger cosmological volumes, as exemplified by MillenniumTNG (MTNG; \citealp{Pakmor2023}), enabling the simultaneous study of galaxy evolution across a wide dynamic range from dwarf galaxies to massive clusters. In parallel, complementary simulation programs—such as Hydrangea, a cluster zoom-in extension of the EAGLE model \citep{Bahe2017}, and the MassiveBlack-I/II \citep{DiMatteo2012,Khandai2015}, BlueTides \citep{Feng2016}, and Astrid \citep{Bird2022,Ni2022} simulations—have explored different regions of parameter space, focusing on aspects such as high-redshift galaxy formation, quasar evolution, reionization, and environmental effects on galaxy populations. These parallel efforts collectively provide a rich foundation for interpreting observations of galaxies across cosmic time.

Such large-volume cosmological boxes, however, sample a limited volume and therefore contain only a handful of the rarest, most massive systems, leaving them not ideally suited to X-ray and Compton-$y$ studies of individual massive galaxy
clusters. This motivated a long line of targeted cluster simulations aimed at predicting these observables at high fidelity: early grid- and SPH-based calculations established quantitative predictions for cluster X-ray emission
\citep[e.g.,][]{Evrard1996, BryanNorman1998, Frenk1999, Borgani2004, Nagai2007} and for the thermal Sunyaev--Zel'dovich / Compton-$y$ signal
\citep[e.g.,][]{daSilva2000, Springel2001, Battaglia2012}; see
\citet{KravtsovBorgani2012} for a review. The TNG-Cluster project \citep{Nelson2024} threads this needle by applying the full TNG galaxy-formation model to a large, targeted sample of massive clusters, enabling high-fidelity, object-by-object predictions of their X-ray and SZ structure. Beyond the bulk
thermodynamics, however, the ICM is a weakly collisional, magnetized plasma whose state is shaped by non-thermal processes: magnetic fields and turbulent dynamo
amplification \citep[e.g.,][]{Dolag2002, Vazza2018, Steinwandel2022, Steinwandel2024}, thermal conduction \citep{Talbot2025}, anisotropic (Braginskii)
viscosity \citep[e.g.,][]{Berlok2020, MarinGilabert2025}, and cosmic rays \citep[e.g.,][]{Pfrommer2007}. To date these are typically modelled at substantially higher resolution but with much simplified galaxy-formation physics, a notable exception being \citet{Talbot2025}, who couple a thermal-conduction solver directly to the full TNG model.

Parallel to these developments, a number of efforts have aimed to generate cosmological simulations constrained by observations of the local Universe. The CLUES and subsequent constrained-realisation projects pioneered by \citet{Sorce2017} demonstrated that large-scale initial conditions can be reconstructed from galaxy surveys to reproduce local structures such as the Virgo Cluster, Coma, and the Local Supercluster. The SLOW programme \citep{Dolag2023,HernandezMartinez2024,Boess2024} subsequently performed hydrodynamical resimulations of such constrained initial conditions to study the interplay of baryons in local environments. In parallel, dark-matter-only work led by \citet{Jasche2010}, \citet{Lavaux2019}, \citet{McAlpine2025}, and the Sibelius project \citep{Sawala2023} has refined Bayesian reconstruction and MCMC-based inference pipelines to produce statistically consistent ensembles of local-universe realisations. Throughout this paper we distinguish between methods for \emph{constructing} constrained initial conditions (e.g.\ the CLUES reverse-Zel'dovich approach and the \texttt{BORG}/\texttt{Manticore} Bayesian forward model) and hydrodynamical \emph{resimulations} of those constrained ICs (such as SLOW, and the present work).
These constrained simulations provide a unique bridge between cosmological theory and the observed properties of nearby galaxy clusters, enabling statistical studies of formation histories and environmental effects that are otherwise inaccessible.
Despite this remarkable progress, prior constrained hydrodynamical work targeting cluster zooms often relies on one preferred model or has employed galaxy formation physics that differs from the IllustrisTNG model \citep[e.g.,][for an excellent line of work on Virgo]{Sorce2016, Sorce2017, Sorce2021, Sorce2026}. The distinctive contribution of the present work is therefore the combination of (i) \texttt{BORG}/\texttt{Manticore} Bayesian inference of the local large-scale structure as the source of the constrained initial conditions, (ii) a posterior ensemble of 50 Coma-targeted realisations rather than a single constrained run, and (iii) the IllustrisTNG galaxy formation model evolved at zoom-in resolution. We adopt the \texttt{BORG}/\texttt{Manticore} posterior chains from the \texttt{Manticore-Local} inference of the local Universe by \citet{McAlpine2025} as the basis for constructing constrained initial conditions, generating zoom-in realisations for 50 posterior samples and evolving them forward with the TNG model. This approach allows us to sample a finite ensemble of Coma analogues drawn from the posterior-constrained initial conditions and evolve each within a self-consistent baryonic framework, capturing the diversity of assembly histories and environmental interactions admitted by the posterior.

Our simulations provide a testbed for connecting theoretical predictions to multiwavelength observations, including optical, X-ray, and Sunyaev--Zel'dovich \citep{Sunyaev1972} measurements. Specifically, we will focus on detailed comparisons with the data obtained with the SRG/eROSITA \citep{Predehl2021,Sunyaev2021} for X-ray radial profiles and the Planck telescope \citep{Planck2013, Planck2014}. Furthermore, our work will serve as a foundation for future studies of cluster physics in the local Universe, including, but not limited to, detailed comparisons to XRISM/Resolve, the kinetic Sunyaev--Zel'dovich effect as well as studies that target radio observations in greater detail.

This paper is structured as follows. In Sec.~\ref{sec:methods}, we describe the generation of the constrained initial conditions and the details of the simulation code. Sec.~\ref{sec:results} presents an overview of the simulations, including general properties and cluster scaling relations, and compares simulated X-ray surface brightness profiles to observations from eROSITA \citep{Churazov2021} and Compton-$y$ measurements from Planck \citep{Planck2014}. Additionally, we investigate the diversity of Coma analogues across the 50-sample posterior ensemble and compare the results to the dark matter–only realisations of \citet{McAlpine2025}. In Sec.~\ref{sec:discussion}, we discuss our results and place them into the context of previous simulation efforts of the Coma cluster. Finally, we identify the best-fitting Coma analogues, summarise their baseline properties in a table, and discuss pathways for immediate and future work for constrained hydrodynamical simulations. In Sec.~\ref{sec:conclusions}, we summarise our main results and present the conclusions from our work.

\section{Numerical Methods}
\label{sec:methods}

\subsection{Initial Conditions Generation}

We have developed a customised pipeline to generate cosmological zoom-in initial conditions (ICs) using the public code \textsc{MUSIC} \citep{Hahn2011}. Before describing the pipeline it is useful to distinguish three resolutions that are sometimes conflated in this kind of work: (i) the resolution of the underlying \emph{constrained reconstruction} of the local large-scale structure, which here is the \texttt{BORG}/\texttt{Manticore} Bayesian inference of \citet{McAlpine2025} and is fixed by that analysis; (ii) the resolution of the \emph{parent simulation} that we evolve forward from each posterior sample to identify the Coma analogue at $z=0$; and (iii) the resolution of the \emph{zoom-in simulation} that we evolve with the TNG galaxy formation model and which sets the resolution of all the science presented in this paper. Only the third of these is the resolution at which our results are reported.

The core workflow has three steps. First, we evolve each of the 50 \texttt{BORG}/\texttt{Manticore} posterior samples forward with \textsc{Gadget-4} \citep{Springel2021} as a low-resolution dark-matter-only parent simulation in a $L=1000$ Mpc domain with $256^3$ particles (dark matter particle mass $\sim10^{12}\,{\rm M}_\odot$); the \texttt{BORG}/\texttt{Manticore} inference itself uses the 2M++ galaxy catalogue to constrain the local density field. Second, in each parent simulation we identify the Coma counterpart at $z=0$ from the posterior reconstruction. We then select all dark matter particles within $6\,R_{200c}$ of this halo at $z=0$ and trace them back to their Lagrangian positions in the initial conditions, defining the high-resolution region as their convex hull. The selected cluster candidates have virial masses in the range $\sim8\times10^{14}$ to $2\times10^{15}\,{\rm M}_\odot$, with at least $\sim 500$ particles inside $R_{200c}$ at $z=0$ --- sufficient for unambiguous target identification, which is the only role of the parent run. Third, we regenerate the IC at high resolution using \textsc{MUSIC}'s standard multi-level refinement scheme: nested refinement levels are added around the Lagrangian region in factors of two in spatial resolution until the desired mass resolution is reached, with progressively coarser particle shells used as boundary padding to ensure smooth force transitions and minimise contamination from low-resolution regions. The displacement field is generated using second-order Lagrangian perturbation theory (2LPT) consistent with the cosmological parameters of the corresponding \texttt{BORG}/\texttt{Manticore} realisation. Long-wavelength modes are inherited from the constrained large-scale phases, while small-scale modes beyond the parent-box Nyquist frequency are newly sampled to provide the additional high-frequency power needed in the zoom region.

A practical aspect of our current pipeline is worth noting. The version of \textsc{MUSIC} we use ingests the \texttt{BORG}/\texttt{Manticore} output as a uniform $256^3$ parent cube, which sets the resolution at which Coma candidates are identified and Lagrangian regions are tagged. The constrained large-scale phases admit, in principle, a higher-resolution parent realisation by adding more small-scale power, and a future iteration of this workflow will incorporate that capability; for the present study the $256^3$ parent run is sufficient because its only role in the pipeline is target identification, while the constrained phases that drive the large-scale environment of the Coma analogue are inherited from the \texttt{BORG}/\texttt{Manticore} inference itself rather than from the parent run. With this pipeline, our Coma zoom realisations reproduce the large-scale environment of the constrained \texttt{BORG}/\texttt{Manticore} fields and recover the posterior peak in mass and position of the corresponding high-resolution simulations \citep{McAlpine2025}. 
\begin{table}
    \centering
    \caption{Numerical setup of the constrained Coma zoom simulations. Particle
    masses are physical; the collisionless softening is the Plummer-equivalent
    value at $z=0$ and the gas softening is the adaptive comoving floor. The
    effective resolution refers to the parent volume sampled at the high-resolution
    element mass.}
    \label{tab:resolution}
    \begin{tabular}{lc}
        \hline\hline
        Parameter & Value \\
        \hline
        Code / physics                          & \textsc{Arepo}, IllustrisTNG model \\
        Number of zoom realisations             & $50$ \\
        Parent box length $L$                   & $1\,\mathrm{Gpc}$ ($681\,h^{-1}\mathrm{Mpc}$) \\
        $h$                                     & $0.681$ \\
        $\Omega_{\rm m},\,\Omega_\Lambda,\,\Omega_{\rm b}$ & $0.306,\,0.694,\,0.0486$ \\
        \hline
        $m_{\rm DM}$ [$M_\odot$]                & $6.0\times10^{7}$ \\
        $m_{\rm gas}$ (target) [$M_\odot$]      & $1.2\times10^{7}$ \\
        $\epsilon_{\rm DM,\star}$ ($z=0$) [kpc] & $1.47$ \\
        $\epsilon_{\rm gas,min}$ [comoving pc]  & $294$ \\
        Effective resolution                    & $8192^{3}$ \\
        \hline
    \end{tabular}
\end{table}

The numerical setup of our constrained Coma realisations is summarised in Table~\ref{tab:resolution}. We evolve $50$ realisations of the same large-scale environment with \textsc{Arepo} and the IllustrisTNG galaxy-formation model,
embedded in a constrained $1\,\mathrm{Gpc}$ BORG/Manticore parent volume \citep{McAlpine2025}. Each high-resolution zoom region reaches a dark-matter particle mass of $m_{\rm DM}\simeq6.0\times10^{7}\,M_\odot$ and a targetbaryonic (gas-cell) mass of $m_{\rm gas}\simeq1.2\times10^{7}\,M_\odot$, with a Plummer-equivalent gravitational softening of $\epsilon_{\rm DM,\star}=1.47$\,kpc (physical) at $z=0$ for the collisionless component and an adaptive gas softening down to a comoving floor of $\sim$0.3\,kpc. This is essentially identical to the
resolution of the TNG-Cluster suite \citep{Nelson2024}, which adopts the same TNG300-1 physics, albeit marginally higher. Sampling the full parent volume at this element mass corresponds to an effective resolution of $\simeq8192^{3}$
resolution elements of the total \texttt{BORG/Manticore} volume.

In the future it might be a good strategy to adopt the Multi-zoom technique presented in Burger et al. (2026, submitted) for a variety of  zoom-in simulations to run more of these local clusters simultaneously in different realisations at high resolution. 

\subsection{Galaxy formation simulations with {\small AREPO} and the TNG model}

Our simulations are performed with the \textsc{Arepo} code \citep{Springel2010}, a quasi-Lagrangian magnetohydrodynamics (MHD) solver that evolves the coupled dynamics of dark matter, gas, stars, and black holes within a cosmological volume. \textsc{Arepo} employs a second-order finite-volume scheme on an unstructured Voronoi mesh, whose cells move with the local flow of matter. This moving-mesh approach provides the adaptivity and Galilean invariance of Lagrangian methods while retaining the accuracy, stability, and shock-capturing properties of Eulerian schemes. As a result, \textsc{Arepo} achieves excellent accuracy in modeling fluid instabilities, shocks, and mixing processes, which are essential for capturing the complex gas dynamics associated with galaxy formation, cluster assembly, and cosmic web evolution. The code also conserves mass, momentum, and energy to machine precision and avoids the advection errors typical of fixed-grid solvers, ensuring accurate tracking of thermodynamic quantities even over long integration times.

The gravitational evolution of dark matter and collisionless components is computed with a hierarchical Tree–Particle-Mesh (TreePM) algorithm \citep{Springel2005_G2}, which combines a long-range particle-mesh solver for large-scale forces with a hierarchical tree algorithm for short-range interactions. This hybrid scheme provides both high accuracy and computational efficiency across a broad range of spatial scales. The baryonic component is evolved hydrodynamically on the mesh, with adaptive refinement and derefinement operations that maintain an approximately constant target gas cell mass. This dynamic resolution adjustment ensures that dense, star-forming regions are spatially well-resolved, while low-density environments such as cosmic voids are treated efficiently. Such adaptivity allows the simulations to simultaneously resolve the internal structure of galaxies and the large-scale distribution of matter within the same cosmological volume.

Magnetic fields are evolved self-consistently using the MHD formulation of \citet{Pakmor2011,Pakmor2013}, with divergence control enforced through a Powell cleaning scheme. The MHD implementation extends the hydrodynamic capabilities of \textsc{Arepo} to magnetized plasmas while preserving numerical stability and accuracy. 

The baryonic physics in our simulations follow the IllustrisTNG (hereafter TNG) galaxy formation model \citep{Weinberger2017,Pillepich2018,Nelson2018,Springel2018,Nelson2019}, an updated and extended version of the original Illustris model \citep{Vogelsberger2014,Genel2014,Sijacki2015}. The TNG model includes the key physical processes required for galaxy formation in a cosmological context, implemented through physically motivated and numerically calibrated subgrid prescriptions. These prescriptions are designed to model unresolved processes—such as star formation, stellar feedback, and black hole growth—in a manner that is both computationally efficient and consistent with observational constraints. The model parameters are calibrated using a small number of low-redshift observables, such as the galaxy stellar mass function and gas fractions, but are not explicitly tuned to reproduce higher-order or redshift-dependent quantities, thereby preserving the model’s predictive power.

Gas cooling and heating are computed for nine individually tracked elements (H, He, C, N, O, Ne, Mg, Si, Fe), accounting for both primordial and metal-line cooling \citep{Wiersma2009}. The cooling rates depend on the local gas metallicity, temperature, and ionization state, with the latter determined under the influence of a spatially uniform, time-evolving ultraviolet background \citep{FaucherGiguere2009}. This treatment allows for self-consistent modeling of gas thermodynamics during reionization and subsequent structure formation. Star formation occurs in dense, cold gas following the effective two-phase interstellar medium (ISM) model of \citet{Springel2003}, in which cold molecular clouds coexist in pressure equilibrium with a hot ambient phase. The star formation rate is proportional to the local gas density above a critical threshold and calibrated to reproduce the empirical Kennicutt–Schmidt relation. This model provides a robust, resolution-independent approach to star formation in cosmological simulations, enabling the conversion of cold gas into stars to proceed at physically motivated rates.

Chemical enrichment and stellar evolution follow the mass and metal return from Type~Ia and Core Collapse supernovae, as well as asymptotic giant branch (AGB) stars, using the yield tables described by \citet{Vogelsberger2013} and \citet{Pillepich2018}. The enrichment model tracks the production and distribution of individual elements, allowing the simulation to follow the evolving metallicity of the interstellar and intergalactic medium. Feedback from stellar processes is implemented through kinetic galactic winds that carry mass, metals, and energy out of star-forming regions. The TNG model introduced a physically motivated wind velocity scaling that depends on both the local dark matter velocity dispersion and the redshift, enabling efficient self-regulation of star formation across halo masses \citep{Pillepich2018}. This refinement improved upon the original Illustris feedback scheme by producing more realistic galaxy stellar mass functions and circumgalactic medium properties.

Supermassive black holes are seeded in haloes above a threshold mass and evolve via gas accretion and mergers \citep{Springel2005_BH}. The TNG model employs a dual-mode active galactic nucleus (AGN) feedback scheme \citep{Weinberger2017}: (i) a high-accretion ``thermal'' mode, where feedback energy is isotropically injected as heat into the surrounding gas, and (ii) a low-accretion ``kinetic'' mode, in which energy is released in the form of directed, jet-like outflows. The thermal mode dominates during rapid black hole growth, while the kinetic mode becomes active in massive, quiescent systems, where it effectively quenches star formation and drives large-scale outflows. This two-mode feedback implementation leads to realistic stellar and gas fractions in massive haloes and reproduces the observed transition from blue, star-forming galaxies to red, passive ones.

Magnetic fields, evolved self-consistently via MHD, play a critical role in shaping the thermodynamic and kinematic properties of the intracluster medium (ICM). The TNG model initializes a uniform comoving seed field of $10^{-14}\,\mathrm{G}$ at high redshift, which is subsequently amplified through gravitational collapse, turbulence, and galactic feedback \citep{Pakmor2017}. This implementation captures the emergence and evolution of cosmic magnetisation and its influence on gas transport, mixing, and thermal conduction. The predicted magnetic field strengths and topologies within galaxies and clusters are consistent with observational constraints from Faraday rotation measures and synchrotron emission, demonstrating the physical realism of the MHD approach \citep{Marinacci2018}.

The combination of \textsc{Arepo}'s moving-mesh hydrodynamics and the TNG galaxy formation model provides a robust framework for modeling the coupled evolution of baryons and dark matter across cosmic time. This framework achieves a useful balance between physical realism, numerical accuracy, and computational efficiency, making it particularly well-suited for large-scale cosmological simulations. The TNG model reproduces a wide range of observational data reasonably well---including the galaxy stellar mass function, cosmic star formation rate density, galaxy colour bimodality, halo gas fractions, and large-scale clustering---without further parameter tuning.

\subsection{Orientation onto the observed Coma sky frame}
\label{sec:rotation}

Each zoom is re-simulated from initial conditions in which \textsc{MUSIC} has translated the Lagrangian patch of the target halo to the centre of the zoom box. This translation removes the cluster's absolute position within the
constrained volume, and with it the on-sky orientation, but it applies no rotation, so the orientation of the cluster relative to the simulation axes is identical in the parent box and in the zoom. We therefore recover the line of
sight from the parent \texttt{BORG/Manticore} box, in which the fiducial observer sits at the box centre $\mathbf{c} = \tfrac{1}{2}L\,(1,1,1)$ with $L = 1\,\mathrm{Gpc}$
(i.e.\ $\mathbf{c} = 0.5\,\mathrm{Gpc}$ on each axis).

For realisation $i$, let $\mathbf{x}_p^{(i)}$ be the parent-box position of the Coma counterpart and $\mathbf{x}_c^{(i)}$ the cluster centre in the zoom. The
unit line-of-sight vector from the observer is
\begin{equation}
  \hat{\mathbf{n}} = \frac{\mathbf{x}_p^{(i)} - \mathbf{c}}
                          {\lvert \mathbf{x}_p^{(i)} - \mathbf{c}\rvert}.
\end{equation}
The \texttt{BORG/Manticore} field is defined with the $z$-axis pointing to the celestial North pole; the \textsc{MUSIC}/\textsc{gadget} pipeline interchanges the $x$ and $z$ axes relative to \texttt{BORG}, so North corresponds to $\hat{\mathbf{N}} = (1,0,0)$ in
the simulation frame. We construct a right-handed orthonormal viewing triad by projecting North into the plane of the sky,
\begin{equation}
  \hat{\mathbf{u}} = \frac{\hat{\mathbf{N}} - (\hat{\mathbf{N}}\!\cdot\hat{\mathbf{n}})\,\hat{\mathbf{n}}}
                          {\lvert \hat{\mathbf{N}} - (\hat{\mathbf{N}}\!\cdot\hat{\mathbf{n}})\,\hat{\mathbf{n}}\rvert},
  \qquad
  \hat{\mathbf{e}} = \hat{\mathbf{u}} \times \hat{\mathbf{n}},
\end{equation}
where $\hat{\mathbf{u}}$ and $\hat{\mathbf{e}}$ point North and East on the sky.
The rotation into the viewing frame is the proper rotation
($\det\mathsf{R}=+1$) with these vectors as rows,
\begin{equation}
  \mathsf{R} =
  \begin{pmatrix} \hat{\mathbf{e}}^{\mathsf{T}} \\
                  \hat{\mathbf{u}}^{\mathsf{T}} \\
                  \hat{\mathbf{n}}^{\mathsf{T}} \end{pmatrix}.
\end{equation}
Every gas cell is rotated about the cluster centre, $\mathbf{x}' = \mathsf{R}\,(\mathbf{x} - \mathbf{x}_c)$, and projected along the new third axis
$x'_3 = \hat{\mathbf{n}}\!\cdot(\mathbf{x} - \mathbf{x}_c)$, i.e.\ the line of sight; the resulting maps are displayed with North ($x'_2$) up and East ($x'_1$) to the left, following the standard astronomical convention. Because the viewing geometry is fixed by the observer--cluster direction, the same $\mathsf{R}$ is applied at every snapshot in the short-baseline time-evolution analysis (snaps~095--099); only the centre $\mathbf{x}_c$ is re-determined per snapshot.

As a consistency check, the equatorial coordinates implied by the parent positions --- $\delta = \arcsin[(\mathbf{x}_p-\mathbf{c})\!\cdot\hat{\mathbf{N}} /
\lvert\mathbf{x}_p-\mathbf{c}\rvert]$ and the corresponding right ascension --- cluster tightly at $(\alpha,\delta)\simeq(193.8^\circ,+29.6^\circ)$ with a mean distance of $\sim$105\,Mpc, close to the true position of the Coma cluster $(194.95^\circ,+27.98^\circ)$.

\section{Results}
\label{sec:results}

In this section we present the results of our 50 constrained Coma analogue simulations. Throughout, the integer index in the simulation name uniquely identifies a posterior sample (Coma\_000 to Coma\_049). The presentation is organised as follows. We first showcase a single representative realisation,
Coma\_025, in Sec.~\ref{sec:coma025_showcase}, to give the reader a concrete visual feel for what one of these constrained Coma analogues looks like before moving to ensemble statistics. Sec.~\ref{sec:overview} gives a tabular and visual overview of all 50 realisations, and Sec.~\ref{sec:scaling_relations}
checks the ensemble against standard galaxy cluster scaling relations as a sanity check on the underlying physics. We then turn to the spatially resolved properties of the intracluster medium: Sec.~\ref{sec:projected_maps} presents projected Compton-$y$ and X-ray surface brightness maps, and
Sec.~\ref{sec:radial_profiles} brings in the direct comparison with the observed Coma cluster through azimuthally averaged radial profiles, from which we identify the 10 best-fitting realisations. The remaining sections interpret this picture:
Sec.~\ref{sec:central_scatter} traces the cluster-to-cluster scatter in the central profiles back to its origin in the assembly history, and Sec.~\ref{sec:time_variability} tests how stable the radial profiles are over the last $\sim$500~Myr of evolution. Finally, Sects.~\ref{sec:bestfit_maps}
and~\ref{sec:mass_distribution} return to the 10 best-fitting realisations, examining their projected maps and the range of halo masses admitted by the posterior.

\begin{table*}
    \centering
    \caption{Key physical properties of the most massive Friends-of-Friends (FoF) halo (the Coma cluster analogue) at $z=0$ across the 50 zoom-in simulations. Each simulation corresponds to one realisation (Coma\_000--Coma\_049) drawn from the 50 \texttt{BORG}/\texttt{Manticore} MCMC chains. Halo properties are derived from the \textsc{Subfind} catalogues using the TNG galaxy formation model. $M_{200m}$ and $R_{200m}$ denote the mass and radius enclosing 200 times the mean matter density. $M_{500c}$ and $R_{500c}$ are defined relative to the critical density. $M_\ast$ is the total stellar mass in the FoF group. $f_{\mathrm{gas}}$ is the gas mass fraction, and $T_{\mathrm{vir}}$ is the virial temperature computed as $kT_{\mathrm{vir}} = \mu m_p G M_{200} / (2 R_{200})$. $M_{\ast,\mathrm{BCG}}$ is the stellar mass of the central subhalo (BCG), and $M_{\bullet,\mathrm{BCG}}$ is the mass of its central supermassive black hole.}
    \label{tab:coma_properties}
    \begin{tabular}{lccccccccc}
        \hline
        \hline
        Simulation & $M_{200,\mathrm{c}}$ & $R_{200,\mathrm{c}}$ & $M_{500,\mathrm{c}}$ & $R_{500,\mathrm{c}}$ & $M_\ast$ & $f_{\mathrm{gas}}$ & $T_{\mathrm{vir}}$ & $M_{\ast,\mathrm{BCG}}$ & $M_{\bullet,\mathrm{BCG}}$ \\
        & [$10^{14}\,{\rm M}_\odot$] & [Mpc] & [$10^{14}\,{\rm M}_\odot$] & [Mpc] & [$10^{12}\,{\rm M}_\odot$] & & [keV] & [$10^{12}\,{\rm M}_\odot$] & [$10^{10}\,{\rm M}_\odot$] \\
        \hline
        Coma\_000 & 10.1 & 2.11 & 6.9 & 1.37 & 24.3 & 0.122 & 6.3 & 2.52 & 4.39 \\
        Coma\_001 & 14.2 & 2.36 & 10.0 & 1.55 & 24.3 & 0.125 & 8.0 & 1.69 & 4.46 \\
        Coma\_002 & 11.8 & 2.22 & 8.7 & 1.48 & 27.9 & 0.121 & 7.0 & 3.94 & 4.82 \\
        Coma\_003 & 10.8 & 2.15 & 7.6 & 1.41 & 20.5 & 0.121 & 6.6 & 1.23 & 3.83 \\
        Coma\_004 & 13.4 & 2.31 & 9.9 & 1.54 & 24.1 & 0.124 & 7.6 & 1.75 & 4.54 \\
        Coma\_005 & 10.6 & 2.15 & 7.9 & 1.43 & 21.3 & 0.130 & 6.6 & 3.68 & 4.36 \\
        Coma\_006 & 12.0 & 2.23 & 9.5 & 1.52 & 27.9 & 0.132 & 7.1 & 1.41 & 5.19 \\
        Coma\_007 & 11.5 & 2.20 & 7.9 & 1.43 & 22.4 & 0.121 & 6.9 & 2.41 & 4.55 \\
        Coma\_008 & 12.3 & 2.25 & 8.6 & 1.47 & 22.8 & 0.123 & 7.2 & 1.36 & 4.33 \\
        Coma\_009 & 12.1 & 2.24 & 8.4 & 1.46 & 23.7 & 0.118 & 7.1 & 1.94 & 4.78 \\
        Coma\_010 & 9.8 & 2.09 & 8.0 & 1.44 & 17.2 & 0.134 & 6.2 & 2.11 & 3.50 \\
        Coma\_011 & 11.0 & 2.17 & 8.2 & 1.45 & 21.9 & 0.129 & 6.7 & 1.95 & 4.08 \\
        Coma\_012 & 12.4 & 2.26 & 8.4 & 1.46 & 22.3 & 0.124 & 7.3 & 1.96 & 3.90 \\
        Coma\_013 & 9.7 & 2.08 & 5.7 & 1.28 & 23.7 & 0.126 & 6.1 & 1.06 & 3.29 \\
        Coma\_014 & 11.8 & 2.22 & 7.6 & 1.41 & 22.9 & 0.119 & 7.0 & 0.95 & 4.20 \\
        Coma\_015 & 13.1 & 2.30 & 9.7 & 1.53 & 22.4 & 0.124 & 7.5 & 1.60 & 4.42 \\
        Coma\_016 & 11.0 & 2.17 & 7.4 & 1.40 & 21.6 & 0.125 & 6.7 & 1.67 & 4.25 \\
        Coma\_017 & 10.4 & 2.13 & 6.9 & 1.37 & 18.7 & 0.121 & 6.4 & 1.83 & 3.35 \\
        Coma\_018 & 12.9 & 2.29 & 9.6 & 1.53 & 21.9 & 0.132 & 7.4 & 2.16 & 4.12 \\
        Coma\_019 & 11.2 & 2.18 & 8.8 & 1.48 & 22.3 & 0.121 & 6.8 & 1.24 & 4.49 \\
        Coma\_020 & 13.7 & 2.33 & 9.7 & 1.53 & 23.5 & 0.125 & 7.7 & 2.00 & 4.81 \\
        Coma\_021 & 12.6 & 2.27 & 9.3 & 1.51 & 24.5 & 0.120 & 7.3 & 1.69 & 4.85 \\
        Coma\_022 & 9.6 & 2.07 & 6.5 & 1.34 & 19.2 & 0.120 & 6.1 & 1.67 & 3.41 \\
        Coma\_023 & 12.1 & 2.24 & 8.1 & 1.44 & 24.7 & 0.122 & 7.1 & 2.25 & 4.11 \\
        Coma\_024 & 11.7 & 2.21 & 7.4 & 1.40 & 22.2 & 0.122 & 7.0 & 2.42 & 3.82 \\
        Coma\_025 & 13.7 & 2.33 & 9.2 & 1.51 & 24.0 & 0.124 & 7.8 & 1.57 & 4.32 \\
        Coma\_026 & 11.8 & 2.22 & 8.5 & 1.47 & 20.6 & 0.132 & 7.0 & 2.28 & 4.31 \\
        Coma\_027 & 13.0 & 2.30 & 7.8 & 1.43 & 21.6 & 0.123 & 7.5 & 1.56 & 4.12 \\
        Coma\_028 & 11.2 & 2.18 & 7.6 & 1.41 & 24.6 & 0.122 & 6.8 & 1.39 & 4.61 \\
        Coma\_029 & 11.7 & 2.21 & 7.9 & 1.43 & 22.4 & 0.124 & 7.0 & 0.81 & 4.13 \\
        Coma\_030 & 10.6 & 2.14 & 7.9 & 1.43 & 23.5 & 0.127 & 6.5 & 1.37 & 4.36 \\
        Coma\_031 & 11.1 & 2.17 & 8.7 & 1.48 & 19.4 & 0.126 & 6.7 & 1.20 & 3.85 \\
        Coma\_032 & 9.1 & 2.04 & 6.8 & 1.36 & 17.0 & 0.124 & 5.9 & 1.09 & 3.16 \\
        Coma\_033 & 11.1 & 2.18 & 7.0 & 1.38 & 21.7 & 0.121 & 6.7 & 1.58 & 4.30 \\
        Coma\_034 & 10.9 & 2.16 & 8.0 & 1.44 & 20.8 & 0.126 & 6.6 & 1.45 & 3.94 \\
        Coma\_035 & 13.4 & 2.32 & 9.1 & 1.50 & 26.6 & 0.119 & 7.6 & 1.49 & 5.33 \\
        Coma\_036 & 10.2 & 2.12 & 7.3 & 1.40 & 18.7 & 0.127 & 6.4 & 1.65 & 3.23 \\
        Coma\_037 & 11.1 & 2.18 & 8.2 & 1.45 & 20.5 & 0.125 & 6.7 & 1.28 & 3.92 \\
        Coma\_038 & 13.7 & 2.33 & 6.8 & 1.36 & 25.3 & 0.120 & 7.7 & 1.54 & 4.88 \\
        Coma\_039 & 11.5 & 2.20 & 8.9 & 1.49 & 22.5 & 0.122 & 6.9 & 1.50 & 4.05 \\
        Coma\_040 & 12.0 & 2.23 & 8.7 & 1.48 & 23.7 & 0.130 & 7.1 & 2.75 & 4.46 \\
        Coma\_041 & 8.7 & 2.01 & 6.4 & 1.33 & 17.4 & 0.126 & 5.7 & 2.66 & 2.82 \\
        Coma\_042 & 10.8 & 2.16 & 7.0 & 1.37 & 24.1 & 0.119 & 6.6 & 1.02 & 4.26 \\
        Coma\_043 & 11.6 & 2.21 & 6.8 & 1.36 & 23.2 & 0.128 & 6.9 & 1.53 & 4.14 \\
        Coma\_044 & 14.5 & 2.38 & 11.2 & 1.61 & 24.4 & 0.127 & 8.0 & 2.26 & 4.88 \\
        Coma\_045 & 14.0 & 2.35 & 10.0 & 1.55 & 24.6 & 0.124 & 7.9 & 2.87 & 4.68 \\
        Coma\_046 & 10.5 & 2.13 & 7.7 & 1.42 & 18.2 & 0.128 & 6.5 & 1.53 & 3.12 \\
        Coma\_047 & 10.9 & 2.16 & 5.9 & 1.30 & 26.9 & 0.119 & 6.7 & 1.56 & 4.29 \\
        Coma\_048 & 13.7 & 2.34 & 9.8 & 1.54 & 23.0 & 0.123 & 7.8 & 1.80 & 4.16 \\
        Coma\_049 & 13.5 & 2.32 & 10.1 & 1.55 & 25.4 & 0.120 & 7.7 & 1.66 & 4.89 \\
        \hline
    \end{tabular}
\end{table*}

\begin{figure*}
    \centering
    \vspace{-0.4cm}
    \includegraphics[width=0.85\textwidth]{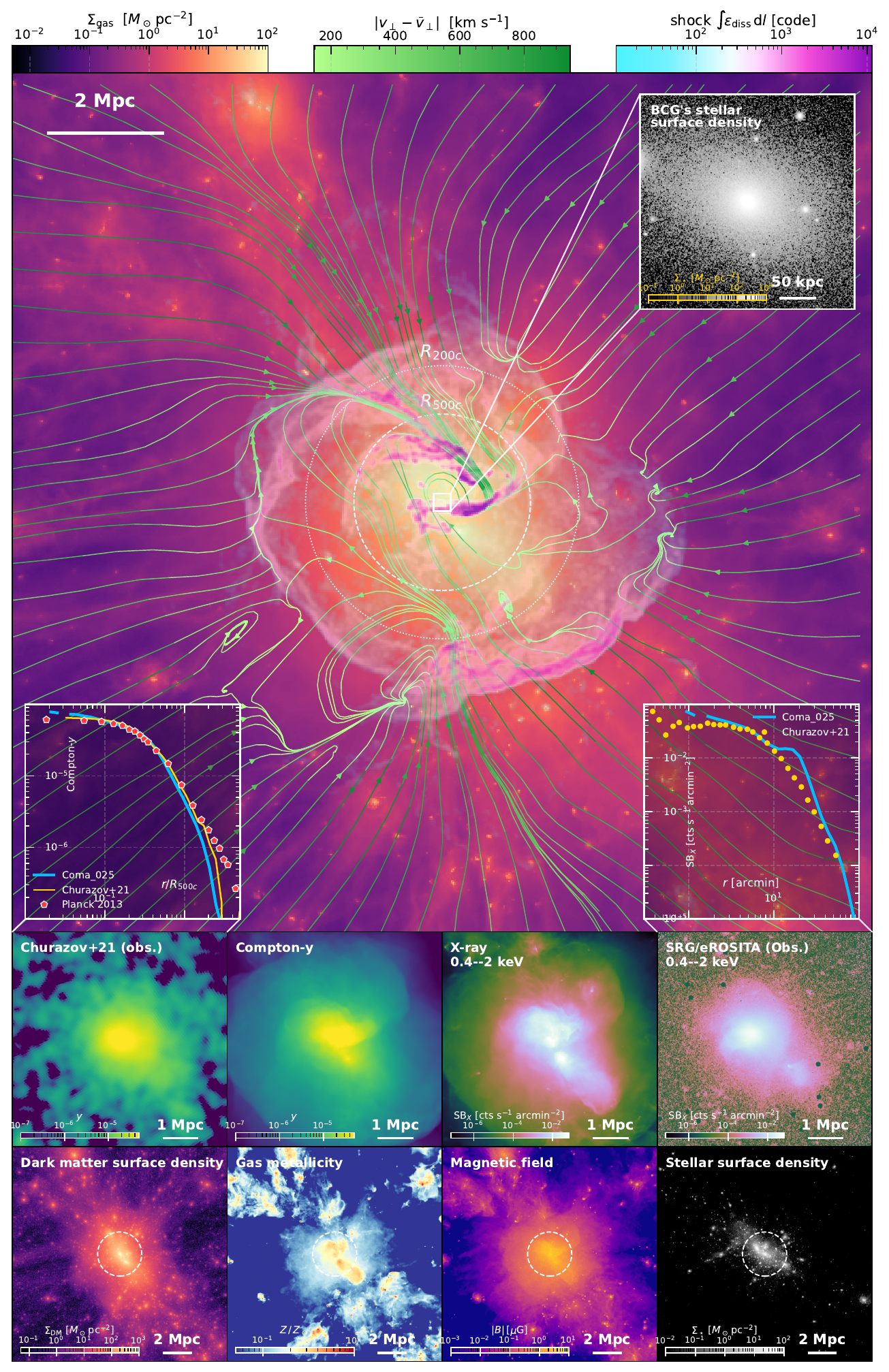}
    \vspace{-0.3cm}
    \caption{Coma\_025 constrained overview. The big top panel shows the gas surface density with in-plane velocity streamlines, the BCG stellar-light zoom-in, and the radial Compton-$y$ and X-ray surface brightness profiles; the middle row compares the observed Planck and SRG/eROSITA maps to the simulation, and the bottom row shows the dark matter, metallicity, magnetic field, and stellar surface density. All panels are oriented so the NGC~4839-analogue companion sits in the lower-right.}
    \label{fig:coma025_showcase}
\end{figure*}

\begin{figure*}
    \includegraphics[width=\textwidth]{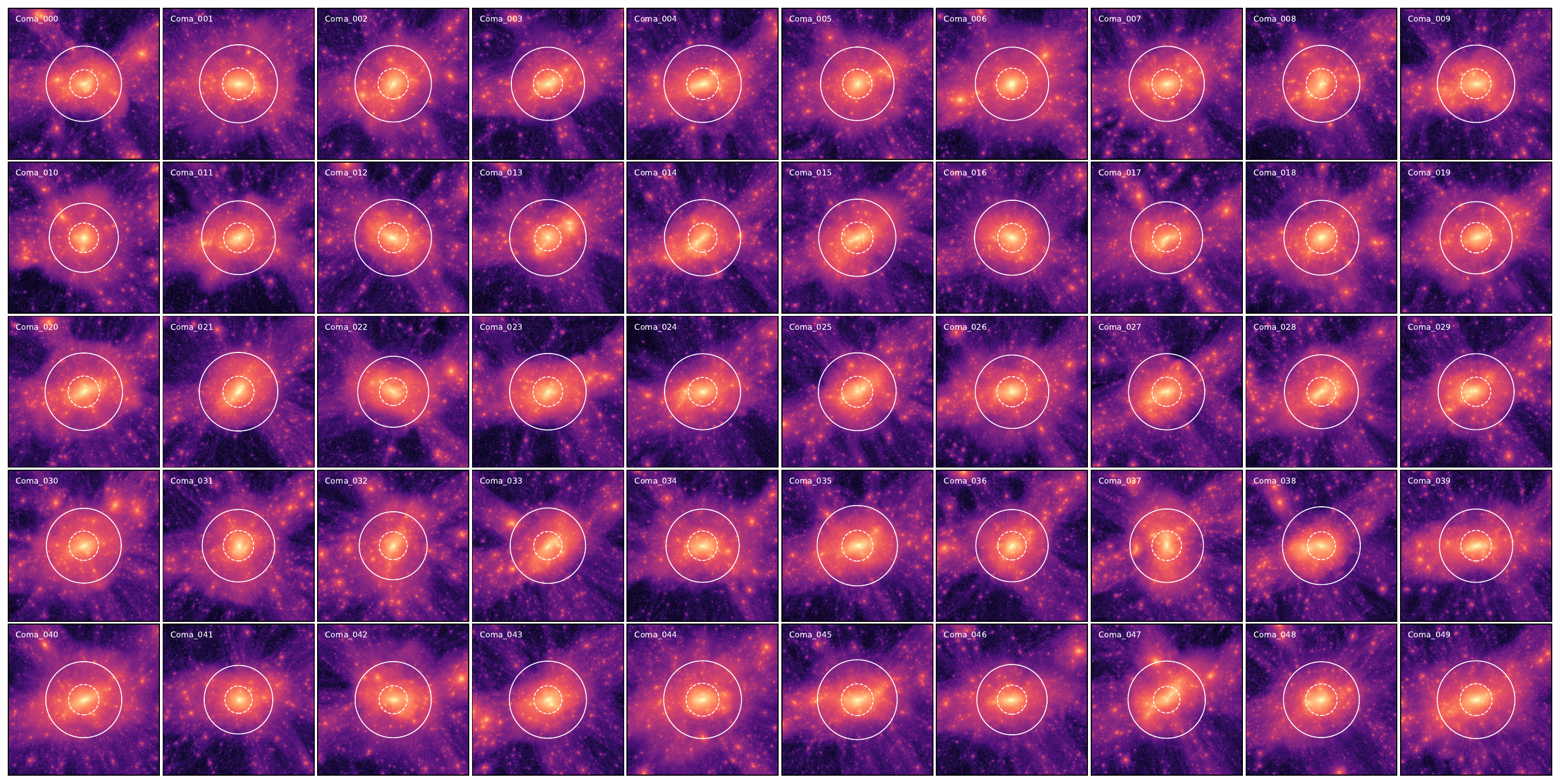}
    \caption{Projected dark-matter surface density maps for all 50 constrained Coma analogues (labelled Coma\_000 to Coma\_049), arranged in a $5\times10$ grid. Each panel spans a $10\times10\;\mathrm{Mpc}$ field of view centred on the Coma analogue. The projection is computed by ray-tracing the Voronoi cell geometry along the $z$-axis. White dashed and solid circles indicate $R_{500c}$ and $R_{200m}$, respectively. The uniform logarithmic colour scale (0.05--$10^3\;{\rm M}_\odot\,\mathrm{pc}^{-2}$) allows direct comparison of the large-scale structure and filamentary environment across realisations.}
    \label{fig:dm_density}
\end{figure*}

\begin{figure*}
    \includegraphics[width=\textwidth]{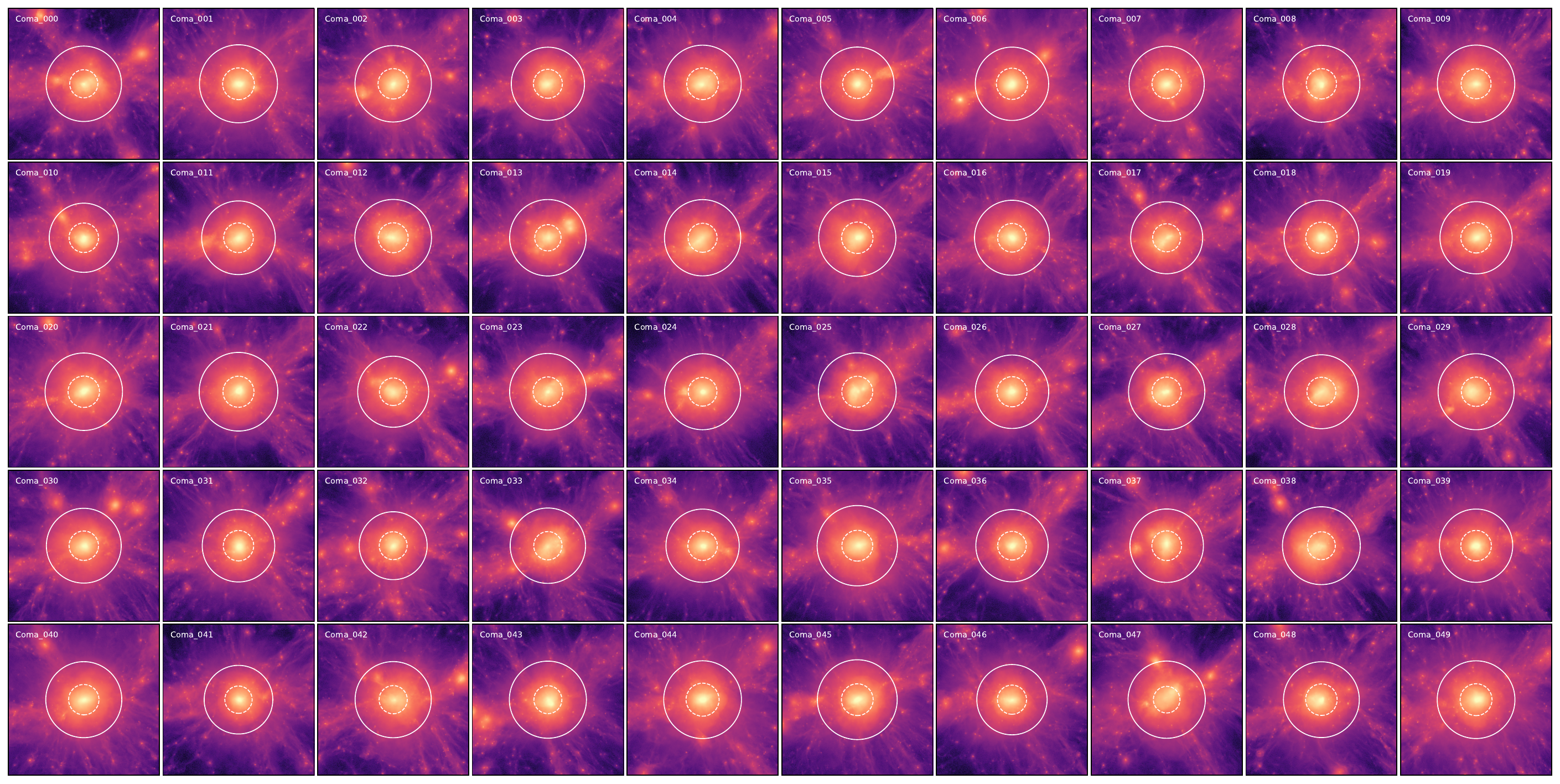}
    \caption{Projected gas surface density maps for all 50 constrained Coma analogues, in the same $5\times10$ layout as Fig.~\ref{fig:dm_density}. The logarithmic colour scale spans $10^{-2}$ to $10^{2}\;{\rm M}_\odot\,\mathrm{pc}^{-2}$. The gas distribution highlights the ICM morphology --- from smooth, centrally concentrated hot atmospheres to disturbed configurations with visible merger-driven substructure --- sampling the full range of dynamical states permitted by the constrained initial conditions at fixed TNG300 physics.}
    \label{fig:gas_density}
\end{figure*}

\begin{figure}
    \includegraphics[width=\columnwidth]{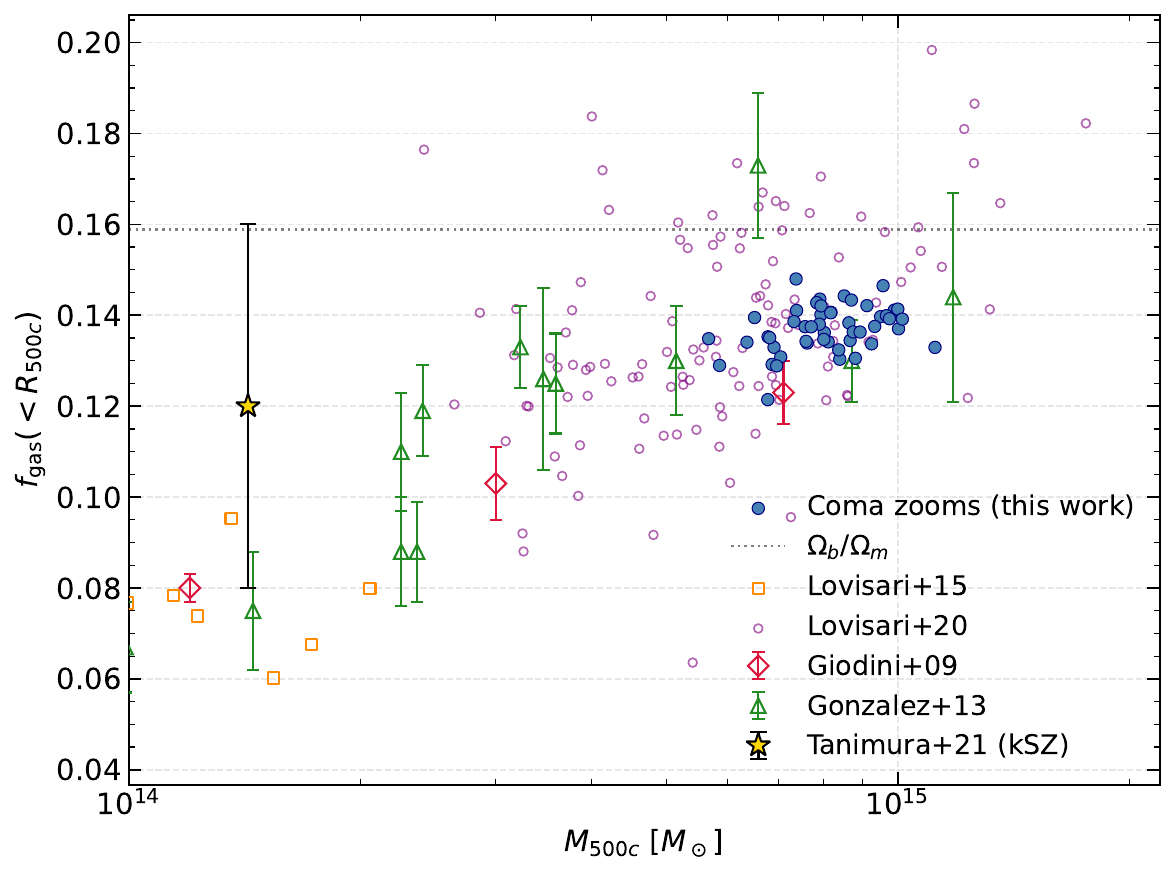}
    \caption{Hot gas fraction $f_{\rm gas} = M_{\rm gas}(<R_{500c})\,/\,M_{500c}$ as a function of total halo mass $M_{500c}$ for the 50 constrained Coma analogues. The simulated clusters are compared with observational estimates from the X-ray group and cluster samples of \citet{Giodini2009}, \citet{Gonzalez2013}, \citet{Lovisari2015}, \citet{Lovisari2020}, and the kSZ stacking analysis of \citet{Tanimura2021}. All masses and gas fractions are evaluated within $R_{500c}$. The dotted grey line marks the cosmic baryon fraction $\Omega_b/\Omega_m \approx 0.159$.}
    \label{fig:fgas}
\end{figure}

\begin{figure}
    \includegraphics[width=\columnwidth]{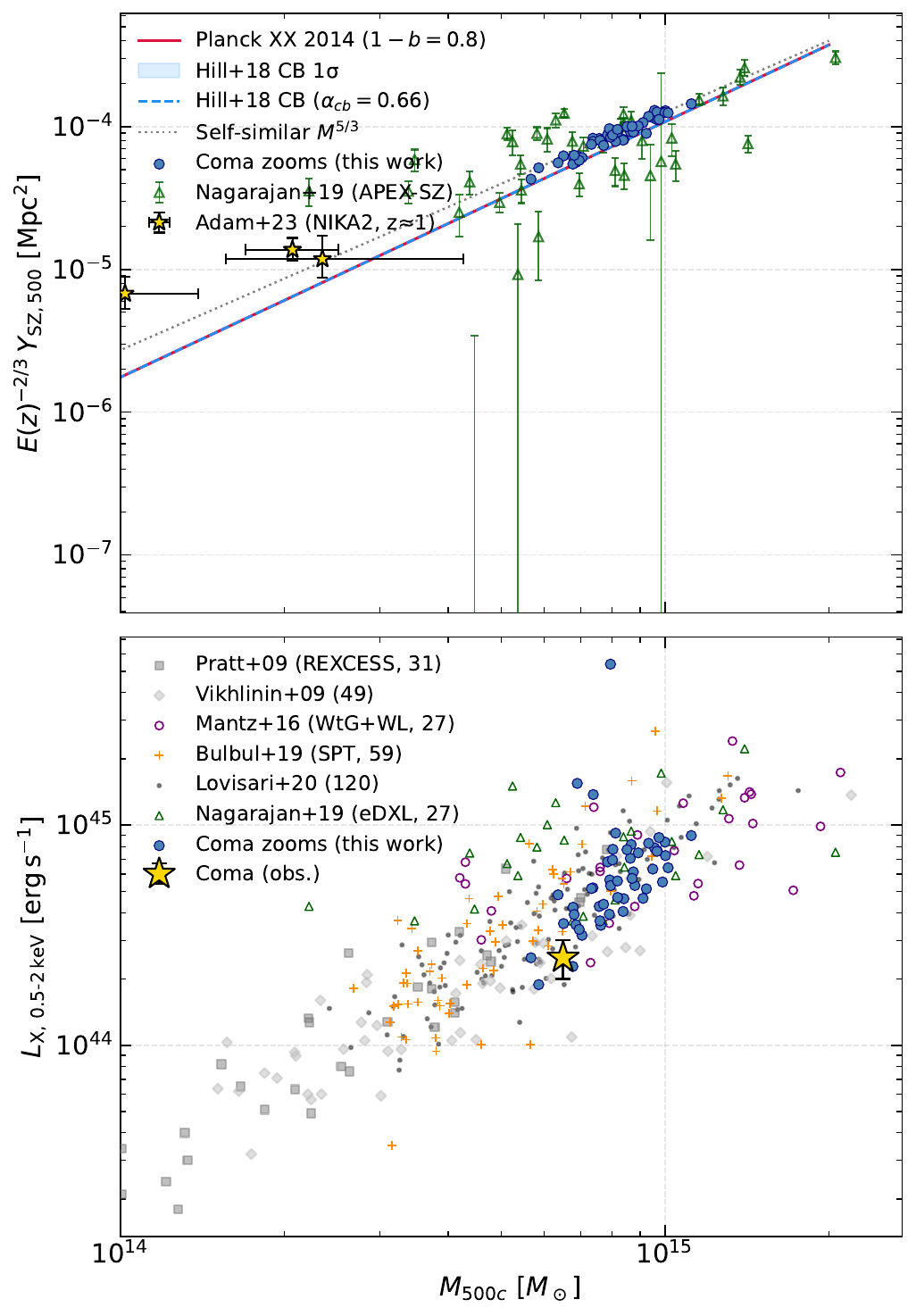}
    \caption{Integrated thermal Sunyaev--Zel'dovich signal $Y_{\rm SZ,500}$ (top) and soft X-ray luminosity $L_X$ in the 0.5--2\,keV band (bottom) as a function of $M_{500c}$ for the 50 constrained Coma analogues. \textit{Top}: observational relations from \citet{Planck2014}, and the central-bias-corrected \citet{Hill2018} model, along with points from \citet{Nagarajan2019} and \citet{Adam2023}. Self-similar $Y \propto M^{5/3}$ is indicated. \textit{Bottom}: $L_X$ comparisons include \citet{Pratt2009}, \citet{Vikhlinin2009}, \citet{Mantz2016}, \citet{Bulbul2019}, \citet{Lovisari2020}, and the eDXL subsample of \citet{Nagarajan2019}.}
    \label{fig:ysz_lx}
\end{figure}

\begin{figure}
    \includegraphics[width=\columnwidth]{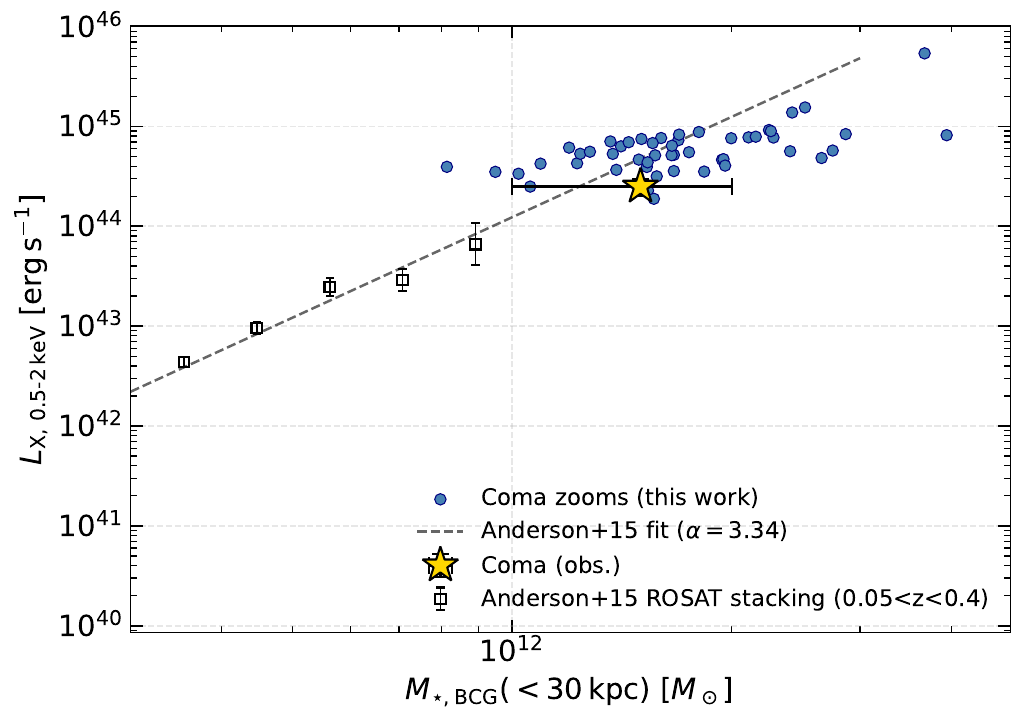}
    \caption{Soft X-ray luminosity $L_X$ (0.5--2\,keV) versus BCG stellar mass $M_{\star,\rm BCG}(<30\,\mathrm{kpc})$ for all 50 constrained Coma analogues, compared with the ROSAT stacking results of \citet{Anderson2015} and their best-fit power law ($\alpha=3.34$). Our simulated clusters occupy the high-stellar-mass end of this relation and are consistent with the \citet{Anderson2015} trend extrapolated to cluster-BCG scales.}
    \label{fig:lx_mstar}
\end{figure}

\begin{figure*}
    \includegraphics[width=\textwidth]{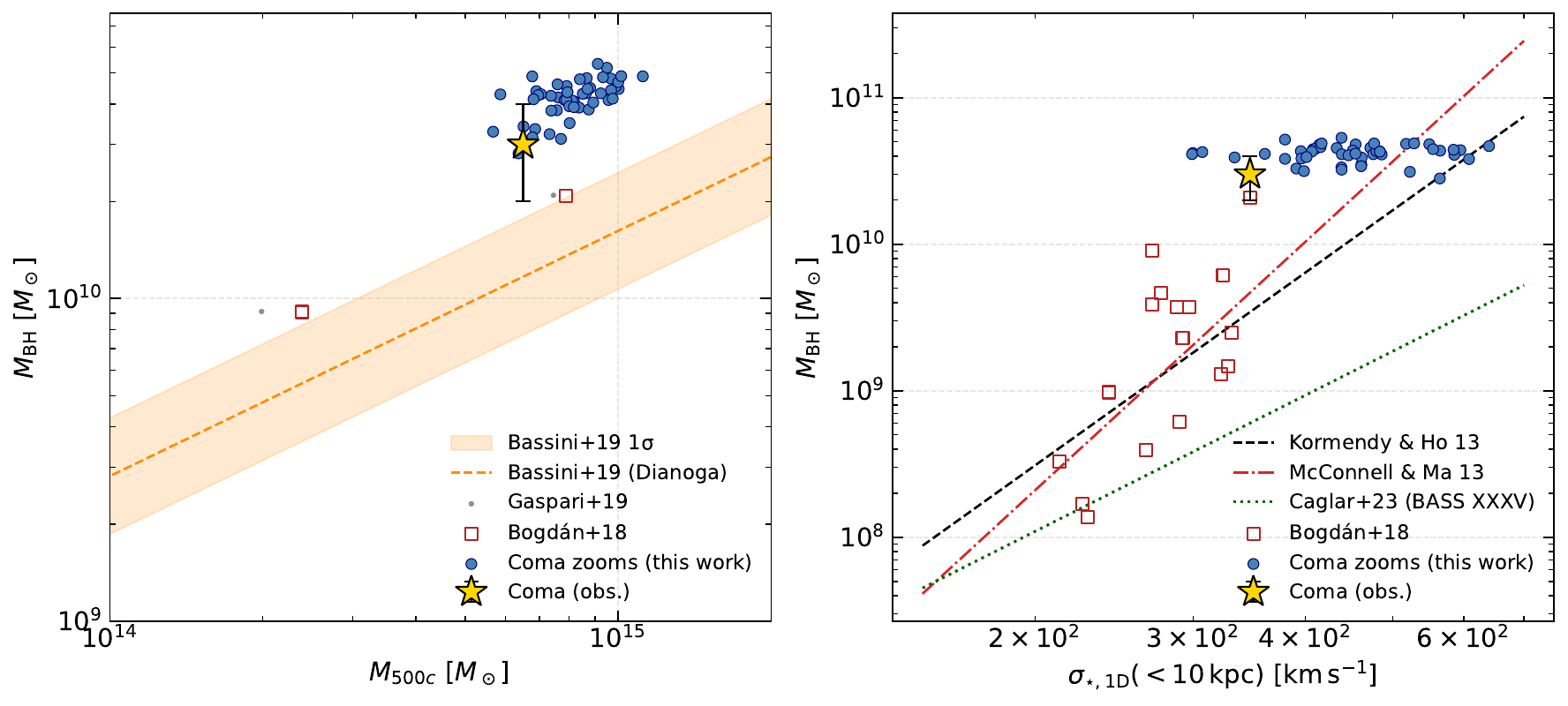}
    \caption{Supermassive black hole scaling relations for the BCG in each of the 50 constrained Coma analogues. \textit{Left:} $M_{\rm BH}$ versus $M_{500c}$, compared with the dynamical measurements of \citet{Bogdan2018}, the \citet{Gaspari2019} sample (tables converted $T_{X,c}$ to $M_{500c}$ via \citealt{Lovisari2015}), and the \citet{Bassini2019} Dianoga best-fit relation with 1$\sigma$ scatter band. \textit{Right:} $M_{\rm BH}$ versus the 1D stellar velocity dispersion $\sigma_{\star,\mathrm{1D}}$ measured within 10\,kpc of the BCG, with the \citet{Kormendy2013}, \citet{McConnell2013}, and \citet{Caglar2023} $M_{\rm BH}$--$\sigma$ relations for reference.}
    \label{fig:smbh}
\end{figure*}

\begin{figure*}
    \includegraphics[width=\textwidth]{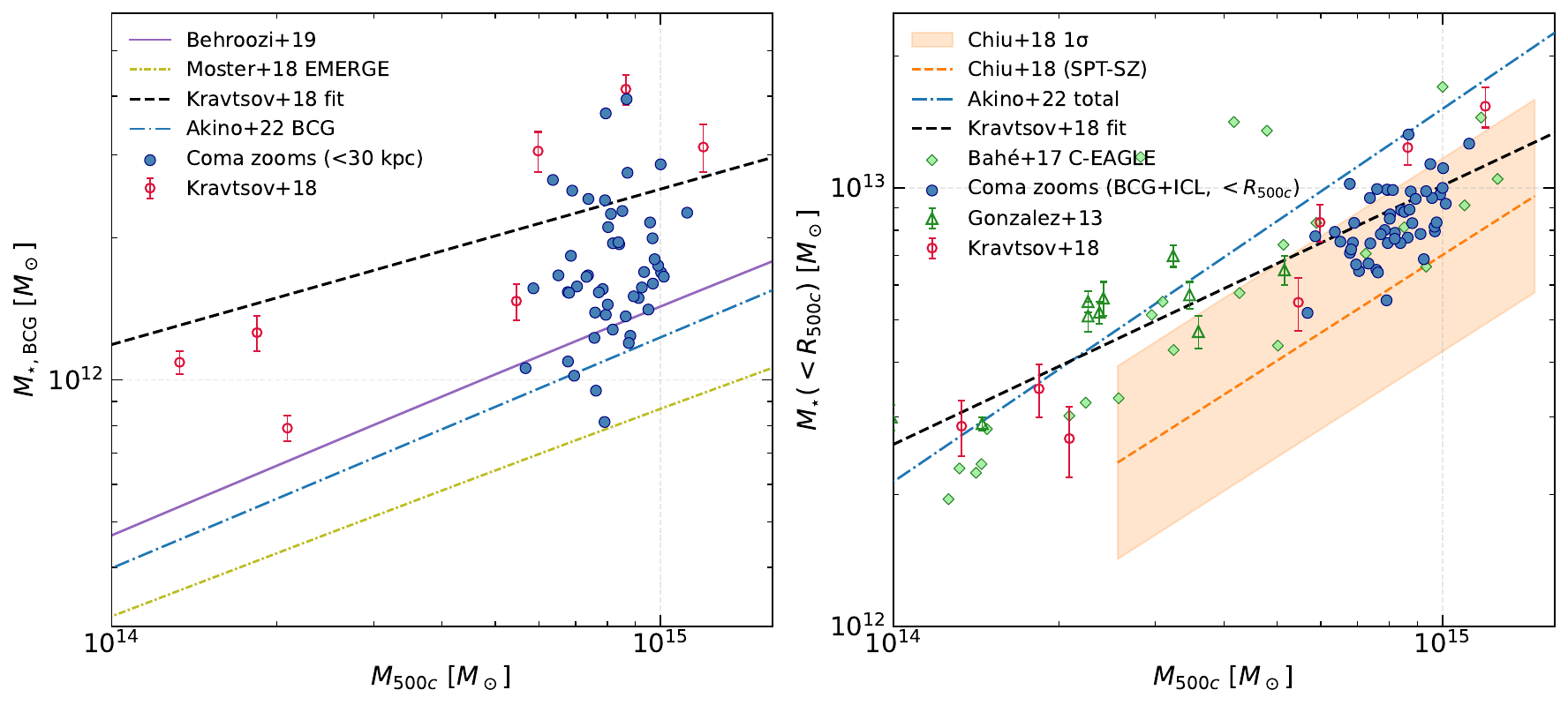}
    \caption{Stellar mass scaling relations for the 50 constrained Coma analogues. \textit{Left:} BCG stellar mass $M_{\star,\rm BCG}(<30\,\mathrm{kpc})$ vs $M_{500c}$, compared with the individual-cluster measurements and best-fit relation of \citet{Kravtsov2018}, the BCG relation of \citet{Akino2022}, and the central-galaxy SHMR curves from \citet{Behroozi2019}, \citet{Moster2018}, and \citet{Leauthaud2012}. Halo-mass definitions from abundance-matching papers have been converted to $M_{500c}$ via a \textsc{colossus} NFW profile transformation. \textit{Right:} Total stellar mass within $R_{500c}$ vs $M_{500c}$, compared with \citet{Kravtsov2018}, \citet{Gonzalez2013}, the \citet{Akino2022} total-$M_\star$ relation, the SPT-SZ $L$--$M$ band from \citet{Chiu2018}, and the 30 C-EAGLE clusters of \citet{Bahe2017}.}
    \label{fig:stellar_mass}
\end{figure*}

\begin{figure}
    \includegraphics[width=\columnwidth]{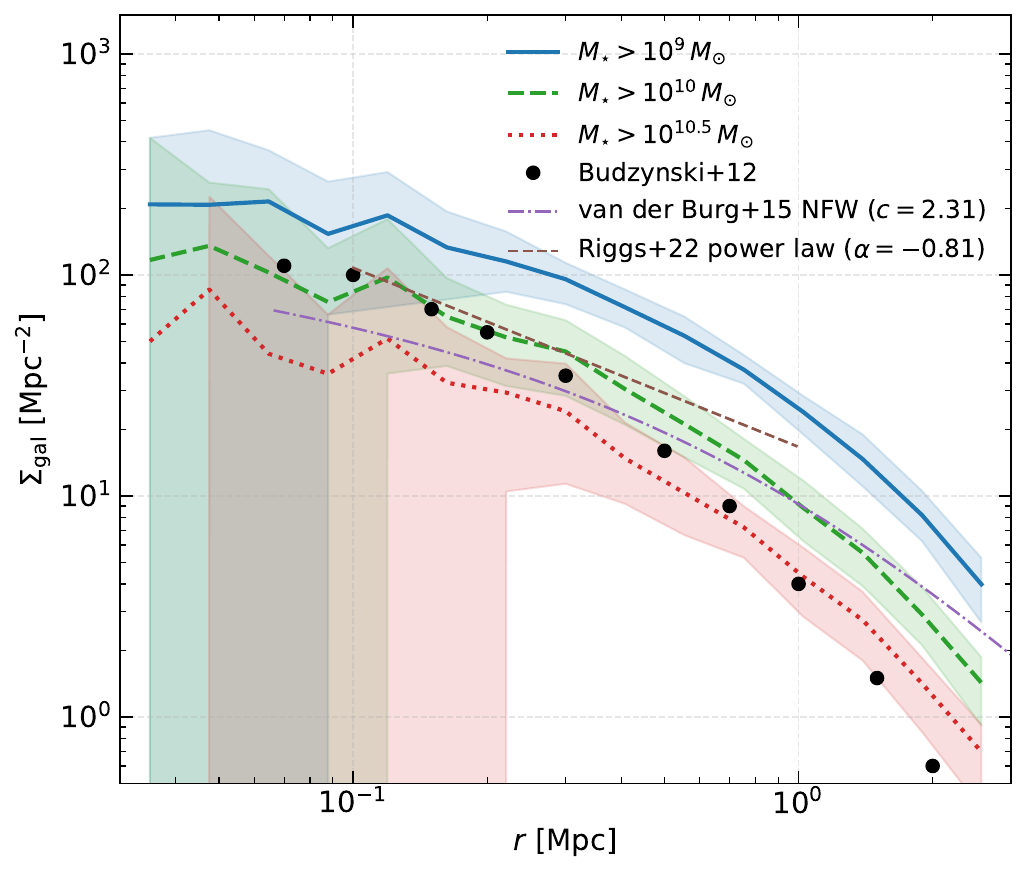}
    \caption{Projected radial surface number density profiles $\Sigma_{\rm gal}(r)$ of satellite galaxies around the Coma analogue in each of the 50 realisations, for three stellar mass thresholds ($M_\star > 10^{9}$, $10^{10}$, and $10^{10.5}\;{\rm M}_\odot$). Solid lines show the mean across all realisations; shaded bands indicate 16--84th percentile scatter. Black circles show \citet{Budzynski2012} (SDSS, $\log M_{500}=14.7$--15.0, $M_r\leq-20.5$). A power-law reference line from \citet{Riggs2022} (slope $-0.81$, valid $0.03<r/R_{200m}<0.3$) and a projected-NFW profile with concentration $c=2.31$ from \citet{vanderBurg2015} are overplotted.}
    \label{fig:satellite_profiles}
\end{figure}

\begin{figure*}
    \includegraphics[width=0.92\textwidth]{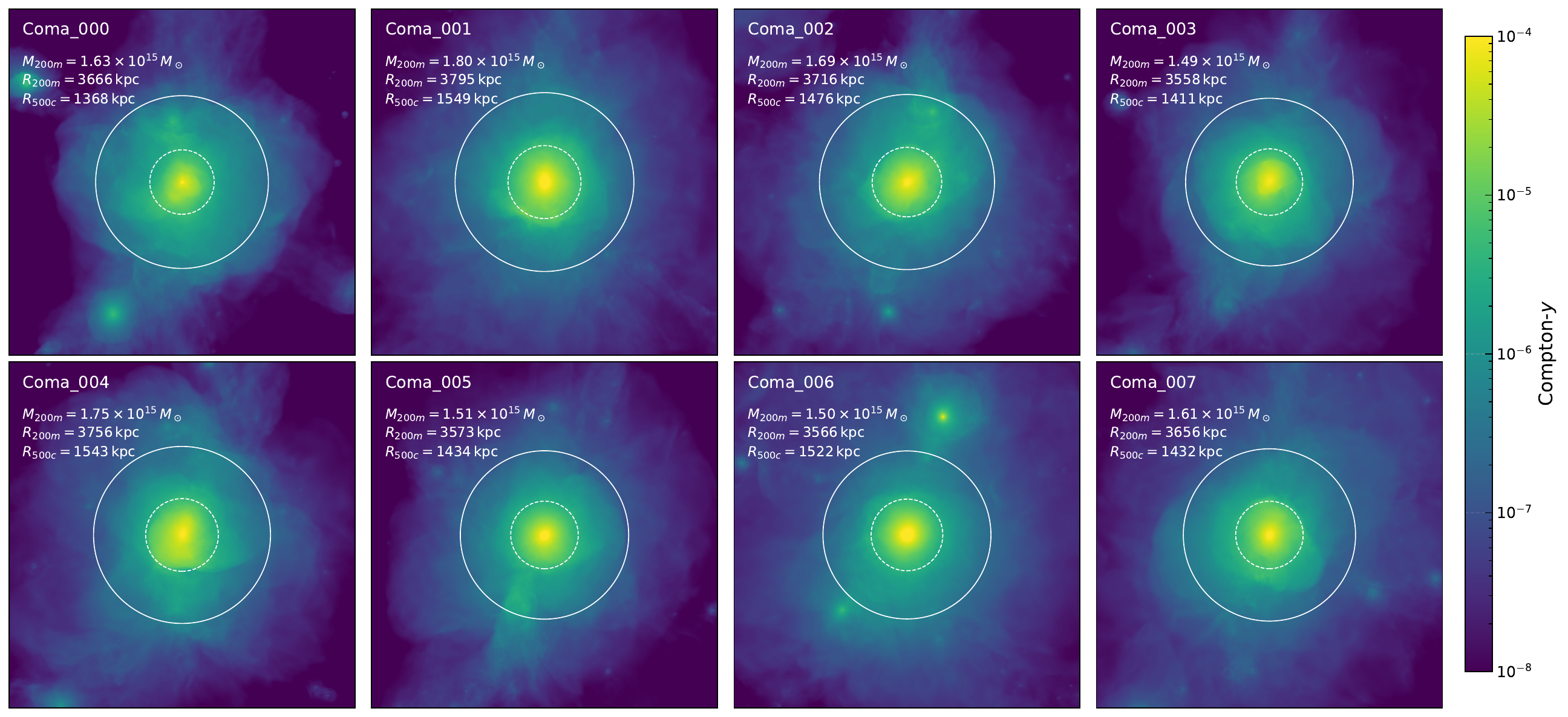}\\[0.2em]
    \includegraphics[width=0.92\textwidth]{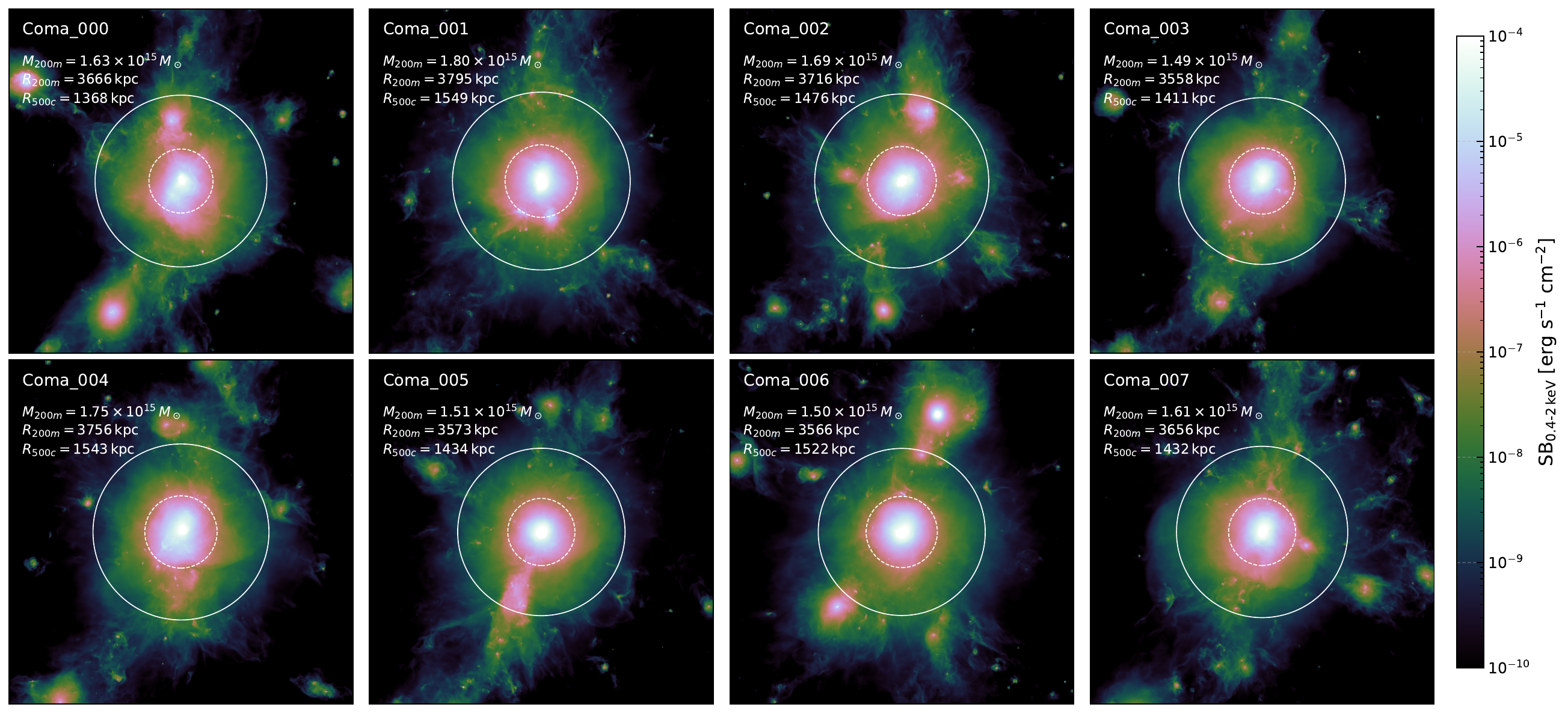}\\[0.2em]
    \caption{Projected maps of the ICM for 8 representative realisations (Coma\_000--Coma\_007), each panel a $10\times10\;\mathrm{Mpc}$ field of view along the $z$-axis. \textit{Top row:} Compton-$y=(\sigma_{\rm T}/m_e c^2)\int n_e\,k_{\rm B}T\,\mathrm{d}l$, log scale $10^{-8}$--$10^{-4}$. \textit{Bottom row:} X-ray surface brightness (0.4--2\,keV) from APEC cooling tables \citep{Smith2001, Foster2012} using per-particle $X_{\rm H}$, $n_e$, $T$; log scale $10^{-10}$--$10^{-4}\,\mathrm{erg\,s^{-1}\,cm^{-2}}$. }
    \label{fig:maps_8}
\end{figure*}

\begin{figure*}
    \includegraphics[width=0.49\textwidth]{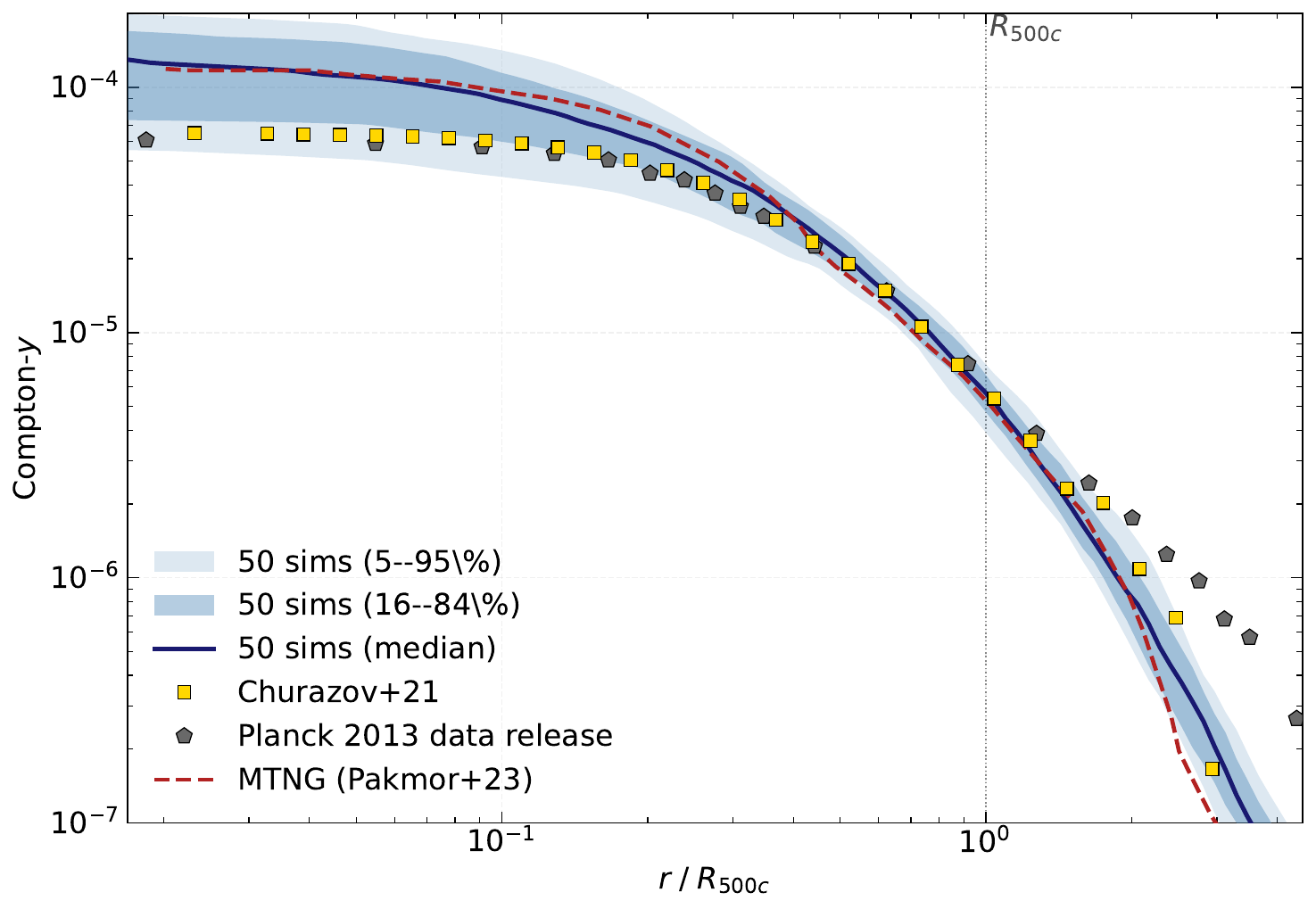}
    \includegraphics[width=0.49\textwidth]{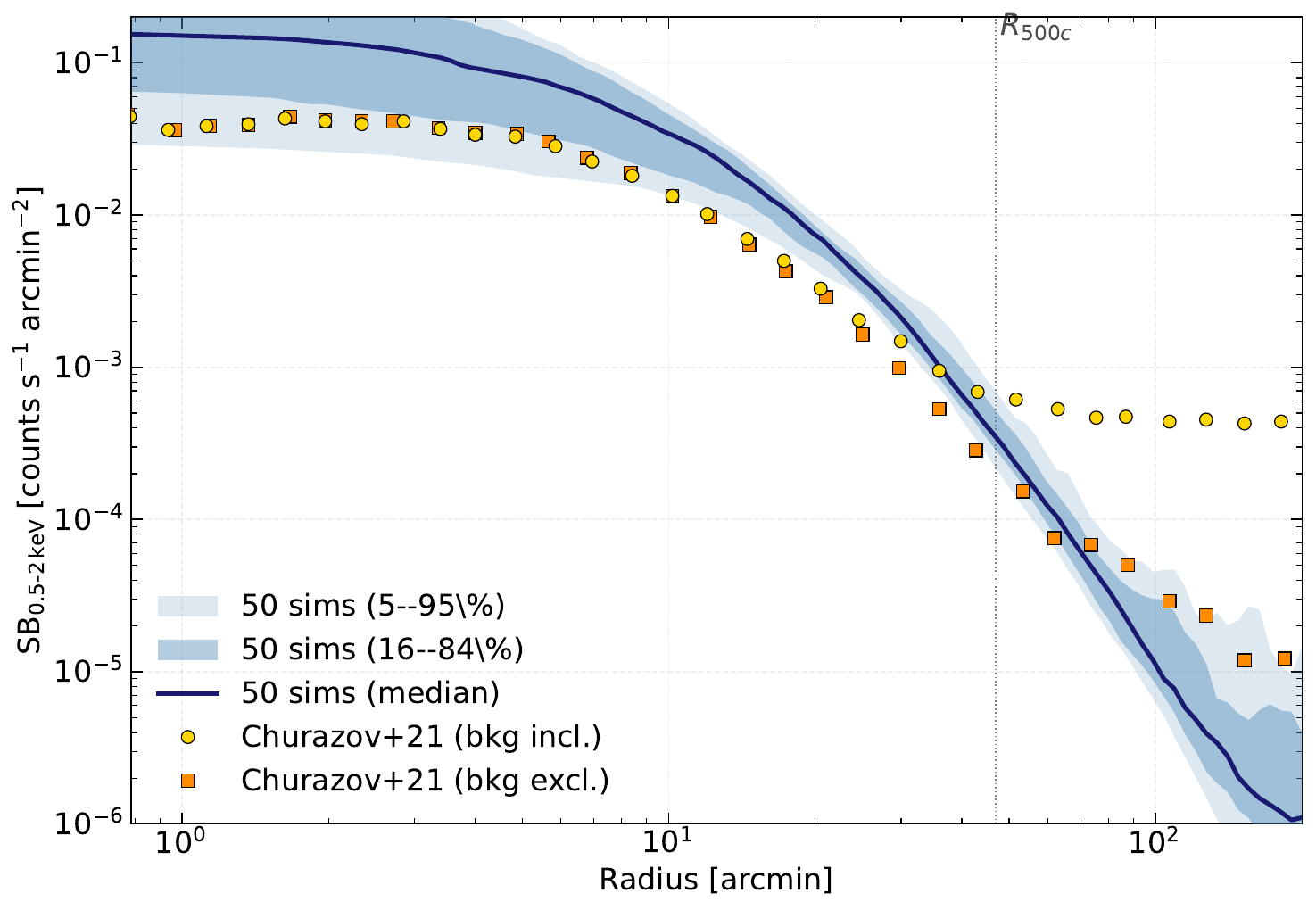}
    \caption{Azimuthally averaged radial profiles for all 50 constrained Coma analogues (green lines). \textit{Left:} Compton-$y$ vs $r/R_{500c}$, compared with \citet[grey pentagons]{Planck2013} and the MillenniumTNG prediction of \citet[solid red]{Pakmor2023}. \textit{Right:} X-ray surface brightness (0.5--2\,keV) vs angular radius (assuming $R_{200c}\approx70$\,arcmin for Coma), compared with SRG/eROSITA observations of \citet{Churazov2021} with (blue) and without (orange) background. The 10 best-fitting realisations selected from this ensemble are highlighted in detail in Fig.~\ref{fig:best10_profiles}.}
    \label{fig:profiles_all50}
\end{figure*}

\begin{figure*}
    \includegraphics[width=\textwidth]{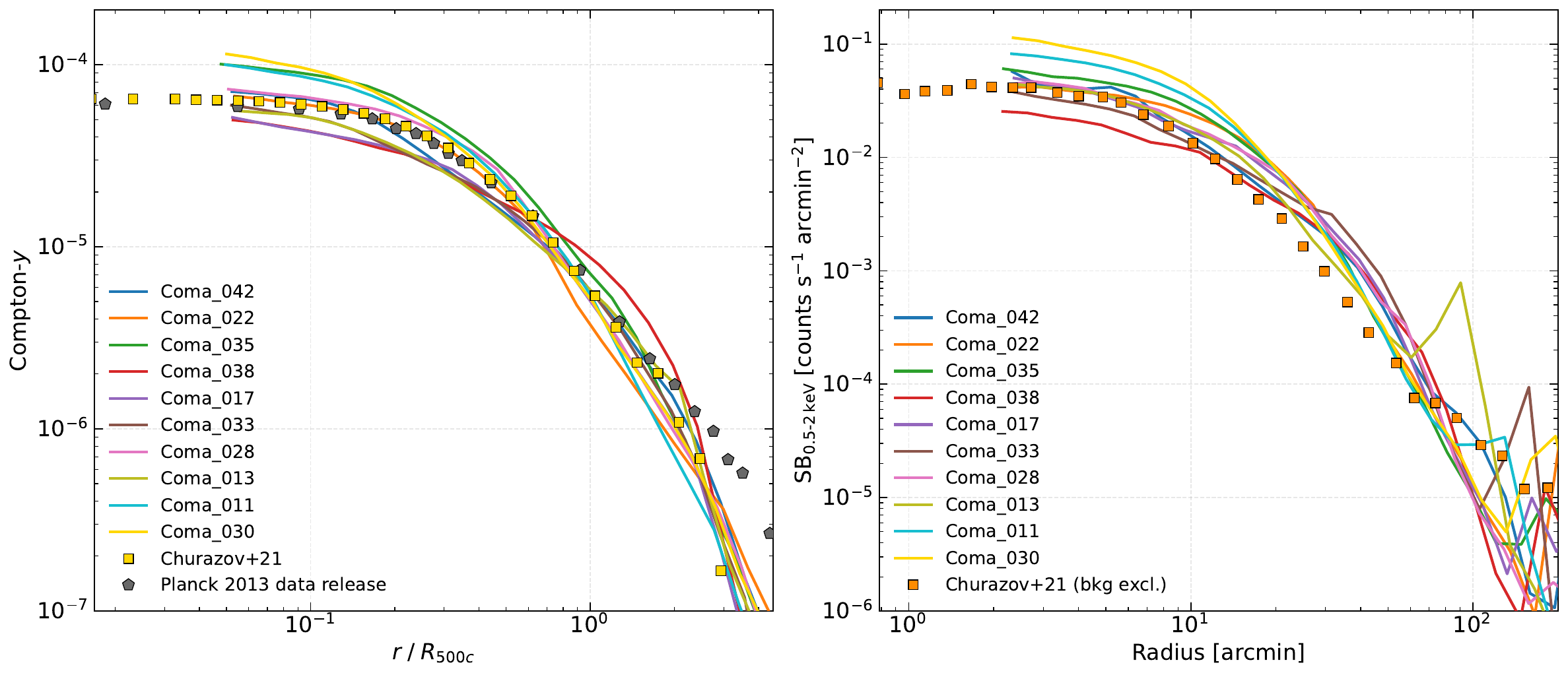}
    \caption{Radial profiles of the 10 best-fitting realisations, selected by minimising the combined mean-squared error in $\log_{10}$-space against both the \citet{Churazov2021} X-ray profile and the \citet{Planck2013} Compton-$y$ profile. \textit{Left:} Compton-$y$ vs $r/R_{500c}$, compared with \citet{Planck2013} and \citet{Churazov2021}. \textit{Right:} X-ray surface brightness vs angular radius, compared with background-subtracted \citet{Churazov2021} data. Matching line colours identify the same realisation across both panels.}
    \label{fig:best10_profiles}
\end{figure*}

\begin{figure}
    \includegraphics[width=\columnwidth]{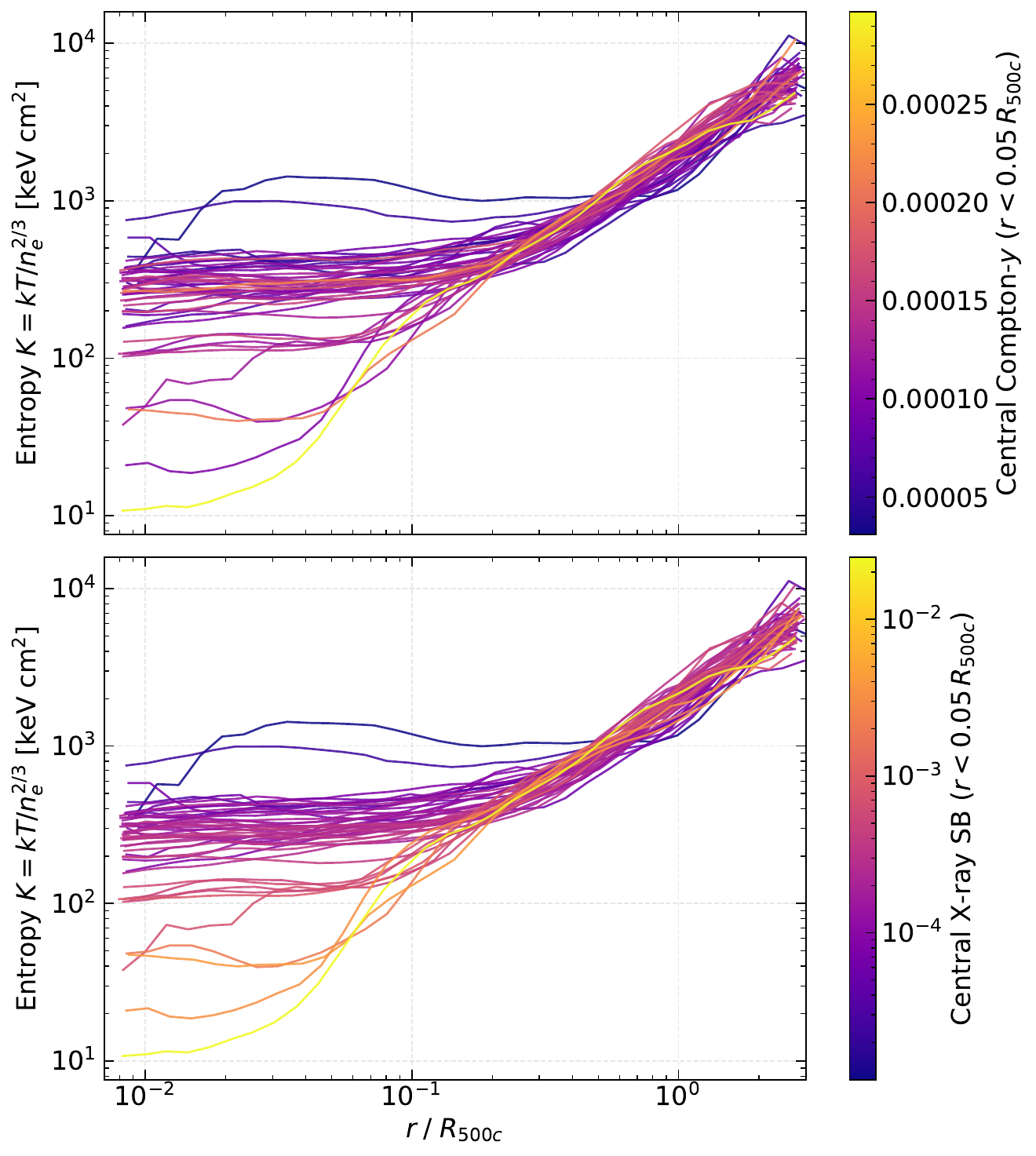}
    \caption{Azimuthally averaged entropy profiles $K(r) = kT/n_e^{2/3}$ for all 50 realisations, plotted against $r/R_{500c}$. \textit{Top:} each profile coloured by the central Compton-$y$ value inside $r<0.05\,R_{500c}$. \textit{Bottom:} same but coloured by central X-ray surface brightness. Cool-core clusters (low central $K$) drive the high-brightness tail in both observables.}
    \label{fig:entropy_vs_cc}
\end{figure}

\begin{figure*}
    \includegraphics[width=\textwidth]{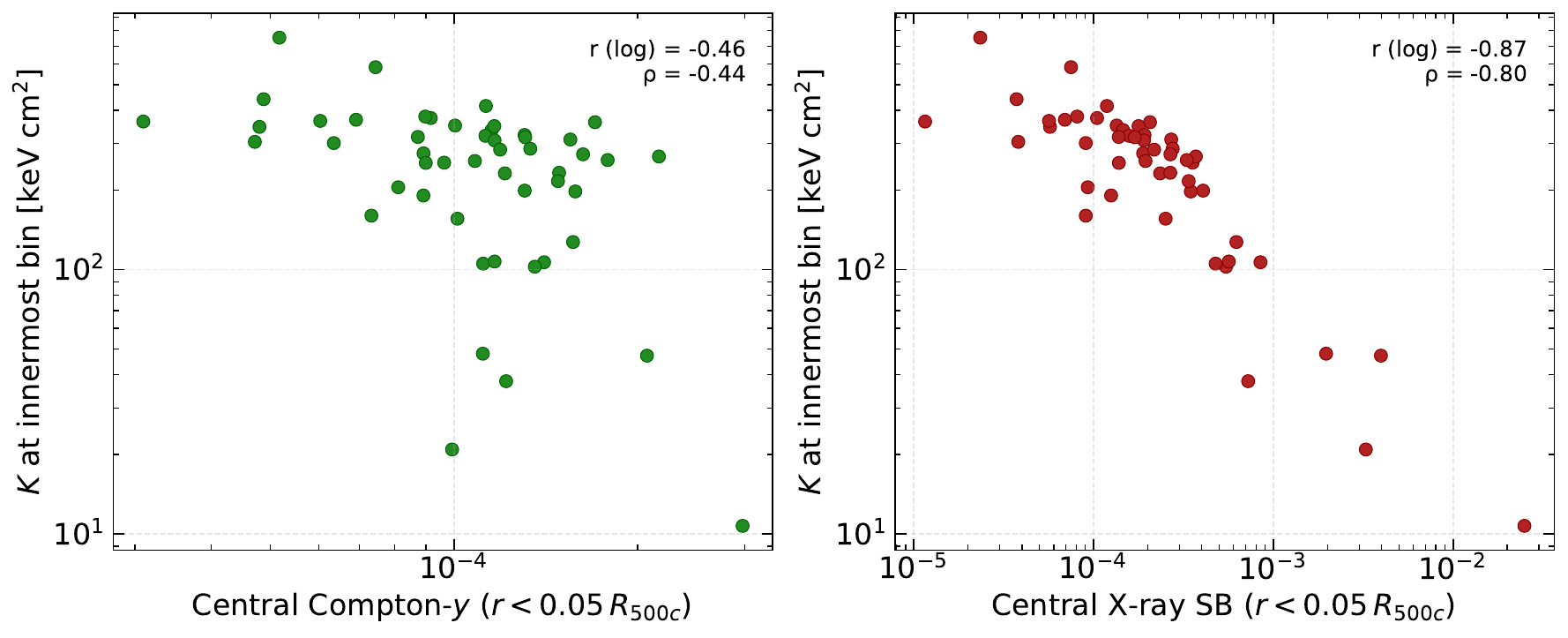}
    \caption{Correlation between the innermost-bin entropy $K_{\rm core}$ ($r/R_{200c}\approx0.005$, $\sim$10\,kpc physical) and central brightness measured inside $r<0.05\,R_{500c}$. \textit{Left:} $K_{\rm core}$ vs central Compton-$y$. \textit{Right:} $K_{\rm core}$ vs central X-ray surface brightness. Text annotations give the Pearson and Spearman correlation coefficients. The $\log K_{\rm core}$--$\log \mathrm{SB}_X$ correlation is particularly strong ($r=-0.87$, $\rho=-0.80$), confirming that central entropy is the primary driver of cluster-to-cluster variance in soft X-ray brightness.}
    \label{fig:kcore_vs_brightness}
\end{figure*}

\begin{figure*}
    \includegraphics[width=0.49\textwidth]{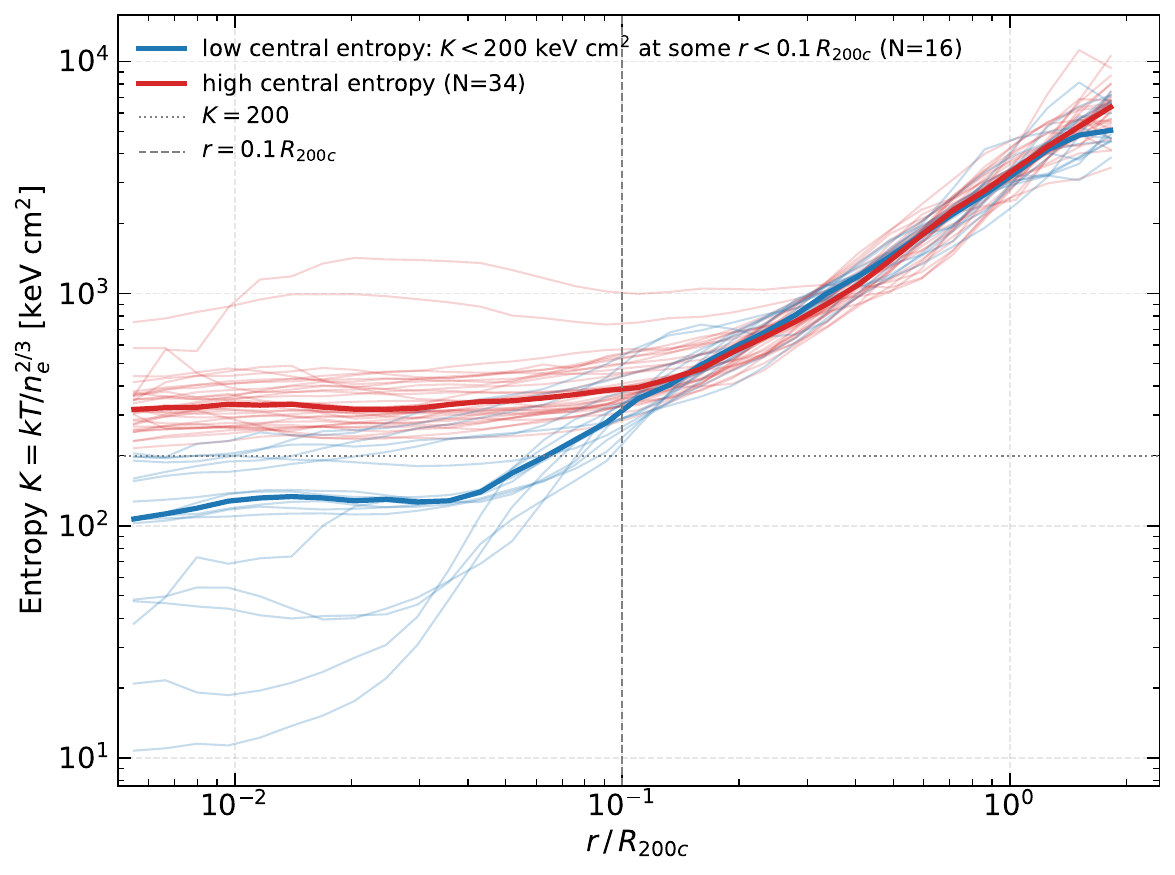}
    \includegraphics[width=0.49\textwidth]{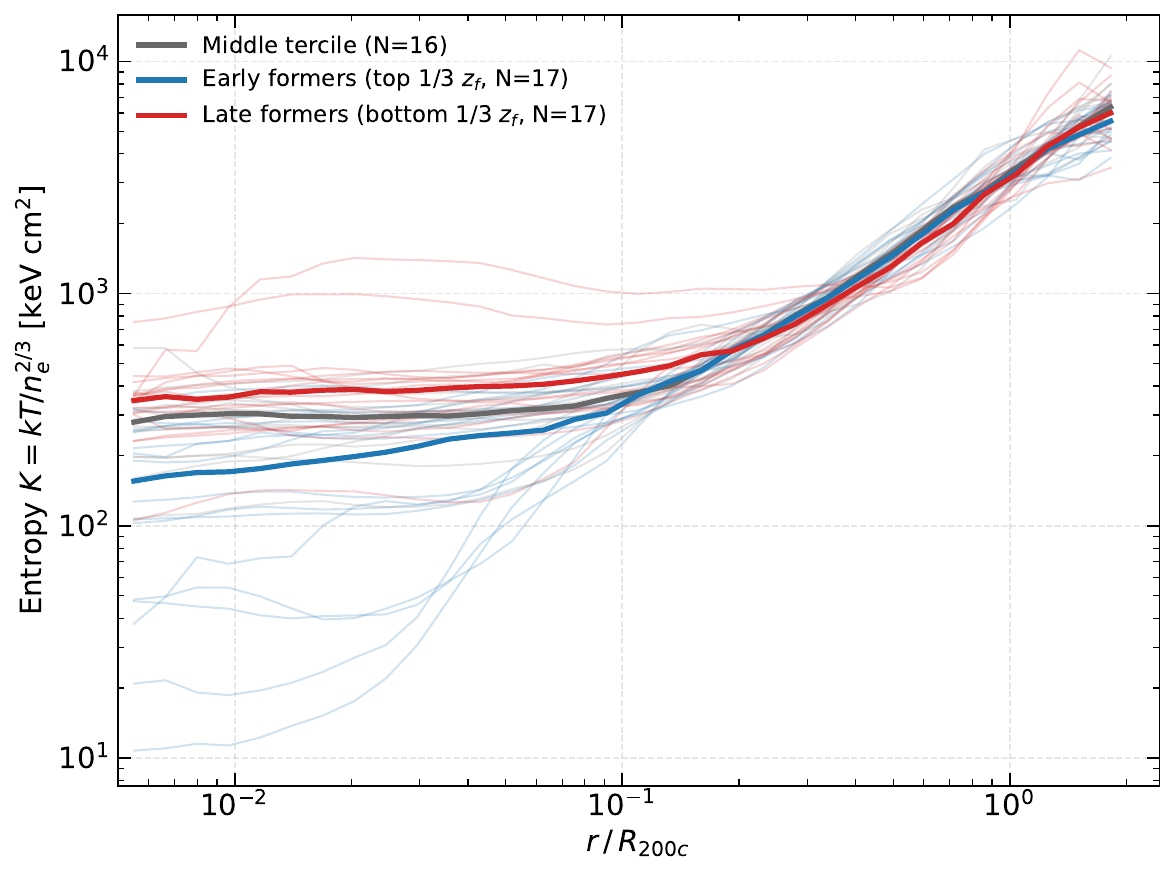}
    \caption{Entropy profiles for all 50 realisations, split two ways. \textit{Left:} physical CC threshold --- CC (blue, $N=16$) = $K<200\,\mathrm{keV\,cm^2}$ at some $r<0.1\,R_{200c}$; non-CC (red, $N=34$) otherwise. \textit{Right:} formation-time terciles by $z_f(10\%)$ (the redshift at which the main progenitor first reached 10\% of its final $M_{200c}$) --- early formers (blue), late formers (red), middle tercile (grey). Individual profiles thin; per-group medians bold.}
    \label{fig:entropy_split}
\end{figure*}

\begin{figure*}
    \includegraphics[width=0.49\textwidth]{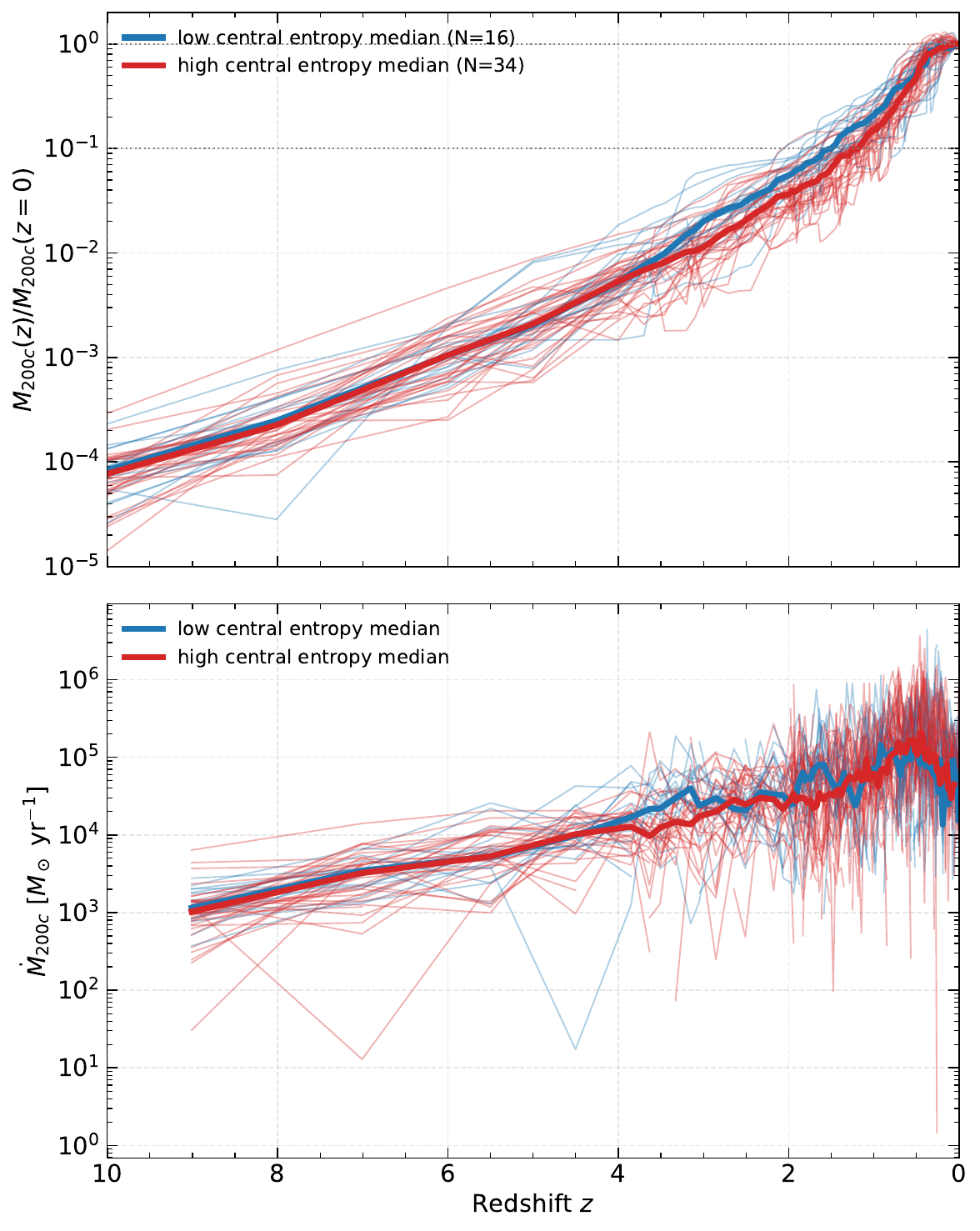}
    \includegraphics[width=0.49\textwidth]{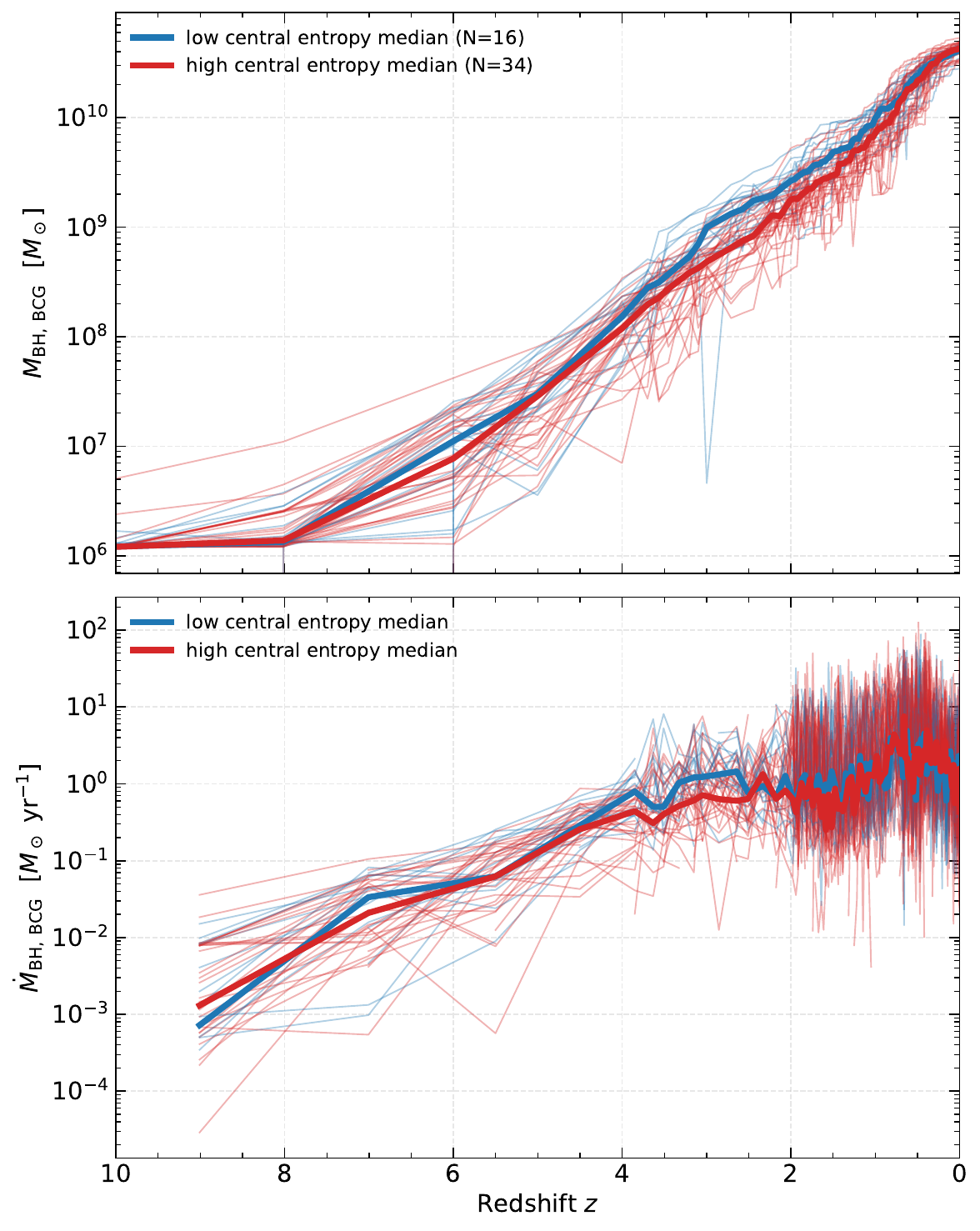}
    \caption{Mass accretion history for the host halo (\textit{left}) and for the central BCG black hole (\textit{right}), split by low-entropy (blue) and high-entropy (red). Top panels show the mass $M(z)$; bottom panels the accretion rate $\dot M(z)$. The thin lines represent individual clusters and the bold lines are the sample medians. Low-entropy progenitors reach 10\% of $M_{200c}(z=0)$ $\sim$1.3\,Gyr earlier than high-entropy progenitors and have more massive central black holes by $z\approx1$ (median $10^{10}\,{\rm M}_\odot$ vs $7\times10^9\,{\rm M}_\odot$), with a slower late-time black hole growth factor ($\sim$4$\times$ vs $\sim$6$\times$). The black hole and halo accretion histories track each other well (Pearson correlation coefficient of $0.79$ in log-log) so the black hole trend is not an independent physical probe. However, it underpins the fact the  formation history of clusters drives the shape of the radial profile evolution for Compton-$y$ and X-ray emission, and not random AGN-feedback episodes.}
    \label{fig:mah_panel}
\end{figure*}

\begin{figure*}
    \includegraphics[width=\textwidth]{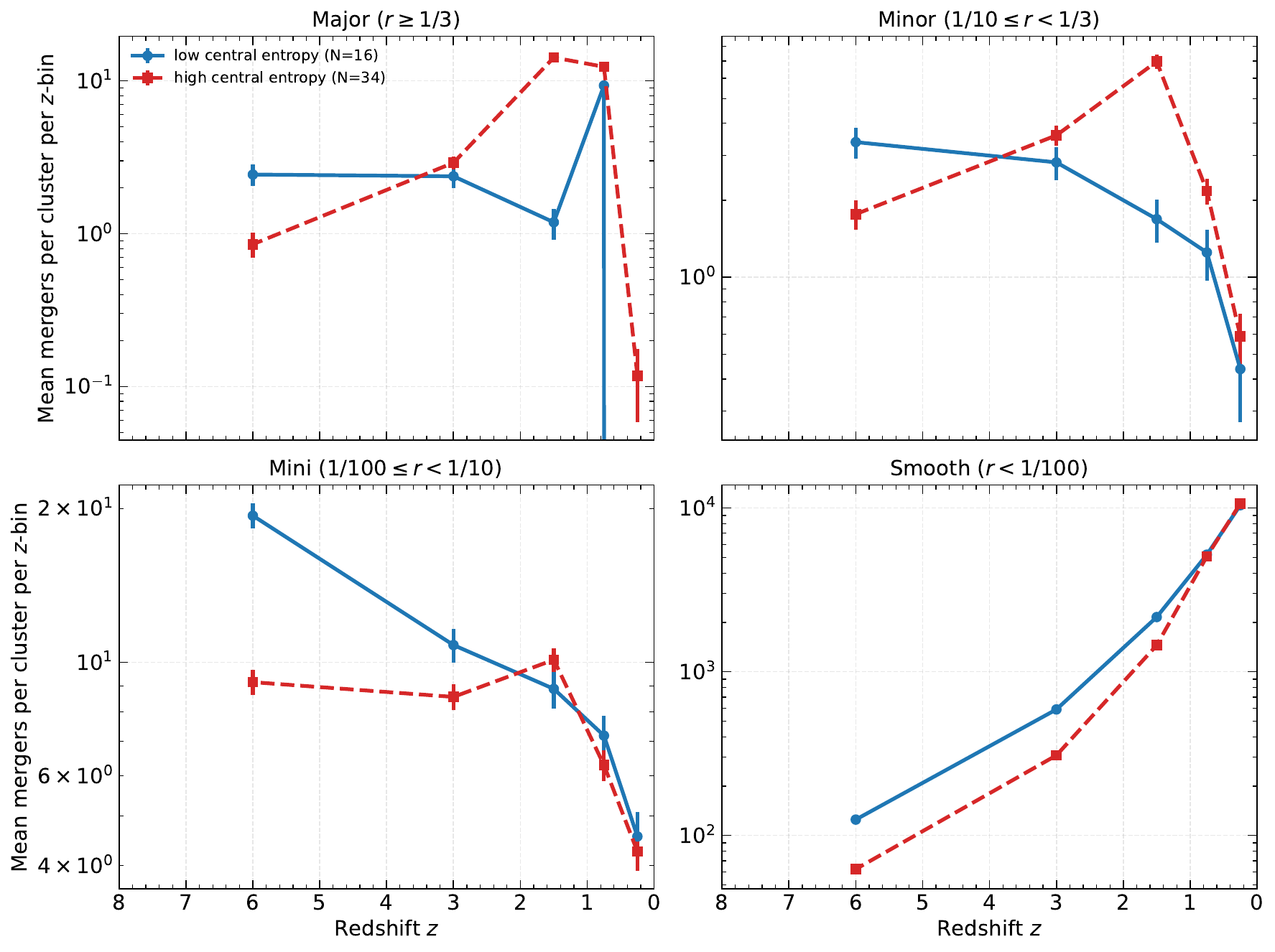}
    \caption{Mean number of FoF mergers per cluster per redshift bin, split into four mass-ratio classes: major ($r\geq1/3$), minor ($1/10\leq r<1/3$), mini ($1/100\leq r<1/10$), and smooth ($r<1/100$). Low-entropy clusters (blue) and high-entropy clusters (red) are shown with Poisson error bars. Redshift bins: $[0,0.5], [0.5,1], [1,2], [2,4], [4,8]$; $z=0$ on the right. The mini-merger channel significantly distinguishes the two populations, while major merger counts alone are statistically indistinguishable.}
    \label{fig:fof_mergers}
\end{figure*}

\begin{figure}
    \includegraphics[width=\columnwidth]{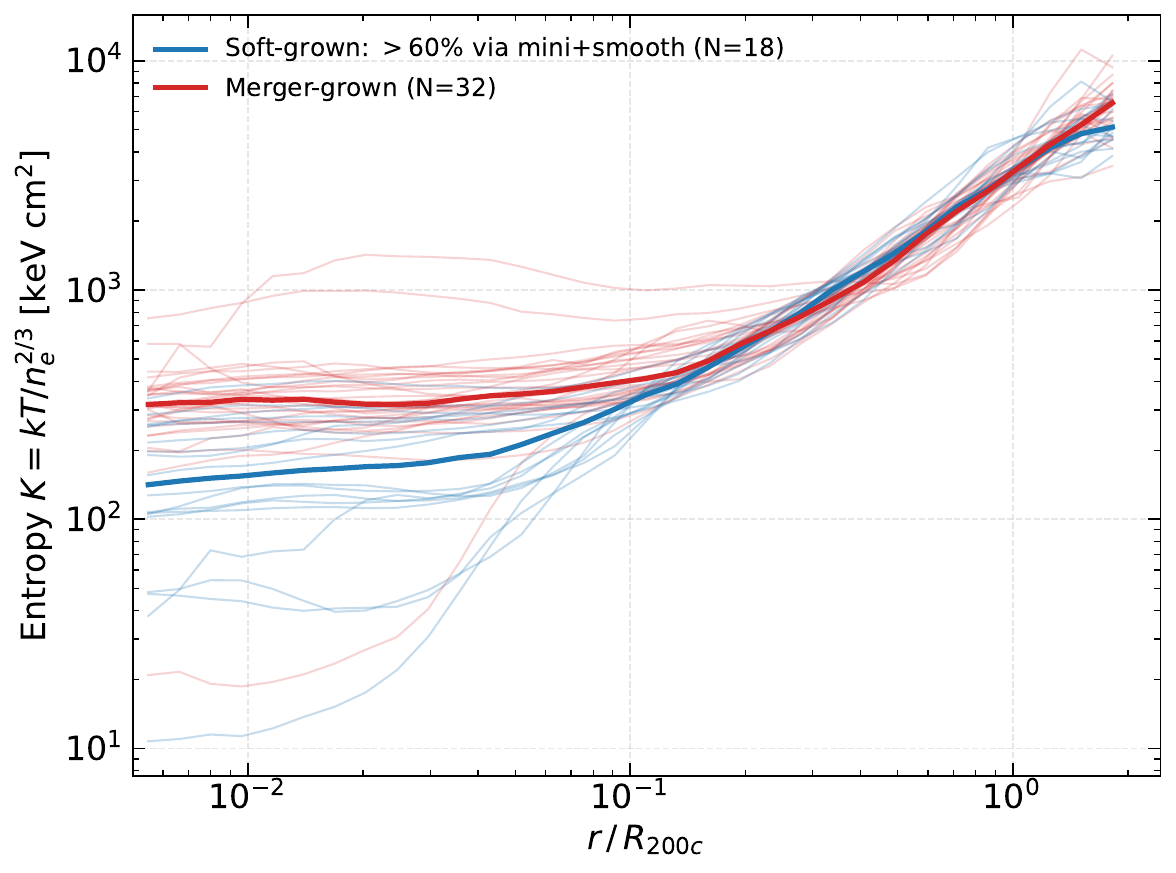}
    \caption{Entropy profiles split by mass-growth mode. Soft-grown clusters (blue, $N=18$) have accreted more than 60\% of their $M_{200c}(z=0)$ through mini mergers ($1/100\leq r<1/10$) plus smooth accretion ($r<1/100$); merger-grown (red, $N=32$) are the rest. The soft-grown subset is 67\% low-entropy clusters (12/18) and covers 75\% of all low-entropy clusters, making this the most effective low-entropy cluster predictor in our simulated data.}
    \label{fig:entropy_mass_budget}
\end{figure}

\begin{figure*}
    \includegraphics[width=\textwidth]{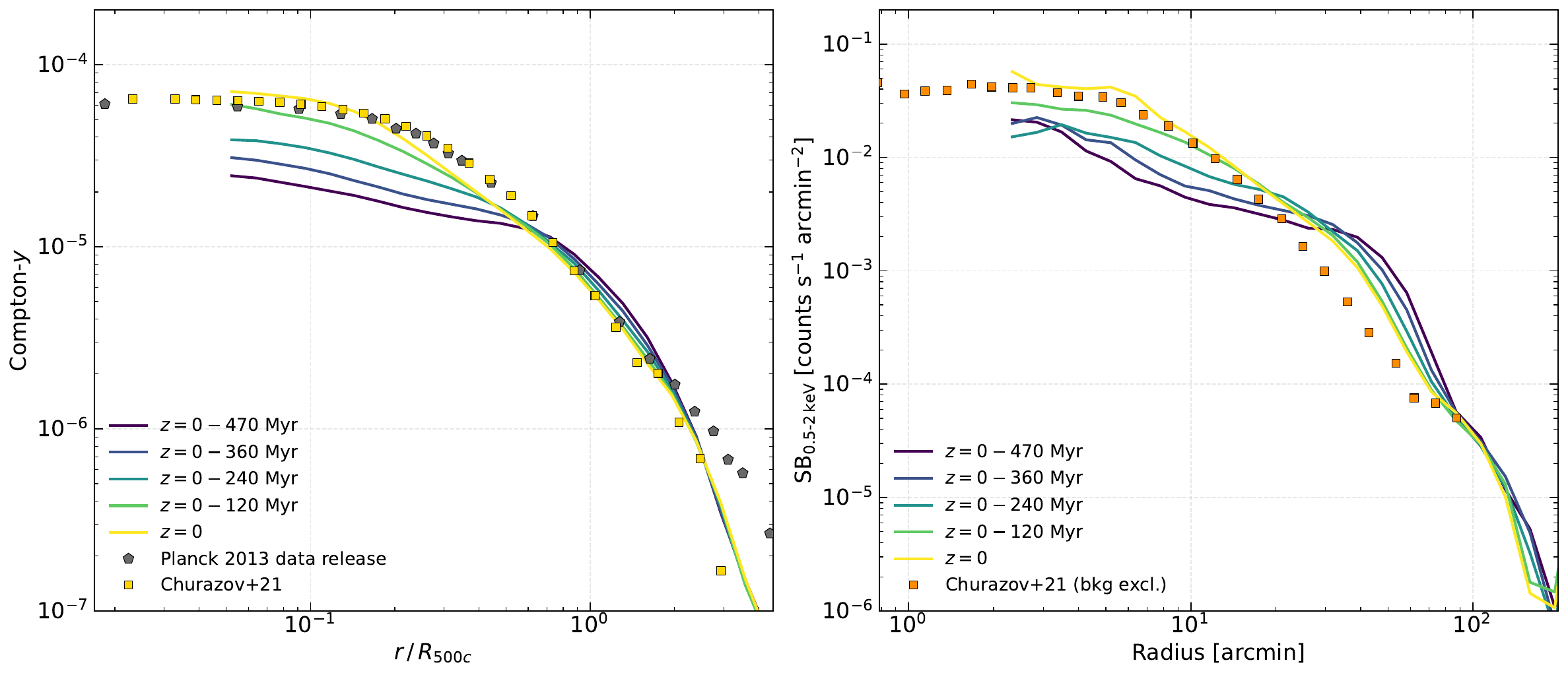}
    \caption{Example cluster that shows rapid change in the radial profiles in the last 500 Myrs through late mass assembly (Coma\_042). \textit{Left:} Compton-$y$ profile vs $r/R_{500c}$ with \citet{Planck2013} data overlaid. \textit{Right:} X-ray surface brightness vs arcmin with \citet{Churazov2021} data. The colour coded lines represent the time evolution of the cluster within the last 500 Myrs. For this particular cluster the late mass accretion history significantly influences the clusters emission. This example is noteworthy, in so far that this is our best fit cluster compared to Coma's radial profile for X-ray and Compton-$y$ emission.}
    \label{fig:step042}
\end{figure*}

\begin{figure*}
    \includegraphics[width=\textwidth]{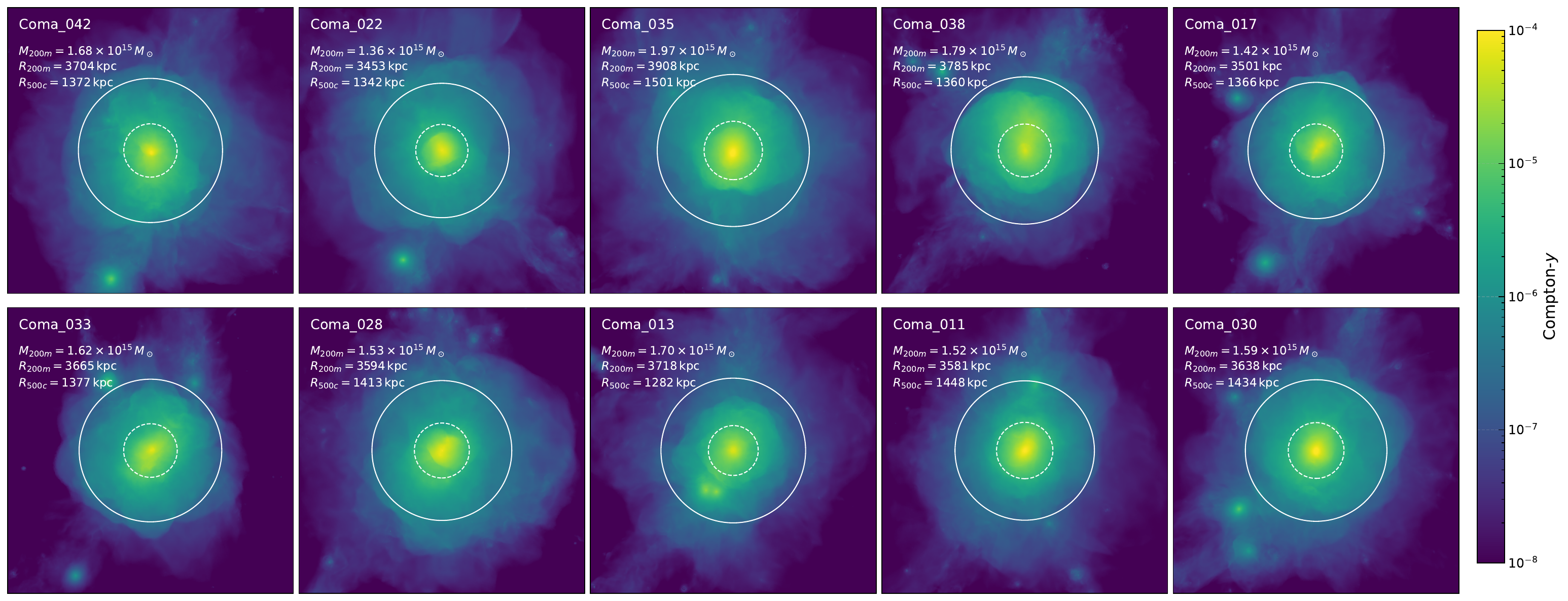}\\[0.2em]
    \includegraphics[width=\textwidth]{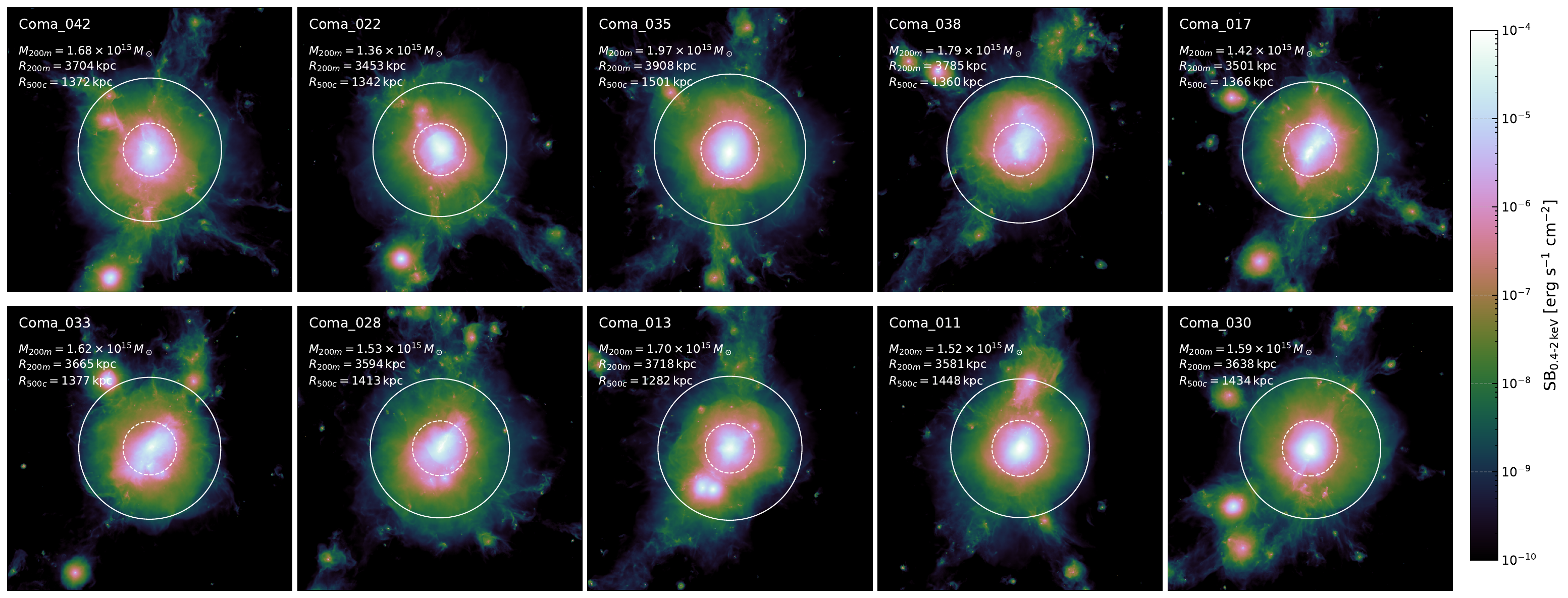}\\[0.2em]
    \caption{Projected maps of the 10 best-fitting realisations (same combined ranking as Fig.~\ref{fig:best10_profiles}), arranged from best fit (top left) to 10th best (bottom right) within each observable. \textit{Top two rows:} Compton-$y$ parameter, logarithmic colour scale from $10^{-8}$ to $10^{-4}$. \textit{Bottom two rows:} X-ray surface brightness in the 0.4--2\,keV band, logarithmic colour scale from $10^{-10}$ to $10^{-4}\;\mathrm{erg\,s^{-1}\,cm^{-2}}$. Each panel shows a $10\times10\;\mathrm{Mpc}$ field of view; dashed and solid white/black circles mark $R_{500c}$ and $R_{200m}$. The morphological diversity across the best-fitting clusters --- from relaxed, centrally peaked systems to disturbed merger remnants --- illustrates the range
    of dynamical states consistent with the observed Coma radial profiles.}
    \label{fig:top10_maps_combined}
\end{figure*}

\begin{figure}
      \includegraphics[width=0.49\textwidth]{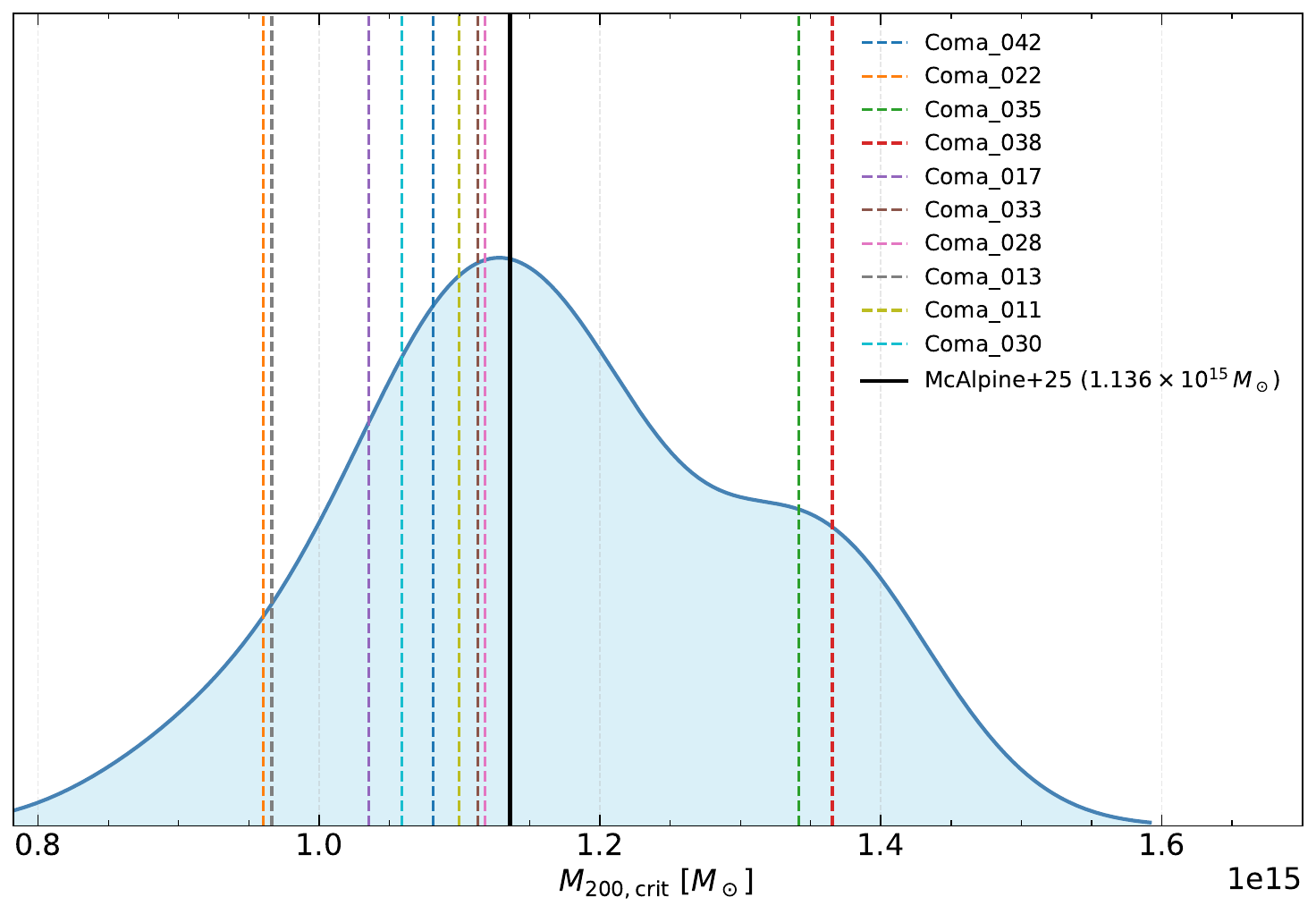}
      \caption{Kernel density estimate of the $M_{200c}$ halo mass distribution across all 50 constrained Coma realisations. The light blue shaded curve shows the full distribution. Coloured dashed vertical lines mark the masses of the 10 best-fitting clusters (same combined ranking as in Fig.~\ref{fig:best10_profiles}), each identified by its Coma\_XXX label in the legend. The solid black line indicates the Coma cluster mass of $1.136\times10^{15}\,{\rm M}_\odot$ from the \texttt{Manticore-Local} posterior of \citet{McAlpine2025}. The best-fitting realisations span a range of halo masses rather than clustering at a single value, indicating that the quality of the radial profile fit is not solely determined by total mass.}
      \label{fig:mass_kde}
\end{figure}

\subsection{A constrained Coma analogue: Coma\_025}
\label{sec:coma025_showcase}

Before turning to the statistical comparison across the full ensemble, we showcase a single representative realisation, Coma\_025, which combines a Coma-analogue total mass with the presence of a massive companion analogous to the NGC~4839 group. Fig.~\ref{fig:coma025_showcase} summarises this realisation in a single multi-panel figure. The large top panel shows the projected gas surface density across a $10\times10\;\mathrm{Mpc^2}$ field of view, with the white dashed circle marking $R_{500c}$. We also colour code the velocity streamlines, as well as the dissipated shock energy to visualise the ongoing cluster evolution. The streamlines show that the system is still strongly accreting, along at least two prominent filaments, and the dissipated energy reveals that there is a prominent accretion shock located at roughly 2 times $R_{200c}$. Furthermore, the dissipated shock energy illuminates a number of stronger internal shocks within $R_{500c}$. The Coma\_025 box has been reoriented using the symmetries of the projection plane to place its NGC~4839 analogue in the lower-right quadrant, matching the south-west sky position of the real NGC~4839 group relative to Coma in the astronomical convention (east-left, north-up).

Overlaid on the gas density are mass-weighted in-plane velocity streamlines, with the bulk halo motion subtracted so that internal flows are visible; the streamlines are coloured by the local in-plane gas speed (cool palette). The inset at the upper right shows the BCG stellar light in a $\pm100\;\mathrm{kpc}$ zoom around the cluster centre, with a thin connector indicating the corresponding region in the parent panel. Two transparent insets at the lower edge of the gas panel show the azimuthally averaged Compton-$y$ profile (left) and the 0.5--2\,keV X-ray surface brightness profile (right) for Coma\_025, compared with three observational reference profiles. The first is a Planck-based Compton-$y$ profile reanalysed by \citet{Churazov2021} with a different background-subtraction procedure than the original Planck pipeline (golden line). The second is a profile we re-extract directly from the publicly released \citet{Planck2013} Compton-$y$ maps (grey line) using the same azimuthal-binning routine that we apply to our simulated maps. Showing both profiles is useful because the two background-subtraction approaches differ on the cluster outskirts, and they bracket the level of systematic uncertainty in the observed Coma tSZ signal at large radius. The third is the background-subtracted SRG/eROSITA X-ray surface brightness profile of Coma by \citet{Churazov2021}, plotted in the right-hand inset.

The middle row of Fig.~\ref{fig:coma025_showcase} compares the simulated and observed sky in the same observable: the leftmost panel is the Planck Compton-$y$ map of the real Coma cluster, the second panel is the Coma\_025 Compton-$y$ map on the same scale, the third panel is the observed X-ray surface brightness from eROSITA, the background excluded image from \citet{Churazov2021} and shares the colour map with the fourth panel, the simulated X-ray surface brightness of Coma\_025. Each of the four small panels has the same field of view ($\pm 2.5\,R_{500c}$). The bottom row shows four further physical quantities of Coma\_025 across the full $10\times10\;\mathrm{Mpc}$ field: dark-matter surface density, mass-weighted gas metallicity (in solar units), mass-weighted gas magnetic-field strength (in $\mu$G), and stellar surface density. Together with the streamline overlay in the top panel, this figure provides a self-contained visual summary of the physical state of Coma\_025: the spatial alignment of dark matter, gas, stars, metals and magnetic fields, the presence and orientation of the NGC~4839 analogue, the bulk and turbulent gas motions, and the level of agreement between the simulation and the existing tSZ and X-ray data sets that we will use to evaluate the entire ensemble in the following sections. The qualitative takeaway from this figure is that Coma\_025 already provides, by eye, a strong analogue of the observed Coma cluster: the projected morphology, the location and orientation of the NGC~4839-like companion, and the broad shape and normalisation of the radial Compton-$y$ and X-ray surface brightness profiles all match the observational reference data sets reasonably well, even before any quantitative ranking against the full ensemble.

\subsection{Overview of the Simulations}
\label{sec:overview}

We start our presentation of the results by showcasing our 50 simulations of the Coma cluster. We list some of the general cluster properties in Table~\ref{tab:coma_properties} including $M_{200c}$, $R_{200c}$, $M_{500c}$, $R_{500c}$, the total stellar mass within $R_{200m}$, the gas fraction within $R_{500c}$, the virial temperature in keV, the stellar mass of the BCG and the mass of its central black hole.
In Fig.~\ref{fig:dm_density} we show the dark matter surface density, and in Fig.~\ref{fig:gas_density} we give overview plots for all of our 50 simulations that are digital twins of the Coma galaxy cluster, for the dark matter surface density and the (total) gas surface density. The realisation-to-realisation differences are dominated by small-scale structure and the configuration of the central merger --- in particular the position, mass, and orbital phase of substructure in the immediate vicinity of the cluster --- while the broader large-scale environment fixed by the constrained posterior remains consistent across the ensemble: the cluster location, the orientation of the dominant filaments, and the presence and approximate geometry of the NGC~4839-like infall companion are all recognisable in every realisation.

\subsection{Galaxy cluster scaling relations}
\label{sec:scaling_relations}

We begin by examining how the 50 constrained Coma analogue realisations compare with established galaxy cluster scaling relations. All realisations are drawn from the same \texttt{BORG}/\texttt{Manticore} posterior of the local large-scale structure and are evolved with identical TNG300 galaxy-formation physics, but each corresponds to a different posterior sample, so the cluster-to-cluster scatter visible in the scaling-relation plots reflects posterior variation across both large and small scales: differences in the surrounding filamentary structure, the orientation and timing of nearby companion infalls, the assembly history of the main halo, and the small-scale perturbations that determine the inner ICM and central-galaxy state. We acknowledge that the choice and presentation of these scaling relations was directly inspired by the TNG-Cluster presentation paper of \citet{Nelson2024}; we follow their selection of relations and observational compilations to enable a like-for-like reading at the cluster mass scale.

Fig.~\ref{fig:fgas} shows the ICM gas fraction $f_{\rm gas}$ within $R_{500c}$ as a function of $M_{500c}$. Our 50 constrained Coma analogues are represented by the blue data points. The red diamonds show the X-ray group and cluster measurements from \citet{Giodini2009}, which span a wide range in halo mass from groups ($\sim 10^{13}\,{\rm M}_\odot$) to massive clusters. The green triangles are taken from the multi-wavelength cluster sample of \citet{Gonzalez2013}. The horizontal grey dotted line marks the cosmic baryon fraction $\Omega_b/\Omega_m \approx 0.159$. The \texttt{Manticore-Local} posterior places relatively tight constraints on the Coma halo mass, so our constrained Coma analogues naturally occupy a narrow high-mass window and do not sample the lower-mass regime covered by the observational compilations. In the high-mass cluster regime where our data points lie, the simulated gas fractions sit comfortably within the spread of observational estimates, falling in the range $f_{\rm gas} \sim 0.10$--$0.15$. Given the substantial object-to-object scatter in the observational compilations themselves at fixed halo mass, the agreement is in fact strong, and is consistent with the expectation that massive clusters retain a near-universal baryon fraction within $R_{500c}$.

Fig.~\ref{fig:ysz_lx} presents the integrated thermal Sunyaev--Zel'dovich signal $Y_{\rm SZ,500}$ (top panel) and the soft X-ray luminosity $L_X$ in the 0.5--2\,keV band (bottom panel), both as a function of $M_{500c}$. In the top panel, our simulations (blue points) are compared with the \citet{Planck2013} $Y$--$M$ scaling relation (solid red line, assuming a hydrostatic mass bias of $1-b = 0.8$) and the self-similar expectation $Y \propto M^{5/3}$ (grey dashed line, normalised to the median of our sample). The simulated $Y_{\rm SZ}$ values lie close to the central \citet{Planck2014} relation across the narrow mass window covered by our ensemble, with no significant systematic offset. In the bottom panel, the X-ray luminosities of our constrained Coma analogues are compared with the REXCESS cluster sample from \citet[][grey squares]{Pratt2009} and the low-redshift sample from \citet[][silver diamonds]{Vikhlinin2009}. Our simulated luminosities sit comfortably within the observed scatter of $L_X$ at the high-mass end. The realisation-to-realisation spread is itself a result of the posterior variation across the constrained ensemble, and the physical origin of this spread --- predominantly the central entropy state of each Coma analogue and its dependence on the long-term assembly history --- is the subject of Sec.~\ref{sec:central_scatter} and is revisited in the discussion (Sec.~\ref{sec:disc_central_variability}). However, for a rough comparison we mark the Coma cluster as a gold star in the bottom panel of Fig.~\ref{fig:ysz_lx}. Its total $0.5$--$2$\,keV luminosity within $R_{500c}$, estimated from the ROSAT and SRG/eROSITA radial profiles \citep[e.g.,][]{Churazov2021}, is $L_X\simeq(2.5\pm0.5)\times10^{44}\,
\mathrm{erg\,s^{-1}}$ at $M_{500c}\approx6.5\times10^{14}\,M_\odot$, with the
dominant uncertainty set by the treatment of the angular wedge towards NGC~4839 \citep[e.g.,][]{Churazov2023}. Coma lies near the lower-$L_X$ edge of our simulated distribution, consistent with its well-known status as a dynamically active, non-cool-core system.

In Fig.~\ref{fig:lx_mstar} we show the soft X-ray luminosity $L_X$ (0.5--2\,keV) as a function of the BCG stellar mass $M_{\star,\rm BCG}$, measured within a $30\,{\rm kpc}$ aperture. Our simulations (blue points) are compared with the ROSAT stacking results of \citet[][black squares]{Anderson2015}, who measured mean X-ray luminosities in bins of stellar mass for $\sim$250\,000 locally brightest galaxies from SDSS. The dashed black line shows their best-fit power-law relation with slope $\alpha = 3.34$. Our simulated Coma clusters occupy the high stellar mass end of this relation ($M_{\star,\rm BCG} \sim 10^{11.91}$--$10^{12.6}\,{\rm M}_\odot$) and their X-ray luminosities are consistent with the extrapolation of the \citet{Anderson2015} relation to the most massive systems. To guide the eye we show Coma as a gold star in Fig.~\ref{fig:lx_mstar}. Its brightest galaxy, NGC~4889, has an estimated central stellar mass of $\sim(1.26\pm0.26)\times10^{12}\,M_\odot$ \citep[e.g.,][]{Bogdan2018}, while the second-brightest galaxy, NGC~4874, is $0.4$\,mag fainter ($\sim(0.9\pm0.2)\times10^{12}\,M_\odot$). Given the long-standing debate over which galaxy is the true BCG (i.e.\ which sits closest to the global potential
minimum) and the uncertain mass of the giant galaxy their eventual merger will form, we bracket the Coma BCG mass as $M_{\star,\rm BCG}\sim(1.5\pm0.5)\times 10^{12}\,M_\odot$ for a qualitative comparison; combined with $L_X\simeq
(2.5\pm0.5)\times10^{44}\,\mathrm{erg\,s^{-1}}$ it falls squarely within the locus traced by our simulated zooms.

Fig.~\ref{fig:smbh} presents the supermassive black hole scaling relations for the BCG in each realisation. The left panel shows the black hole mass $M_{\rm BH}$ as a function of $M_{500c}$, compared with the dynamical black hole mass measurements in groups and clusters from \citet[][open black squares]{Bogdan2018}. Our simulated BCGs harbour black holes with masses $M_{\rm BH} \sim 10^{10.45}$--$10^{10.73}\,{\rm M}_\odot$, broadly consistent with the observed trend at the Coma mass scale. The right panel shows $M_{\rm BH}$ versus the one-dimensional stellar velocity dispersion $\sigma_{\star}$, measured from star particles within $10\, {\rm kpc}$ of the BCG centre (mass weighted). The dashed black line shows the $M_{\rm BH}$--$\sigma$ relation of \citet{McConnell2013} and the red dash-dotted line the relation from \citet{Kormendy2013}. The simulated black holes fall between the two empirical relations, with velocity dispersions in the range $\sigma_\star \sim 299$--$640$\,km\,s$^{-1}$, as is typical of massive BCGs. 
We additionally overplot the observed Coma cluster (gold star) in both panels of Fig.~\ref{fig:smbh}. The dynamical SMBH mass of NGC~4889 is $M_{\rm BH}=(2.1\pm1.6)\times10^{10}\,M_\odot$ \citep[Table~1 of][]{Bogdan2018}, and a scaling-relation estimate for NGC~4874 gives $M_{\rm BH}\sim1.7\times
10^{10}\,M_\odot$; we therefore bracket the Coma value as $M_{\rm BH}\sim (3\pm1)\times10^{10}\,M_\odot$, plotted against $M_{500c}\approx6.5\times 10^{14}\,M_\odot$ (left) and the NGC~4889 stellar velocity dispersion $\sigma_\star\approx347\,\mathrm{km\,s^{-1}}$ \citep{Bogdan2018} (right). This bracket lies at the low-mass end of, but consistent with, our simulated BCG black holes.

Fig.~\ref{fig:stellar_mass} shows the stellar mass content of the simulated clusters. The left panel displays the BCG stellar mass $M_{\star,\rm BCG}$, measured within a $30\,{\rm kpc}$ aperture, as a function of $M_{500c}$. Our simulations (blue points) are compared with the individual cluster measurements from \citet[][red circles]{Kravtsov2018} and their best-fit power-law relation (dashed black line). The right panel shows the total halo stellar mass within $R_{500c}$ as a function of $M_{500c}$, where the ``total halo stellar mass within $R_{500c}$'' is the sum of the stellar masses of all subhaloes that (i) are members of the same Friends-of-Friends (FoF) group as the Coma analogue and (ii) have their centre of mass inside the $R_{500c}$ spherical aperture; this includes the BCG and all bound satellite galaxies whose centres lie inside the aperture, but excludes diffuse intracluster light not bound to any subhalo. The simulation values are compared with the deprojected stellar-mass estimates of \citet[][green triangles]{Gonzalez2013} and the total stellar-mass measurements of \citet[][red circles]{Kravtsov2018}. In both panels, our simulated clusters are in reasonable agreement with the observations, though we note a mild tendency for the simulated BCG stellar masses to fall slightly below the \citet{Kravtsov2018} relation, which may reflect the choice of aperture ($30\,{\rm kpc}$ vs.~their extrapolated total light).

Finally, Fig.~\ref{fig:satellite_profiles} shows the projected radial surface number density profiles $\Sigma_{\rm gal}(r)$ of satellite galaxies around the Coma analogue in each realisation. Profiles are computed for three stellar mass thresholds: $M_\star > 10^{9}\,{\rm M}_\odot$ (blue solid), $M_\star > 10^{10}\,{\rm M}_\odot$ (green dashed), and $M_\star > 10^{10.5}\,{\rm M}_\odot$ (red dotted). Solid lines show the mean across the 50 realisations and shaded bands indicate the 16th--84th percentile scatter. The black circles show the observational data from the SDSS-based cluster stacking analysis of \citet{Budzynski2012} for their highest mass bin ($\log M_{500} = 14.7$--$15.0$, $M_r \leq -20.5$). For a cluster-mass passive galaxy population, the \citet{Budzynski2012} luminosity cut $M_r \leq -20.5$ corresponds approximately to a stellar-mass threshold of $M_\star \gtrsim 10^{10.5}\,M_\odot$ \citep[adopting a typical $M_\star/L_r$ for red-sequence galaxies; e.g.][]{Bell2003}, so the most relevant comparison is with our intermediate stellar-mass-threshold curve rather than with the lowest threshold. The simulated galaxy density profiles are broadly consistent with the observed data, reproducing the steep decline in satellite number density from the cluster core outward. The scatter across realisations is relatively modest, suggesting that the radial distribution of satellite galaxies is primarily determined by the halo mass rather than the details of the assembly history.

\subsection{Projected maps of the ICM}
\label{sec:projected_maps}

Having established that the 50 constrained Coma analogues reproduce the key cluster scaling relations, we now turn to the spatially resolved properties of the intracluster medium. For each realisation we compute projected 2D maps along the $z$-axis through a $10 \times 10\;\mathrm{Mpc}$ field of view centred on the Coma analogue. The projection integrates through the full depth of the box along the line of sight using all non-star-forming gas cells; the dense, cold star-forming phase --- which in TNG is described by the subgrid two-phase ISM model of \citet{Springel2003} and is therefore an effective-EoS artefact rather than a faithful representation of the local thermal state of the gas --- is excluded so that the maps trace only the volume-filling, hot, diffuse ICM, following standard practice in the cluster literature. Two observables are computed: the thermal Sunyaev--Zel'dovich (tSZ) Compton-$y$ parameter and the X-ray surface brightness in the 0.4--2\,keV band.

Fig.~\ref{fig:maps_8} presents the two projected maps --- Compton-$y$ (top row) and X-ray surface brightness (bottom row), for 8 representative realisations (Coma\_000--Coma\_007) in a single combined figure. The Compton-$y$ maps in the top row are computed as $y = (\sigma_{\rm T} / m_e c^2) \int n_e\,k_{\rm B} T\,\mathrm{d}l$ and shown on a logarithmic colour scale from $10^{-8}$ to $10^{-4}$. All 8 realisations show a bright, centrally concentrated tSZ signal extending beyond $R_{500c}$; despite sharing the same large-scale environment, the detailed morphology varies considerably. Some realisations display a smooth, relaxed appearance (e.g.\ Coma\_000), while others exhibit elongated or asymmetric features indicative of recent or ongoing mergers (e.g.\ Coma\_005). Since each realisation corresponds to a different posterior sample of the local-universe inference, this morphological diversity reflects posterior variation in the surrounding large-scale environment, the filamentary structure, the merger history, and the present-day dynamical state of the Coma analogue --- rather than a range of dynamical states sampled within a single constrained volume.

The bottom row shows the projected X-ray surface brightness maps in the 0.4--2\,keV band. The emissivity is computed using APEC cooling tables \citep{Smith2001, Foster2012} under the assumption of collisional ionisation equilibrium, using the per-particle hydrogen fraction, electron density, and temperature, with the logarithmic colour scale spanning $10^{-10}$ to $10^{-4}\;\mathrm{erg\,s^{-1}\,cm^{-2}}$. The X-ray emission is more centrally concentrated than the tSZ signal, reflecting its stronger dependence on gas density ($\propto n_e^2$ rather than $\propto n_e T$). Substructure --- infalling groups, merging subclusters, and filamentary accretion streams --- is clearly visible in several realisations, particularly beyond $R_{500c}$.

Taken together, the two rows of Fig.~\ref{fig:maps_8} demonstrate that the constrained Coma analogues span a plausible and representative range of ICM morphologies and dynamical states for clusters of Coma's mass: relaxed and disturbed configurations both appear, the substructure population looks reasonable, and we do not find any obviously pathological behaviour in the maps (e.g.\ runaway central cooling, extreme gas-mass loss, or dramatic dark-matter--gas offsets). This is a useful sanity check before turning to the quantitative radial-profile comparison with the observed Coma cluster in the following sections.

\subsection{Radial profiles and best-fitting realisations}
\label{sec:radial_profiles}

To quantitatively compare the simulated clusters with observations of the real Coma cluster, we compute azimuthally averaged radial profiles of the Compton-$y$ parameter and the X-ray surface brightness for all 50 realisations.

Fig.~\ref{fig:profiles_all50} shows the radial profiles for the full set of 50 simulations. In the left panel, the Compton-$y$ profiles are plotted as a function of projected radius normalised by $R_{500c}$. Each green line corresponds to one realisation, with increasing opacity to aid visual distinction. The grey pentagons show the observed Coma Compton-$y$ profile from \citet{Planck2013}, and the solid red line shows the lightcone-stacked Compton-$y$ profile of Coma-mass ($M_{200c}\sim 10^{15}\,{\rm M}_\odot$) clusters in the MillenniumTNG simulation \citep{Pakmor2023}, providing a complementary reference point from a much larger random-IC sample at the same mass scale. The ensemble of simulated profiles brackets the Planck data points, with the inner profiles ($r \lesssim 0.5\,R_{500c}$) showing good agreement and increasing scatter at larger radii where the contribution from infalling substructure and the accretion shock become significant. In the right panel, the X-ray surface brightness profiles in the 0.4--2\,keV band are shown as a function of angular radius in arcmin, assuming $R_{200c} \approx 70$\,arcmin for Coma at a distance of $\sim$100\,Mpc. Blue and orange circles show the SRG/eROSITA observations from \citet{Churazov2021}, with and without background subtraction, respectively. The simulated profiles span a wide range in normalisation, reflecting the strong sensitivity of $L_X \propto n_e^2$ to the central gas density and cool-core properties. The observed \citet{Churazov2021} profile falls well within the envelope of simulated profiles. 

We caution that the observed X-ray image, and hence the radial profile beyond $r=60'$, is likely affected by the far wings of eROSITA's PSF arising from singly scattered (``stray-light'') photons \citep[as illustrated in Fig.~4 of][]{Churazov2023}. Although this contribution can be modelled and approximately corrected for \citep{Churazov2023}, the complex spatial
distribution of the primary X-ray emission in Coma leads to significant uncertainty in the observed profile at these radii.

Within this ensemble we define a subset of 10 best-fitting realisations which we examine in detail in this section and continue to follow through Sects.~\ref{sec:bestfit_maps} (projected maps of the best-fitting clusters) and \ref{sec:mass_distribution} (their halo-mass distribution); the same subset reappears in the time-variability and central-scatter analyses of Sects.~\ref{sec:time_variability} and \ref{sec:central_scatter}.
To identify which realisations best reproduce the observed Coma cluster, we rank all 50 simulations by their combined goodness of fit to both the \citet{Planck2013} Compton-$y$ profile and the \citet{Churazov2021} X-ray profile (background-subtracted). The ranking is based on the sum of the mean squared errors (MSE) computed in $\log_{10}$-space, interpolating each simulated profile onto the observed radial sampling points. 

Fig.~\ref{fig:best10_profiles} shows the radial profiles of the 10 best-fitting realisations in a two-panel layout. The left panel shows the Compton-$y$ profiles compared with
\citet{Planck2013} and the \citet{Churazov2021} reanalysis, while the right panel displays the X-ray surface brightness profiles compared with the background-subtracted \citet{Churazov2021} data (orange circles). Matching line colours identify the same realisation across both panels. The best-fitting clusters reproduce the observed profiles over more than two decades in radius, from the bright cluster core out to $\sim$3--4\,$R_{500c}$ in Compton-$y$ and $\sim$100\,arcmin in X-ray surface brightness. The top-ranked realisation (step\_042) provides an excellent match to both observables.

\subsection{Investigating the scatter in the centre}
\label{sec:central_scatter}

The large ensemble scatter in the inner radial profiles motivates a closer examination of the central thermodynamics and its link to the observed Compton-$y$ and X-ray brightness. We compute the azimuthally averaged gas entropy profile $K(r) = kT / n_e^{2/3}$ for each realisation from the three-dimensional pressure and density profiles.

Fig.~\ref{fig:entropy_vs_cc} shows $K(r)$ as a function of $r/R_{500c}$ for all 50 realisations. Each profile is colour-coded by the central observable brightness measured inside a $r<0.05\,R_{500c}$ aperture: the central Compton-$y$ parameter in the top panel, and the central X-ray surface brightness in the bottom panel. Cool-core clusters --- defined here as those with low central entropy --- populate the high-brightness tail in both observables, while clusters with flat, elevated entropy cores have correspondingly low central brightness. Fig.~\ref{fig:kcore_vs_brightness} quantifies this correlation by plotting the innermost-bin entropy $K_{\rm core}$ (evaluated at $r/R_{200c}\approx0.005$, corresponding to $\sim$10\,kpc physical for our clusters) against the central Compton-$y$ (left) and central X-ray surface brightness (right). Text annotations in each panel give the Pearson and Spearman correlation coefficients; the $\log K_{\rm core}$--$\log \mathrm{SB}_X$ correlation is particularly strong ($r=-0.87$, $\rho=-0.80$), confirming that central entropy is the primary driver of the cluster-to-cluster variance in soft X-ray brightness. This is an expected result given the long-established understanding of low central cluster entropies in which the central entropy structure of the ICM regulates the central density and hence the soft X-ray surface brightness \citep[e.g.][]{Voit2005,Cavagnolo2009,Pratt2010}; the present analysis recovers the same trend within a single ensemble of constrained-IC realisations of one cluster.

We split our cluster sample in two categories based on their central entropy in the innermost radial bin: a cluster is low-entropy if its entropy drops below $K(r)<200\;\mathrm{keV\,cm^2}$ at some $r<0.1\,R_{200c}$, and high-entropy otherwise\footnote{We specifically do not label them cool core and non cool core given the more stringent definitions in that particular literature. This is a deliberately inclusive threshold relative to the strong-CC and weak-CC entropy cuts ($K\lesssim 30$--$50\,\mathrm{keV\,cm^2}$) proposed by \citet{Hudson2010} on the basis of the ACCEPT entropy-profile catalogue \citep{Cavagnolo2009}. We relax the threshold to capture weak cool cores at our zoom-in resolution and to remain consistent with recent TNG-Cluster analyses \citep[e.g.][]{Lehle2024}}. With this criterion we find $N_{\rm lowK}=16$ and $N_{\rm highK}=34$ in our sample. Fig.~\ref{fig:entropy_split} shows the entropy profiles split two ways. The left panel is split by this physical threshold: lowK in blue, highK in red, with individual thin lines and bold per-group medians. The lowK population shows the canonical inner low-entropy dip while the two groups converge at the outskirts. The right panel splits the same profiles into terciles of formation redshift $z_f(10\%)$, defined as the redshift at which the main progenitor first reached 10\% of its $z=0$ mass: early formers (top tercile, blue), late formers (bottom tercile, red), and the middle tercile (grey). Nearly all lowK clusters fall in the top two terciles of $z_f(10\%)$. This is consistent with the long-standing theoretical expectation that clusters which finish their main accretion phase early give the central ICM time to develop and protect a low entropy state from violent disruption, while those that continue accreting at late times have their cool cores disrupted by merger heating and turbulent mixing \citep[e.g.][]{Burns2008,McCarthy2008,Rasia2015,Barnes2018}. That we recover that general trend within a single ensemble of constrained-IC realisations of one cluster is a useful confirmation of the established picture.

Fig.~\ref{fig:mah_panel} shows the halo mass accretion history $M_{200c}(z)$ and the central BCG black-hole mass history $M_{\rm BH}(z)$ in parallel, with $\dot M$ shown in the lower subpanel of each side. The left panels show the halo mass accretion history split by low entropy (blue) and high entropy (red), the thin lines correspond to individual clusters, and the bold lines represent the medians. Low entropy progenitor clusters appear to reach 10\% of $M_{200c}(z=0)$ approximately 1.3\,Gyr earlier than high entropy progenitor clusters, and their accretion rate drops off at late times while high entropy clusters continue assembling mass through $z<1$. The right panels in Fig.~\ref{fig:mah_panel} show the corresponding black hole mass history from the main-progenitor tree of each BCG, demonstrating that low entropy clusters had more massive central black holes already by $z=1$ (median $10^{10}\,M_\odot$ vs $7\times10^9\,M_\odot$) and grew only $\sim$4$\times$ from $z=1$ to today, while high entropy cluster black holes grew $\sim$6$\times$. The black hole growth trend correlates tightly with the halo mass accretion history trend, so the two histories convey largely overlapping information.

Fig.~\ref{fig:fof_mergers} presents the mean number of merger events per cluster per redshift bin, split into four mass-ratio classes: major ($r\geq1/3$), minor ($1/10\leq r<1/3$), mini ($1/100\leq r<1/10$), and smooth ($r<1/100$). Merger trees are built using the \textsc{Arepo} built-in implementation of the Millennium-style merger-tree scheme of \citet{Springel2005}: \textsc{Subfind} subhaloes are identified at each snapshot, FoF groups are constructed around them, and descendant pointers are computed between successive snapshots. A merger event is registered when a satellite \textsc{Subfind} subhalo enters the main FoF group of the cluster's main progenitor; the mass ratio $r$ for that event is computed from the FoF host masses of the two progenitors at the snapshot immediately before they coalesce. We refer to these events as ``FoF mergers'' throughout this section to emphasise that the mass ratio is host-mass-based rather than subhalo-mass-based. Redshift bins span $[0,0.5]$, $[0.5,1]$, $[1,2]$, $[2,4]$, $[4,8]$, with $z=0$ on the right of each panel. Low entropy clusters (blue) and high entropy clusters (red) are plotted with Poisson error bars. Only the mini-merger channel shows a statistically significant difference between the populations, where low-entropy clusters have roughly $\sim$2$\times$ as many mini merger events, while major-merger counts alone are statistically indistinguishable. To complement this count-based view with a mass-budget view, we sum the accreted secondary mass $M_{\rm sec}$ across all events in each ratio class and express the contribution as a fraction of the final $M_{200c}(z=0)$.

Fig.~\ref{fig:entropy_mass_budget} shows the entropy profiles split according to the resulting mass-growth mode. Soft-grown clusters (blue, $N=18$) are those whose combined mini plus smooth channels account for more than 60\% of their $M_{200c}(z=0)$; merger-grown clusters (red, $N=32$) are the rest. The soft-grown subset contains 12 of our 16 low-entropy clusters ($67\%$ purity, covering 75\% of all low-entropy clusters), making the mass-growth-mode split the single most effective merger-tree-based low-entropy predictor in our sample. The two most dramatic outliers (Coma\_000 and Coma\_028) in that metric built their mass predominantly through major mergers yet retained a cool core, and six high entropy clusters (Coma\_008, 011, 000, 036, 044, 045, 046) satisfy the soft-growth criterion yet failed to develop a cool core. These are the stress-test cases for the merger-history picture and point to additional core-scale regulation (AGN duty cycle, progenitor core gas density) not visible in the halo merger tree. A more detailed investigation of the dynamical state of those systems would be required to address this properly, but is beyond the scope of this work that aims to address the comparability of our simulation sample with important galaxy cluster observables.

\subsection{Radial profile time variability}
\label{sec:time_variability}

Finally, we want to briefly address the variability of the cluster radial profiles during late stage assembly of our massive galaxy clusters. Generally speaking the Compton-$y$ and X-ray radial profiles are remarkably constant through the last 500 Myrs of evolution of the cluster assembly, where most clusters that fit well at redshift zero, also fit well half a Gyr ago and vice versa. However, there are a few outliers to this rule and we show one prominent example for this in Fig.~\ref{fig:step042} for Coma\_042, the top-ranked best-fitting realisation, which happens to undergo rapid mass accretion in the late stage of its formation. The left panel shows its Compton-$y$ profile versus $r/R_{500c}$ with \citet{Planck2013} overlaid, and the right panel its X-ray surface brightness versus angular radius with \citet{Churazov2021} overlaid. The five colour-coded lines per panel correspond to snapshots 095--099 that trace roughly 500 Myrs, 400 Myrs, 300 Myrs, 200 Myrs and 100 Myrs prior to redshift 0. Both profiles visibly reshape within $\sim$500\,Myr, illustrating that a best-fitting match to the observed Coma profile can be transient in a cluster currently experiencing rapid mass growth. We note this is an interesting example of the radial profile variability due to the formation history of individual objects. What makes this example striking is the fact that this is our combined best fit cluster which was not the case 0.5 Gyrs ago making this target an interesting object for re-simulation at higher spatial and time resolution to study the origin of this object in greater detail.

\subsection{Maps of the best-fitting realisations}
\label{sec:bestfit_maps}

Fig.~\ref{fig:top10_maps_combined} presents the projected Compton-$y$ (top row) and X-ray surface brightness (bottom row), for the 10 best-fitting realisations identified in Sec.~\ref{sec:radial_profiles}, arranged from the best fit (top left of each row) to the 10th best (bottom right).

The Compton-$y$ maps in the top row reveal that even among the best-fitting realisations there is considerable morphological diversity. Some clusters (e.g.\ Coma\_042 and Coma\_022) display a smooth, nearly spherical tSZ signal, while others (e.g.\ Coma\_035 and Coma\_037) exhibit more elongated or irregular morphologies indicative of recent merger activity. This demonstrates that a good match to the azimuthally averaged radial profile does not uniquely determine the two-dimensional structure of the ICM: realisations with very different morphologies can produce similar one-dimensional profiles.

The X-ray maps in the bottom row show even greater morphological variation than the tSZ maps, owing to the stronger density dependence of the X-ray emissivity. Centrally peaked, relaxed systems coexist with highly disturbed configurations featuring multiple brightness peaks and asymmetric emission. The range of X-ray morphologies among the best-fitting clusters illustrates the diversity of dynamical states consistent with the observed radial profile of the Coma cluster.

\subsection{Halo mass distribution of the best-fitting realisations}
\label{sec:mass_distribution}

Fig.~\ref{fig:mass_kde} shows the distribution of halo masses across the 50 constrained realisations. The light blue shaded curve shows the KDE of $M_{200c}$ for the full sample of 50
clusters. Coloured dashed vertical lines mark the masses of the 10 best-fitting realisations (same combined ranking as in Fig.~\ref{fig:best10_profiles}), with each cluster identified by its step number in the legend. The solid black vertical line indicates the Coma cluster mass of $1.136\times10^{15}\,{\rm M}_\odot$ from the \texttt{Manticore-Local} posterior of
\citet{McAlpine2025}.

The mass distribution is broadly unimodal, with the peak near the expected Coma mass. However, the 10 best-fitting realisations do not cluster at a single mass value. Instead, they span a substantial fraction of the full mass distribution, indicating that the quality of the radial profile fit to observations is not solely determined by the total halo mass. Other factors --- including the merger history, the dynamical state at $z=0$, and the detailed thermodynamic structure of the ICM --- play an equally important role in shaping the observable radial profiles. This finding underscores the value of the constrained simulation approach: by sampling 50 realisations of the same large-scale environment, we can disentangle the effects of total mass from those of assembly history and dynamical state on the observable properties of the cluster. Nevertheless, it is interesting to note that most of the clusters that fit the Compton-$y$ and X-ray radial profile data are consistent with a lower mass estimate for Coma based on their location within the posterior distribution.

\section{Discussion}
\label{sec:discussion}

\subsection{Comparison to observations and implications}
\label{sec:disc_observations}

\paragraph*{Baseline galaxy cluster scaling relations.}
Before moving to the spatially resolved comparison with Coma, we briefly assess how the 50 constrained Coma analogues populate the standard galaxy cluster scaling relations summarised in Figs.~\ref{fig:fgas}--\ref{fig:satellite_profiles}. Generally speaking, our ensemble fares well across this set of observational relations. We stress that the scaling relations shown here are by no means a complete validation suite, but they correspond to those put forward by \citet{Nelson2024} for the TNG-Cluster sample; since we run a galaxy formation model that is very close to (although not strictly identical with\footnote{The sub-grid physics --- the dual-mode AGN feedback scheme of \citet{Weinberger2017}, the wind model of \citet{Pillepich2018}, the MHD treatment, and the cooling and chemical-enrichment tables --- is the same as in \citet{Nelson2024}. The differences relative to the production TNG-Cluster runs are practical: we use the \textsc{Arepo-2} code base described in Sec.~\ref{sec:methods}, run on a different set of (constrained) initial conditions, and at a somewhat different mass resolution. Any small numerical differences from these choices contribute, together with the difference in initial conditions, to the offsets between our ensemble and the TNG-Cluster sample discussed below.}) the TNG model used in \citet{Nelson2024}, we should, at minimum, reproduce these baseline trends. Our 50 constrained Coma analogues span a narrow window with $\log_{10}(M_{200c}/{\rm M}_\odot) \approx 14.94$--$15.16$ (median $15.07$). This narrow range is the natural consequence of focusing the suite on Coma analogues; the present comparison is therefore not an attempt to constrain galaxy-cluster scaling relations across the full mass spectrum but a focused check that our Coma realisations occupy a physically plausible location at the high-mass end of those relations, where individual cluster measurements are sparsest.

The most physically pointed offset across the scaling-relation comparisons is in the integrated thermal SZ signal (Fig.~\ref{fig:ysz_lx}, top panel): the simulated $Y_{\mathrm{SZ},500}$ values track the slope of the \citet{Planck2014} relation but with a systematic positive offset in normalisation, and our points lie close to the central-bias-corrected \citet{Hill2018} model that we re-derive in this work. Since $Y_{\mathrm{SZ},500}$ is the shell-integrated thermal energy of the ICM, a positive offset implies that we retain a slightly higher hot-gas mass inside $R_{500c}$ than the Planck normalisation prefers; the gas-fraction comparison (Fig.~\ref{fig:fgas}) tells the same story, with our points lying at the upper end of the observed range despite the substantial object-to-object scatter in the observational compilations themselves. This is consistent with a well-known characteristic of the TNG model at the cluster scale --- relative to the FLAMINGO and SLOW suites that have explicitly retuned cluster-scale AGN feedback, TNG retains more baryons inside $R_{500c}$ --- and the open question this raises (whether the TNG cluster-scale gas content is correctly calibrated, or alternatively whether the inferred Planck normalisation is driven by the assumed hydrostatic-bias factor $1-b$) is what we connect to FLAMINGO and SLOW in Sec.~\ref{sec:disc_previous_work}. We caution that a definitive statement from $Y_{\mathrm{SZ},500}$ alone is hard to make, because the inferred shell-integrated $Y$ depends sensitively on the mass-bias and pressure-profile assumptions made when comparing simulations to Planck-derived relations.

In the soft-band $L_X$--$M_{500c}$ panel (Fig.~\ref{fig:ysz_lx}, bottom), all but three of our clusters fall within the observed scatter of \citet{Pratt2009}, \citet{Vikhlinin2009}, \citet{Mantz2016}, \citet{Bulbul2019}, \citet{Lovisari2020}, and the eDXL subsample of \citet{Nagarajan2019}. The three high-$L_X$ outliers also host the three most massive BCGs in the sample (Fig.~\ref{fig:lx_mstar}), and as we show in Sec.~\ref{sec:disc_central_variability} they correspond to the strongest cool cores of the ensemble. Read this way, the apparent ``$L_X$ spread'' is not a calibration issue with the underlying physics but a direct posterior-driven signal: the constrained-IC ensemble admits a range of long-term assembly histories at fixed $z=0$ mass, and that range is reflected in the central entropy state and hence the soft X-ray luminosity of each Coma analogue. The $L_X$ scatter at fixed mass is therefore best interpreted as evidence that the central thermodynamic state of Coma carries information about its assembly history --- an observation we develop further in Sec.~\ref{sec:disc_central_variability}.

The remaining scaling-relation comparisons --- the BCG and total stellar mass content within $R_{500c}$ (Fig.~\ref{fig:stellar_mass}), the projected satellite-galaxy density profile (Fig.~\ref{fig:satellite_profiles}), and the supermassive black hole scaling relations (Fig.~\ref{fig:smbh}) --- all sit comfortably within the observational scatter and are consistent with what TNG-Cluster reports at the same mass scale; we do not see, in the high-mass-Coma regime, any single offset large enough to force a reconsideration of the underlying physics. Two narrow caveats are worth flagging. First, our BCGs sit on the upper envelope of the abundance-matching SHMR curves of \citet{Behroozi2019} and \citet{Moster2018} but on top of the cluster-BCG measurements of \citet{Kravtsov2018}; the latter comparison is the more diagnostic one because \citet{Kravtsov2018} report directly in $M_{500c}$, while the former requires a $M_{\rm vir}$- or $M_{200m}$-to-$M_{500c}$ conversion via \textsc{colossus} \citep{Diemer2018} that introduces an additional $\sim$0.05--0.1 dex systematic depending on the assumed concentration--mass relation. The apparent slight excess relative to \citet{Behroozi2019,Moster2018} should therefore not be over-interpreted. Second, the $M_{\rm BH}$--$\sigma_{\star,\mathrm{1D}}$ scatter (Fig.~\ref{fig:smbh}, right) at fixed BH mass is dominated by the orbital structure of the central galaxy and the recent merger history rather than by the dynamical influence of the BH itself, since our 50 realisations span less than a factor of two in $M_{\rm BH}$ at the BCG (cf.\ Tab.~\ref{tab:coma_properties}). The $M_{\rm BH}$--$\sigma$ comparison is therefore inherently less informative for our ensemble than for samples spanning many decades in $M_{\rm BH}$, and we read the agreement at the level of being ``in the right place'' rather than as a quantitative test.

In summary: at the $M_{\rm Coma}$ mass scale, the 50 constrained Coma analogues populate the standard galaxy cluster scaling relations as expected for a TNG-physics ensemble, with a single coherent thread of systematic offset --- a slight excess of hot gas content inside $R_{500c}$ that shows up consistently in $f_{\rm gas}$ and in $Y_{\mathrm{SZ},500}$ --- and a posterior-driven $L_X$ spread that is not a calibration issue but a signal carrying information about the assembly history. The first thread connects to the broader open question of cluster-scale AGN feedback calibration in TNG vs.\ FLAMINGO/SLOW (Sec.~\ref{sec:disc_previous_work}); the second thread connects to the central thermodynamics of Coma and is the subject of Sec.~\ref{sec:disc_central_variability}.

\paragraph*{Spatially resolved radial profiles.}
Beyond the integrated-quantity scaling relations, the constrained-IC nature of our suite enables a direct, spatially resolved comparison with the real Coma cluster. For each realisation we generate projected maps along the $z$-axis using a Voronoi ray-tracing scheme (Sec.~\ref{sec:projected_maps}); the same machinery produces Compton-$y$ and soft-band X-ray surface brightness.

In Fig.~\ref{fig:profiles_all50} we show the azimuthally averaged Compton-$y$ and X-ray surface brightness profiles for all 50 realisations, overlaid on these two reference data sets. While the official \citet{Planck2013} release (grey pentagons in Fig.~\ref{fig:profiles_all50} and in Fig.~\ref{fig:coma025_showcase}) is our primary tSZ reference, we additionally compare to the alternative Planck-based Compton-$y$ profile presented by \citet{Churazov2021}, which uses a different background-subtraction procedure (see their paper for the methodological details). Two qualitative trends are visible across the ensemble. First, for the Compton-$y$ profile, our simulated profiles tend to lie systematically above the Planck data points in the cluster centre and below them in the outskirts. The agreement with the alternative \citet{Churazov2021} reduction is in fact better in the inner $\sim 0.3\,R_{500c}$ region, suggesting that part of the central tension is driven by Planck's low angular resolution ($\sim$10\,arcmin FWHM versus $R_{500c}\sim 47$\,arcmin for Coma), which redistributes flux from the cluster centre to larger radii. The smoothing effect is directly visible in the first panel of the middle row of Fig.~\ref{fig:coma025_showcase}, where the observed Planck Compton-$y$ map appears noticeably more blurred than the matched-resolution Coma\_025 simulation panel next to it.
The same qualitative trend holds for the soft-band X-ray surface brightness profile compared to \citet{Churazov2021}: the simulated central surface brightness is on average higher than the observed Coma value, while the outskirts agree well with the background-excluded eROSITA data. The X-ray case is, however, considerably noisier than the tSZ case because of the $n_e^2$ weighting. 

Importantly, these are statements about the \emph{average} of the ensemble: a substantial fraction of our 50 realisations reproduce both profiles essentially within the observational uncertainties, while a minority produce the high central values that drive up the ensemble mean. The physical mechanism that distinguishes the well-fitting realisations from the bright-centre outliers is the central entropy and its dependence on the late-time merger history; we discuss this in detail in Sec.~\ref{sec:disc_central_variability}.

\subsection{Variability of the radial profiles in the centre}
\label{sec:disc_central_variability}

We now address the question that was deferred from the previous subsection: which physical mechanism explains why a fraction of our 50 realisations sit substantially above the observed Compton-$y$ and X-ray surface brightness profiles in the cluster centre, while the majority of clusters reproduce the observations within their scatter? A natural first hypothesis is that the over-bright cores reflect short-timescale dynamical or AGN-driven fluctuations: a central density (and hence X-ray) excess could in principle be a transient feature triggered by a recent cooling pulse, an AGN duty-cycle minimum, or a passing subhalo. We note that we tested the time variability of the radial profiles in the last 500 Myr of evolution for each cluster.

The result is unambiguous: for the vast majority of the 50 realisations, both the Compton-$y$ and the soft X-ray surface brightness profiles are remarkably stationary on this $\sim$500\,Myr baseline. Clusters that already overshoot the Planck and Churazov data in the centre at $z=0$ also overshoot them by a comparable amount 475\,Myr earlier, and the same is true in reverse for the well-fitting realisations. By visual inspection of the panel grid, $\sim$16 clusters exhibit a clear central X-ray excess relative to \citet{Churazov2021}, but in essentially all of these cases the excess was already established at the earliest snapshot we examined. We find only a small number of clusters where the central profile evolves visibly within this short window; the most striking example is Coma\_042, shown in Fig.~\ref{fig:step042}, whose central Compton-$y$ and X-ray profiles are reshaped within $\sim$500\,Myr from a sub-fitting to a near-best-fitting state. We emphasise that this evolution is \emph{not} driven by a major merger: across the full sample, major-merger events are essentially absent below $z\approx 0.3$ (cf.\ Fig.~\ref{fig:fof_mergers}), so what is happening in Coma\_042 in the last $\sim$500\,Myr is a late minor-merger / sloshing episode rather than a $z<0.3$ major-merger encounter. Crucially, this kind of late, sub-major merger-driven evolution is the exception rather than the rule. The implication is that the central excess seen in part of our ensemble is not a transient feature of the moment we observe Coma --- and, by extension, not a phenomenon that a randomly timed AGN outburst could plausibly explain. If late-time AGN activity were responsible, we would expect short-timescale variability to dominate the central profile; instead, the central state appears to be set on much longer timescales than the snapshot cadence we sample.

Having ruled out short-timescale variability, the remaining explanation must lie in the long-term thermodynamic state of the central gas. A higher central X-ray surface brightness directly implies a higher central electron density, which (for fixed temperature) implies a lower central entropy $K = kT/n_e^{2/3}$. This motivates Fig.~\ref{fig:entropy_vs_cc}, which colour-codes the radial entropy profiles of all 50 realisations by the central Compton-$y$ and X-ray surface brightness measured inside $r<0.05\,R_{500c}$. The picture is consistent: clusters with low central entropy populate the high-brightness tail of both observables, while flat, high-entropy cores are systematically faint. In other words, the centrally over-bright realisations are simply the low-entropy clusters of our ensemble. Fig.~\ref{fig:kcore_vs_brightness} quantifies this correlation by plotting the innermost-bin entropy $K_{\rm core}$ against the central Compton-$y$ (left) and against the central X-ray surface brightness (right). The X-ray correlation is very strong (Pearson $r=-0.87$, Spearman $\rho=-0.80$) and substantially tighter than the Compton-$y$ correlation. The reason is structural: the X-ray emissivity scales as $n_e^2 \Lambda(T)$, so any runaway in the central density --- the defining feature of a cool core --- is amplified, whereas the tSZ signal scales linearly in $n_e$ and hence responds only proportionally.

The next question is what \emph{causes} a subset of the constrained-IC ensemble to host cool cores at $z=0$ even though all 50 realisations share the same large-scale environment and the same TNG300 sub-grid physics. Following our results in Sec.~\ref{sec:central_scatter}, we adopt the physical low-entropy criterion $K(r)<200\,\mathrm{keV\,cm^2}$ at some $r<0.1\,R_{200c}$, which yields $N_{\rm lowK}=16$ and $N_{\rm highK}=34$. The left panel of Fig.~\ref{fig:entropy_split} shows that the low-entropy and high-entropy populations are well separated in the inner entropy profile, as expected from the classification, but converge at $r\gtrsim 0.5\,R_{500c}$. The right panel shows the same set of profiles split by terciles of the formation redshift $z_f(10\%)$, defined as the redshift at which the main progenitor first reached 10\% of its $z=0$ mass: early formers (top tercile, blue), late formers (bottom tercile, red), and the middle tercile (grey). Almost all of the low-entropy clusters fall in the early-forming tercile: the link between cool core and early assembly is one of the cleanest signals in our analysis.

This formation-time signal is corroborated by the mass accretion histories themselves (Fig.~\ref{fig:mah_panel}). The low-entropy progenitors assemble mass somewhat faster than the high-entropy progenitors during $z\approx 4\to 1$, and then taper off at late times, while the high-entropy progenitors continue to accrete substantially more mass through $z<1$. In quantitative terms, the median $z_f(10\%)$ is $1.59$ for low-entropy vs.\ $1.15$ for high-entropy clusters, a difference of $\sim$1.3\,Gyr in lookback time. The central black holes track the same trend: The BCGs of low-entropy clusters already harboured a more massive central BH by $z=1$ and grew their BH only by a factor $\sim$4 to $z=0$, while the BCGs of high-entropy clusters grew theirs by $\sim$6$\times$ over the same window. We do not, however, treat the black hole mass accretion history as an independent line of evidence: the halo and black hole accretion histories are tightly correlated in our model (log--log Pearson coefficient $0.79$), so the two largely encode the same information about early assembly. In fact, it is almost the other way round and the strong correlation of these two metrics actually implies that the X-ray and Compton-$y$ profiles are more sensitive to the exact accretion history of an individual massive cluster, rather than to random AGN-feedback episodes. 

The natural follow-up question is which channel of mass accretion drives the early-formation difference. Fig.~\ref{fig:fof_mergers} shows the mean number of FoF merger events per cluster per redshift bin in four ratio classes (major, $r\geq 1/3$; minor, $1/10\leq r<1/3$; mini, $1/100\leq r<1/10$; smooth, $r<1/100$). The major-merger counts alone are statistically indistinguishable between the low-entropy and high-entropy populations; the channel that does discriminate is the mini-merger channel, with low-entropy clusters experiencing roughly twice as many mini-merger events as high-entropy clusters. Translated into a mass budget, this means that high-entropy clusters preferentially assemble their mass through the combined ``mini + smooth'' channel rather than through major mergers. Fig.~\ref{fig:entropy_mass_budget} therefore separates the 50 realisations into a ``soft-grown'' set ($N=18$), defined as those that have built more than 60\% of their final $M_{200c}$ from the combined mini + smooth channel, and a ``merger-grown'' set ($N=32$). 12 of our 16 low-entropy clusters are soft-grown (75\% of all low-entropy clusters).

It is, however, important to be precise about the strength of this connection. The soft-grown criterion is not a pure low-entropy selector: of the 18 soft-grown clusters, 12 are of low central entropy and 6 are of high central entropy, so applying the criterion to the full sample picks up all 12 soft-grown low central entropy clusters but also 6 contaminating high central entropy clusters. Conversely, four of the 16 low-entropy clusters (notably Coma\_000 and Coma\_028 as the largest outliers) were built primarily through major mergers and yet host a cool core at $z=0$; these are the cases where late-time core regulation has overruled the major-merger heating that would normally erase a cool core. The picture that emerges from Figs.~\ref{fig:entropy_split}--\ref{fig:entropy_mass_budget} is therefore that early-formation through smooth and mini-merger accretion is a strongly favoured but not strictly necessary route to a cool core in our ensemble. Bearing in mind these caveats, it remains the case that the cool cores in our sample come predominantly from soft-growth assembly histories (Fig.~\ref{fig:entropy_mass_budget}), and we believe this is the central physical message that we attempt to convey by this analysis: the diversity of central radial profiles seen across the constrained ensemble is not driven by short-timescale fluctuations or by AGN bursts, but by the long-term assembly mode of the cluster, and  specifically, by how much of the final mass was deposited in the central region through a continuous, low-perturbation channel before $z\approx 1$.

\subsection{Implications for constrained-inference likelihoods}
\label{sec:disc_inference_likelihood}

A natural way to read the diversity described above is in the context of the inference framework that produced our initial conditions. The \texttt{Manticore-Local} posterior of \citet{McAlpine2025} is constrained by present-day galaxy-field information from the 2M++ survey, which tightly fixes the mass, location, and large-scale environment of Coma but does not directly probe its formation history. \citet{McAlpine2025} explicitly demonstrated that constrained clusters in the \texttt{Manticore-Local} suite carry some formation-history constraining power relative to a random-IC sample of the same mass --- the assembly histories are not uniquely fixed but they are statistically narrower than for random clusters at the same $z=0$ mass. The 50-sample posterior ensemble we present here is the natural continuation of that picture: the posterior fixes the mass and environment well enough that all 50 realisations look like Coma at $z=0$ in their bulk properties, but the residual freedom in the assembly history is large enough to produce, at fixed mass and environment, a low-entropy fraction of $32\%$, a soft-band $L_X$ spread of factor $\gtrsim 5$, and a central-Compton-$y$ spread of factor $\sim 3$ across the ensemble.

What Sec.~\ref{sec:disc_central_variability} adds to this picture is the observation that these spreads are not random nuisance noise on top of a fixed Coma mass: they are tightly correlated with the long-term assembly history of each Coma analogue, through the central-entropy state of the ICM. Concretely, central X-ray surface brightness, central Compton-$y$, and the central entropy floor all encode information about the formation redshift of the main progenitor and about the relative balance of major-merger versus mini-merger and smooth-accretion mass growth. These quantities are, in principle, directly observable for the real Coma cluster --- the eROSITA X-ray surface brightness profile of \citet{Churazov2021}, the Planck thermal-SZ profile, and the inferred central entropy from joint X-ray/SZ analyses are all well measured.

This points to a concrete extension of the constrained-inference programme. The current \texttt{Manticore-Local} likelihood is primarily a galaxy-field likelihood and does not exploit cluster-scale thermodynamic observables at all. Adding a likelihood term that compares model-predicted Compton-$y$ profiles or soft-band X-ray surface brightness profiles to their observed Coma counterparts --- using the present suite (or a future expansion of it) as a forward-model emulator basis --- would in principle narrow the posterior in two complementary ways: (i) tightening the constraint on Coma's $z=0$ thermodynamic state, and (ii) tightening the posterior on the assembly history through the central-entropy / formation-history connection identified in Sec.~\ref{sec:disc_central_variability}. The expected payoff is a posterior that selects realisations consistent not only with Coma's present-day mass and large-scale environment but also with its present-day central thermodynamic state, and by extension with a tighter assembly-history range than is currently admitted. We see this as a natural avenue for follow-up work, and the constrained-IC suite presented here provides a starting point for the forward-model emulation that such an extended likelihood would require.

\subsection{Comparison to previous work}
\label{sec:disc_previous_work}

Random-IC large-box hydrodynamical simulations --- TNG300 \citep{Pillepich2018,Nelson2019}, MillenniumTNG \citep{Pakmor2023}, FLAMINGO \citep{Schaye2023,Kugel2023}, Magneticum \citep{Hirschmann2014} and EAGLE \citep{Schaye2015} --- deliver Coma-mass clusters as a by-product of sampling the full mass function; the stacked Coma-mass Compton-$y$ profile from MillenniumTNG \citep{Pakmor2023} is the red reference line in Figs.~\ref{fig:profiles_all50} and \ref{fig:best10_profiles}. The most directly comparable big-box effort is, however, the SLOW (Simulating the LOcal Web) project of \citet{Dolag2023}, which evolves a $(500\,h^{-1}\,\mathrm{Mpc})^3$ constrained-IC volume with P-Gadget3 ($2\times 1536^3$ particles, gas mass $4.6\times 10^{8}\,h^{-1}M_\odot$, radiative cooling, star formation, AGN feedback and cosmic rays). The SLOW initial conditions are built from CosmicFlows-2 peculiar velocities using the Hoffman--Ribak constrained-realisation algorithm with reverse Zel'dovich progenitor relocation, and the resulting volume recovers the Virgo void ($\sim 50\%$ underdensity within 16\,Mpc), the Virgo supercluster overdensity, an excess of $M_{\rm vir}>10^{15}\,{\rm M}_\odot$ clusters within 200\,Mpc, and an overall $\sim 5\%$ large-scale underdensity --- all of which are features that our own \texttt{Manticore}-derived IC also reproduces and that we therefore inherit by construction. The SLOW Coma analogue, identified by \citet{HernandezMartinez2024} as one of 45 local cluster analogues in the box, carries $M_{500c}\simeq 9.6\times 10^{14}\,{\rm M}_\odot$ (against the X-ray-derived $9.95^{+2.10}_{-2.99}\times 10^{14}\,{\rm M}_\odot$) and $M_{\rm vir}\simeq 1.8\times 10^{15}\,{\rm M}_\odot$ (against the observed $\sim 1.4\times 10^{15}\,{\rm M}_\odot$). Its ICM temperature is $7.06$\,keV (consistent with the observed $8.07\pm 0.29$\,keV) but its bolometric X-ray luminosity and Compton-$y$ overshoot the observed values by factors of $\sim 1.5$ and $\sim 3$ respectively --- the same central-excess pattern that we identify across the brighter end of our own ensemble (Sec.~\ref{sec:disc_central_variability}). \citet{Groth2025} have used the same SLOW Coma analogue as the simulation counterpart for XRISM line-broadening predictions: their Coma counterpart contains two BCG analogues and requires a two-temperature \textsc{bapec} model ($\sigma_{z,1}=77$, $\sigma_{z,2}=668\,\mathrm{km\,s^{-1}}$) where the actual XRISM Coma spectrum is well fit with a single Gaussian at $\sigma_z=208\pm 12\,\mathrm{km\,s^{-1}}$ \citep{XRISM2025_sims}, even though the SLOW central filtered turbulent velocity ($\sim 600\,\mathrm{km\,s^{-1}}$) is in good agreement with the XRISM-derived value. The discrepancy is most plausibly that the single SLOW Coma counterpart is in a more disturbed late-merger configuration than the actual cluster --- a sensitivity to the single-realisation choice from which a posterior-sampled ensemble like ours is by construction protected.

When competitive resolution at the cluster scale is required, the standard alternative to large boxes is the zoom-in resimulation of cluster-mass haloes drawn from a parent box. C-EAGLE / Hydrangea \citep[30 clusters,][]{Bahe2017,Barnes2017}, the Three Hundred Project \citep[324 clusters,][]{Cui2018,Cui2022} and the recent TNG-Cluster suite \citep[352 clusters with the IllustrisTNG model,][]{Nelson2024} are the cluster-zoom samples most relevant for the present work. None of these use constrained ICs and so contain no specific Coma counterpart; we therefore use them only as scaling-relation validation sets, with the TNG-Cluster sample of \citet{Nelson2024} serving as the closest baseline against which to check our own ensemble in Sec.~\ref{sec:disc_observations}. Since our 50 realisations run a galaxy-formation model that is essentially the IllustrisTNG model, the satisfactory agreement we find with the TNG-Cluster scaling relations is a sanity check rather than a new result.

On the local-IC side, the CLUES collaboration \citep{Gottloeber2010,Yepes2014} pioneered the dark-matter-only branch of constrained local-Universe simulations: a peculiar-velocity catalogue (variants of \textsc{cosmicflows}; \citealt{Tully2013}) is Wiener-filtered to obtain a noise-free linear density-field reconstruction and propagated backwards via the reverse Zel'dovich approximation \citep[RZA;][]{Doumler2013,Sorce2016} to produce constrained initial conditions. The IC step of the SLOW programme described above is methodologically a member of this same CLUES family. The CLUES-style ICs have also been used by Sorce and collaborators to seed full hydrodynamical zoom resimulations of a Virgo counterpart with the \textsc{ramses} adaptive-mesh-refinement code (CLONE: Constrained LOcal and Nesting Environment; \citealt{Sorce2021,Sorce2026}), down to a minimum cell size of $\sim 350$\,pc with AGN and supernova feedback. The CLONE strategy is to build 200 peculiar-velocity-constrained IC realisations of the local Universe and pick the single realisation whose Virgo counterpart most closely matches the average properties of the 200-member ensemble for the actual hydrodynamical zoom. The resulting Virgo has $M_{\rm vir}\simeq 5\times 10^{14}\,{\rm M}_\odot$ and $R_{\rm vir}\simeq 2$\,Mpc, reproduces the late-time minor-merger history that explains Virgo's $z=0$ relaxed state, and has been used to forward-model the observed Virgo galaxy population at the satellite-by-satellite level \citep{Sorce2026}. CLONE and our work are conceptually close (a hydrodynamical resimulation of a specific named local cluster) but methodologically different: CLONE selects a single best-fitting realisation, while we run all 50 posterior samples end-to-end.

The constrained-IC suite presented here sits on the other methodological branch of the local-Universe programme: full Bayesian field-level inference, initiated with the \texttt{BORG} algorithm of \citet{Jasche2013,Jasche2019}. \texttt{BORG} couples a non-linear gravitational solver, a calibrated galaxy bias model and physics-informed priors directly to the observed galaxy catalogue and produces a \emph{posterior} over initial-condition realisations of the local Universe, each of which is internally self-consistent with $\Lambda$CDM and with the data. The DM-only SIBELIUS-DARK simulation of \citet{McAlpine2022} demonstrated that forward-evolving constrained ICs of this type recovers the observed local volume out to $\sim 200$\,Mpc with the most massive nearby clusters reproduced on an object-by-object basis, and exhibits the $\sim 5\%$ large-scale underdensity that is also seen in SLOW. The most recent member of the lineage, \texttt{Manticore} \citep{McAlpine2025}, performs the full \texttt{BORG} field-level inference on the 2M++ catalogue inside an $L=1000$\,Mpc box at $256^3$ grid resolution (DES~Y3 cosmology; $h=0.681$, $\Omega_m=0.306$, $\sigma_8=0.807$), and identifies $14$ named local clusters --- including Coma --- at high posterior significance. \texttt{Manticore-Local} achieves the highest Bayesian evidence among all current local-Universe reconstructions when scored against five independent peculiar-velocity data sets, and produces an ensemble of posterior white-noise realisations that can be forward-evolved with any chosen $N$-body or hydrodynamical code; the Coma posterior mass $M_{200c}=1.136\times 10^{15}\,{\rm M}_\odot$ that we use as the reference value in Fig.~\ref{fig:mass_kde} comes from this reconstruction.

The 50 realisations of the present work are drawn as independent samples from this same \texttt{BORG} posterior, each forward-evolved at fully-resolved zoom resolution with the TNG300 galaxy-formation model. The resulting $M_{200c}$ scatter of $0.87$--$1.45\times 10^{15}\,{\rm M}_\odot$ (median $1.16\times 10^{15}\,{\rm M}_\odot$; Fig.~\ref{fig:mass_kde}) is not a calibration uncertainty but the physical scatter that the local-Universe data actually leave unfixed at the Coma scale, and brackets the \texttt{Manticore-Local} point estimate exactly. Where CLONE and SLOW each commit to a single best-fitting cluster counterpart per object, we resample the constrained-IC posterior 50 times and run each sample through the same galaxy-formation model and forward-modelling pipeline. This is what lets us cleanly separate the part of the Coma observables that is determined by the constrained local environment from the part that is set by the residual stochastic and assembly-history scatter.

\section{Conclusions}
\label{sec:conclusions}

We have presented a suite of 50 constrained-IC zoom-in simulations of the Coma galaxy cluster, drawn as independent samples from the \texttt{BORG/Manticore} posterior of the local Universe and run end-to-end with a galaxy-formation model that closely tracks IllustrisTNG. The main findings are:

\begin{enumerate}
    \item The 50 realisations populate a $M_{200c}$ range of $0.87$--$1.45\times 10^{15}\,{\rm M}_\odot$ with median $1.16\times 10^{15}\,{\rm M}_\odot$, bracketing the \texttt{Manticore-Local} posterior mass for Coma ($1.136\times 10^{15}\,{\rm M}_\odot$; Fig.~\ref{fig:mass_kde}). The corresponding $M_{500c}$ range $5.7$--$11.2\times 10^{14}\,{\rm M}_\odot$ is consistent with the X-ray-derived Coma mass.
    \item The ensemble satisfies the standard cluster scaling-relation tests at the Coma mass scale (Figs.~\ref{fig:fgas}--\ref{fig:satellite_profiles}), with the only systematic offset being a mild excess in the shell-integrated $Y_{\mathrm{SZ},500}$ relative to the \citet{Planck2014} normalisation, consistent with the documented TNG--FLAMINGO gas-fraction offset at the cluster scale.
    \item Azimuthally averaged Compton-$y$ and soft-band X-ray surface brightness profiles (Fig.~\ref{fig:profiles_all50}) bracket the \citet{Planck2013} and \citet{Churazov2021} Coma profiles, with a subset of realisations overshooting both observables in the inner $r\lesssim 0.3\,R_{500c}$ and the remainder agreeing within the observational scatter. The top-ranked realisation, Coma\_042, matches both profiles simultaneously.
    \item The central excess in the bright-core realisations is not a transient feature: profiles are stable over the last $\sim 500$\,Myr in essentially all clusters, ruling out short-timescale AGN-burst or sloshing-driven variability as the cause.
    \item The bright-core realisations are the low-entropy clusters of the ensemble. The innermost-bin entropy $K_{\rm core}$ correlates tightly with the central X-ray surface brightness (Pearson $r=-0.87$) and more weakly with the central Compton-$y$, as expected from the $n_e^2$ vs.\ $n_e$ weighting (Fig.~\ref{fig:kcore_vs_brightness}). Sixteen of the 50 realisations satisfy the $K(r)<200\,\mathrm{keV\,cm^2}$ low-entropy criterion.
    \item Low-entropy clusters form earlier than high-entropy clusters: median $z_f(10\%)=1.59$ vs.\ $1.15$ (a $\sim 1.3$\,Gyr difference in lookback time). The corresponding central black holes in low-entropy BCGs were already more massive by $z=1$ and grew only $\sim 4\times$ to $z=0$, against $\sim 6\times$ for high-entropy BCGs; the black hole and halo accretion histories are tightly correlated (log--log Pearson coefficient $0.79$).
    \item The accretion channel that appears to mark formation of low-entropy clusters from high-entropy clusters is predominant mass accretion through small mergers and smooth accretion. The most predictive low-entropy cluster classifier we find is the soft-growth criterion: clusters that built more than 60\% of their final $M_{200c}$ from the combined mini-plus-smooth channel host 12 of the 16 low-entropy clusters (Fig.~\ref{fig:entropy_mass_budget}). Two low-entropy cluster outliers (Coma\_000 and Coma\_028) built their mass primarily through major mergers and yet retained a cool core, indicating that late-time core regulation can change the picture in individual cases.
    \item The 10 realisations that best fit the observed Coma radial profiles (Fig.~\ref{fig:best10_profiles}) span essentially the full $M_{200c}$ range of our ensemble (Fig.~\ref{fig:mass_kde}), so the goodness of profile fit is not determined by total halo mass alone. The same 10 realisations exhibit a wide range of two-dimensional Compton-$y$ and X-ray morphologies (Fig.~\ref{fig:top10_maps_combined}), so a good one-dimensional radial-profile match does not uniquely determine the two-dimensional structure of the ICM.
    \item The 50-realisation mass scatter is a physical posterior scatter, not a calibration uncertainty: it reflects the residual freedom in the local-Universe data at the Coma scale once the constrained large-scale modes have been imposed. Properties that converge across the 50 realisations are strongly environment and accretion history driven.
\end{enumerate}

Two natural extensions of this work are already underway. The companion Paper II forward-models the XRISM Resolve pipeline onto all 50 realisations and uses the resulting posterior set of line-broadening predictions to interpret the observed Coma line widths in the context of the cluster's assembly history. Furthermore, the kSZ maps of the 50 realisations, will be analysed against upcoming high-angular-resolution CMB experiments in a future paper.

\section*{Acknowledgements}

We thank the members of the LtU collaboration for useful input on the progression of this project. UPS thanks Tom Abel, Klaus Dolag, Laura Sommovigo and Romain Teyssier for useful comments. We thank the eROSITA consortium for allowing the usage of the X-ray map from \citet{Churazov2021}. The simulations have been carried out on the Max Planck Computing and Data Facility (MPCDF) machines raven and viper in Garching and we acknowledge computing time granted by these resources and their Staff for keeping the machines tidy. RS acknowledges financial support from the CCA Pre-doctoral Program, STFC Grant No. ST/X508664/1, the Snell Exhibition of Balliol College, Oxford, and a Hintze Fellowship at the Oxford Centre for Astrophysical Surveys, funded through generous support from the Hintze Family Charitable Foundation. IK was supported by the Simons Foundation via the Simons Investigator Award to A. A. Schekochihin. GLB acknowledges support from the NSF (AST-2307419), NASA TCAN award 80NSSC21K1053, and the Simons Foundation through the Learning the Universe Collaboration. The authors used OpenAI GPT5.5 and Claude Opus 4.8 to refine code written for this work and portions of the draft itself. The authors take full responsibility for the final content.

\section*{Data Availability}

The data will be made available based on reasonable request to the corresponding author.



\bibliographystyle{mnras}
\bibliography{example} 




\appendix

\section{Contamination statistics and the worst-case cluster}
\label{app:contamination}

\begin{figure}
    \centering
    \includegraphics[width=0.49\textwidth]{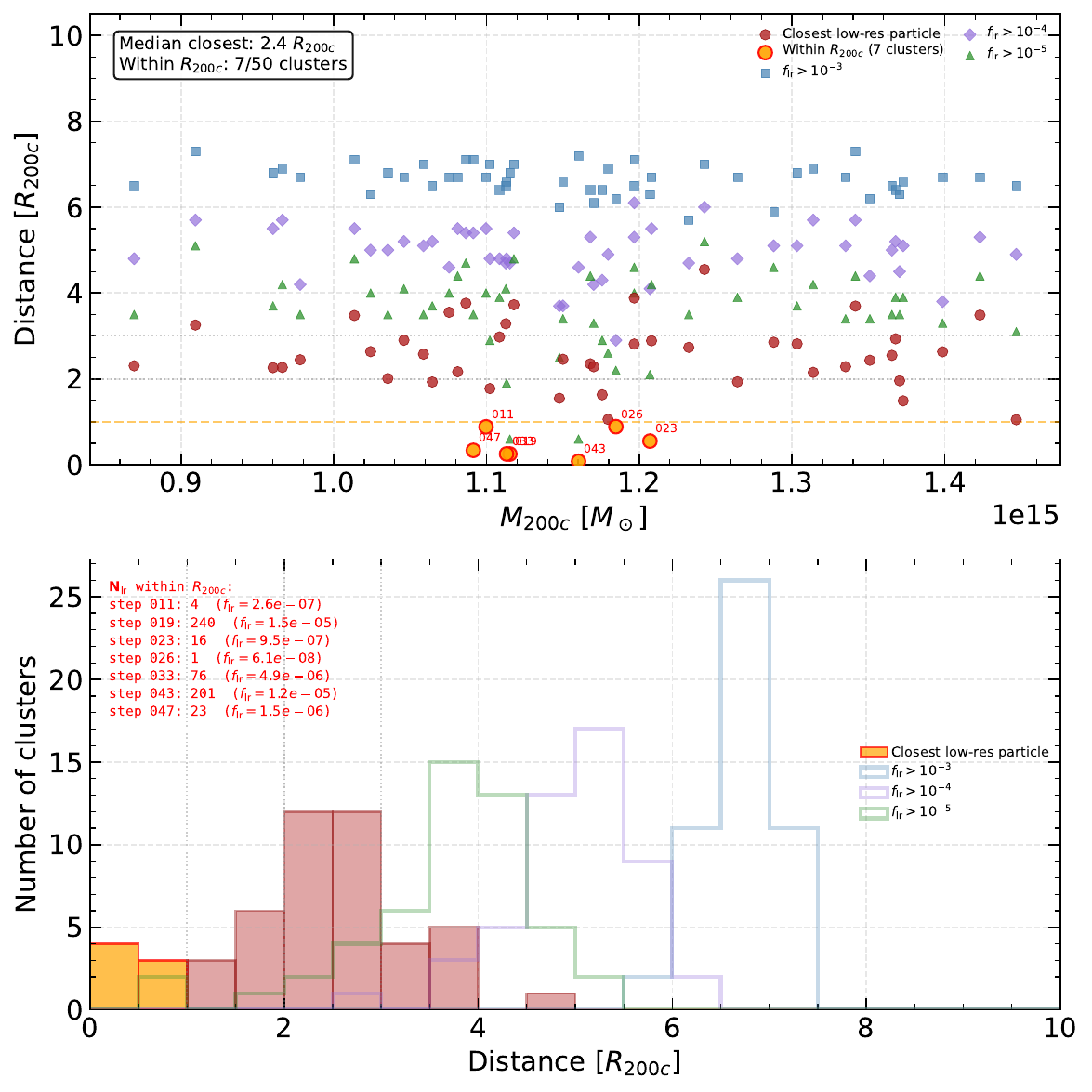}
    \caption{Contamination diagnostic for the 50 constrained Coma realisations. \textit{Top:} distance from the halo centre to the nearest low-resolution (parent-box) particle, in units of $R_{200c}$, plotted as a function of $M_{200c}$. Marker colour encodes the low-resolution mass fraction $f_{\rm lr}$ inside $R_{200c}$; clusters with a low-resolution particle inside $R_{200c}$ are highlighted in orange and labelled by cluster ID. The horizontal dashed line marks the $R_{200c}$ threshold. \textit{Bottom:} histogram of the same nearest-low-resolution-particle distance, with the seven $R_{200c}$-contaminated clusters listed by ID together with their particle count and $f_{\rm lr}$.}
    \label{fig:contamination}
\end{figure}

\begin{figure}
    \centering
    \includegraphics[width=0.49\textwidth]{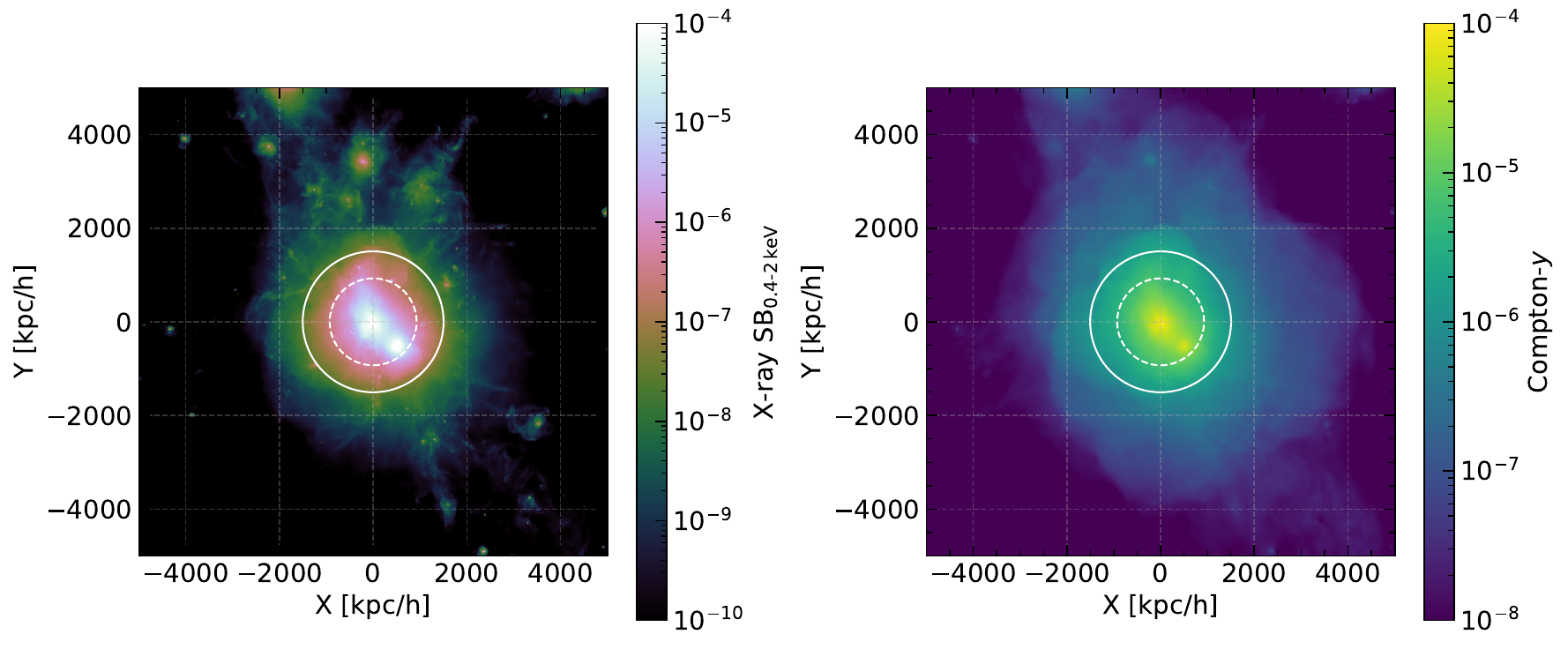}
    \caption{Projected maps of the worst-contamination realisation Coma\_043 ($f_{\rm lr}=1.5\times 10^{-4}$, 291 low-resolution particles inside $R_{200c}$, active major merger with mass ratio $\sim 0.22$). \textit{Left:} X-ray surface brightness in the 0.4--2\,keV band. \textit{Centre:} Compton-$y$ parameter. Each panel spans a $10\times 10\,\mathrm{Mpc}$ field of view; the white circle marks $R_{500c}$. The contaminating particles sit well outside the central observable region and do not affect the ICM diagnostics presented in the main text.}
    \label{fig:worst_case_cluster}
\end{figure}

Because the 50 Coma realisations are zoom-in resimulations carved from the parent \texttt{BORG/Manticore}-derived volume, the high-resolution Lagrangian region is in principle vulnerable to mixing with the surrounding low-resolution particles. We quantify this contamination by tagging every parent-box (low-resolution) particle in the final snapshot and computing, for each cluster, both the distance from the halo centre to the nearest low-resolution particle and the low-resolution mass fraction $f_{\rm lr}$ inside $R_{200c}$. Across the 50 realisations the median nearest-low-resolution-particle distance is $2.4\,R_{200c}$, and only 7 of 50 clusters have any low-resolution particle inside $R_{200c}$ at all (Fig.~\ref{fig:contamination}). The seven clusters are Coma\_011, 013, 018, 026, 028, 040 and 043; in all seven the contaminating mass fraction is $f_{\rm lr}\lesssim 1.5\times 10^{-4}$ and the absolute count of low-resolution particles inside $R_{200c}$ is at most $\sim 300$. None of the seven cases is large enough to affect the integrated cluster properties at a level relevant for any of the analyses in the main text. The worst case, Coma\_043 ($f_{\rm lr}=1.5\times 10^{-4}$, 291 particles inside $R_{200c}$), is an active major-merger configuration (mass ratio $\sim 0.22$); we show its projected X-ray and Compton-$y$ in Fig.~\ref{fig:worst_case_cluster} to demonstrate that even in this extreme case the contaminating particles sit well outside the central observable region and do not corrupt the ICM diagnostics.

\bsp    
\label{lastpage}
\end{document}